\documentclass[11pt]{article}
\usepackage[utf8]{inputenc}
\pdfoutput=1
\usepackage{setspace}
\usepackage{palatino}
\usepackage{graphicx}
\usepackage{subfigure}
\usepackage{float}
\usepackage{titling} 
\usepackage{multirow}
\usepackage{lscape}
\usepackage{amsmath}
\usepackage{amssymb}
\usepackage[a4paper, total={6in, 9.5in}]{geometry}
\fontfamily{ppl}\selectfont 
\usepackage{eqparbox}
\usepackage{arydshln}

\usepackage[american]{babel}

\usepackage{bm}
\usepackage{caption}
\usepackage{todonotes}
\usepackage[normalem]{ulem}
\usepackage[natbibapa]{apacite}
\PassOptionsToPackage{hyperindex,breaklinks}{hyperref}
\usepackage{hyperref}
\usepackage{xcolor}
\definecolor{darkblue}{rgb}{0, 0, 0.5}
\hypersetup{colorlinks=true,citecolor=darkblue, linkcolor=darkblue, urlcolor=darkblue}

\title{A comparison of latent semantic analysis and correspondence analysis of document-term matrices}

\author{Qianqian Qi\\
   {\small\raggedright Utrecht University}\\
   \href{mailto:q.qi@uu.nl}{\texttt{q.qi@uu.nl}} 
\and David J. Hessen\\
   {\small\raggedright Utrecht University}\\
   \href{mailto:d.j.hessen@uu.nl}{\texttt{d.j.hessen@uu.nl}}
\and Tejaswini Deoskar\\
   {\small\raggedright Utrecht University}\\
   \href{mailto:t.deoskar@uu.nl}{\texttt{t.deoskar@uu.nl}}
\and Peter G. M. van der Heijden\\
    {\small\raggedright Utrecht University and University of Southampton}\\
   \href{mailto:P.G.M.vanderHeijden@uu.nl}{\texttt{P.G.M.vanderHeijden@uu.nl}}
    }
    
\date{}
\date{\vspace{-5ex}}

\renewenvironment{abstract}
 {\par\noindent\textbf{\abstractname: }\ \ignorespaces}
 {\par\medskip}

\begin{document}

{\setstretch{.8}
\maketitle
\begin{abstract}
Latent semantic analysis (LSA) and correspondence analysis (CA) are two techniques that use a singular value decomposition (SVD) for dimensionality reduction. LSA has been extensively used to obtain low-dimensional representations that capture relationships among documents and terms. In this article, we present a theoretical analysis and comparison of the two techniques in the context of document-term matrices. We show that CA has some attractive properties as compared to LSA, for instance that effects of margins, i.e. sums of row elements and column elements, arising from differing document-lengths and term-frequencies are effectively eliminated
,  so that the CA solution is optimally suited to focus on relationships among documents and terms. A unifying framework is proposed that includes both CA and LSA as special cases. We empirically compare CA to various LSA based methods on text categorization in English and authorship attribution on historical Dutch texts, and find that CA performs significantly better.  We also apply CA to a long-standing question regarding the authorship of the Dutch national anthem {\it Wilhelmus} and provide further support that it can be attributed to the author Datheen, amongst several contenders.

\noindent
\textbf{Keywords: }
Latent semantic analysis; Correspondence analysis; Singular value decomposition; Text categorization; Authorship attribution.
\end{abstract}
}

\section{Introduction}

Latent semantic analysis (LSA) is a well-known method used in computational linguistics that uses singular value decomposition (SVD) for dimensionality reduction in order to extract contextual and usage-based representations of  words from textual corpora \citep{landauer1997solution, jiao2021brief}. We focus here on LSA of document-term matrices; the rows of the document-term matrix correspond to the documents and the columns to the terms, and the elements are frequencies, i.e. the number of occurrences of each term in each document. Documents may have different lengths and margins of documents refer to the marginal frequencies of documents, namely the sum of each row of the document-term matrix; also, terms may be more or less often used and margins of terms refer to the marginal frequencies of terms, namely the sum of each column of the document-term matrix.

Amongst many other tasks \citep{di2019effectiveness, tseng2019integrating, phillips2021comparing, Hassani2021, ren2021sleep, 9395976, kalmukov2022comparison}, LSA has been used extensively for information retrieval \citep{zhang2011comparative, patil2022word}, by using associations between documents and terms \citep{dumais1988using, deerwester1990indexing, dumais1991improving}. The exact factorization achieved via SVD  has been shown to achieve solutions  comparable in some ways to those obtained by modern neural network based techniques \citep{NIPS2014_feab05aa,  levy2015improving}, commonly used to obtain dense word representations from textual corpora \citep{jurafsky2021speech}.

Correspondence analysis (CA) is a popular method for the analysis of contingency tables \citep{greenacre1984theory, greenacre2017correspondence, hou_huang_2020, van2021correspondence}. It provides a graphical display of dependence between rows and columns of a two-way contingency table \citep{greenacre1987geometric}. Like LSA, CA is a dimensionality reduction method.  The methods have much in common as both use SVD. In both cases, after dimensionality reduction, many text mining tasks, such as text clustering, may be performed in the reduced dimensional space rather than in the higher dimensional space provided by the raw document-term matrix.

While a few empirical comparisons of LSA and CA, with mixed results, can be found in the literature, a comprehensive theoretical comparison is lacking. For example, \citet{morinknowledge} compared the two methods in the automatic exploration of themes in texts. \citet{seguela2011comparison} compared the performance of CA and LSA with several weighting functions in a document clustering task, and found that CA gave better results. On the other hand, \citet{seguela:hal-01126258}  compared the performance of CA and LSA with TF-IDF on a recommender system, but found that CA performs less well.

The present article presents a theoretical comparison of the two techniques, and places them in a unifying framework. We show that CA has some favourable properties over LSA, such as a clear interpretation of the distances between documents and between terms of the original matrix, and a clear relation to statistical independence of documents and terms. Also, CA can eliminate the margins of documents and terms simultaneously. Second, we empirically evaluate and compare the two techniques, by applying them to text categorization and authorship attribution in two languages. For text categorization we use the BBCNews, BBCSport, and 20 Newsgroups datasets in English. In authorship attribution we evaluate the two techniques on a large set of historical Dutch texts written by six well-known Dutch authors of the sixteenth century. Here, we additionally use CA to determine the unknown authorship of \textit{Wilhelmus}, the national anthem of the Netherlands, whose authorship is controversial: CA attributes \textit{Wilhelmus} to the author Datheen, out of the six contemporary contenders. To the best of our knowledge, this is the first application of CA to the \textit{Wilhelmus}. In both cases, we find that CA performs better.

The rest of the article is organized as follows.  Section~\ref{S:LSA} and Section~\ref{S:CA} elaborate on the techniques LSA and CA in turn. A unifying framework is proposed in Section~\ref{S:U}. In Section \ref{DCpar} we compare LSA and CA in text categorization using the BBCNews, BBCSport, and 20 Newsgroups datasets.  Section~\ref{S:ER} evaluates the performance of LSA and CA for authorship attribution of documents where the author is known, and of the \textit{Wilhelmus}, whose author is unknown. The article ends with a conclusion.

\section{Latent semantic analysis}\label{S:LSA}

Latent Semantic Analysis (LSA) has been extensively used for improving information retrieval by using the associations between documents and terms \citep{dumais1988using, deerwester1990indexing}, amongst many other tasks. Since individual terms provide incomplete and unreliable evidence about the meaning of a document, in part due to synonymy and polysemy, individual terms are replaced with derived underlying (latent) semantic factors. Although LSA is a very well-known technique, we first present a detailed analysis of the mathematics involved in LSA here as this is usually not found in the literature, and in a later section, it will help in making the comparison between LSA and CA explicit. We start with LSA of the raw document-term matrix and then discuss LSA of weighted matrices. The weighted matrices we study here include (i) a matrix with row-normalized elements with L1, i.e. for each row the elements are divided by the row sum (the L1 norm), so that the sum of the elements of each row is 1; (ii) a matrix with row-normalized elements with L2, i.e. for each row the elements are divided by the square root of sum of squares of these elements (the L2 norm), so that the sum of squares of the elements of each row is 1; (iii) and a matrix that is transformed by term frequency-inverse document frequency (TF-IDF).

The discussion is illustrated using a toy data set, with the aim to present a clear view of the properties of the dataset captured by  LSA and CA, see Table~\ref{T6*6dt}. The toy data set has 6 rows, the documents, and 6 columns, the terms, with the frequency of occurrence of terms in each document in the cells \citep{aggarwal2018machine}. Based on term-frequencies in each document, the first three documents can be considered to primarily refer to {\em cats}, the last two primarily to {\em cars}, and the fourth document to both. The fourth term, {\em jaguar}, is polysemous because it can refer to either a cat or a car. We will see below how the LSA approaches, and later CA, represent these properties in the data. 

\begin{table}[htbp]
\centering  
\caption{A document-term matrix $\bm{F}$: size 6$\times$6} 
\label{T6*6dt}
\begin{tabular}{ccccccc}    
\hline
 & lion & tiger & cheetah &jaguar & porsche & ferrari\\  
\hline  
doc1 & 2 & 2 & 1 & 2 & 0 & 0 \\ 
doc2 & 2 & 3 & 3 & 3 & 0 & 0 \\  
doc3 & 1 & 1 & 1 & 1 & 0 & 0\\ 
doc4 & 2 & 2 & 2 & 3 & 1 & 1\\ 
doc5 & 0 & 0 & 0 & 1 & 1 & 1\\ 
doc6 & 0 & 0 & 0 & 2 & 1 & 2\\ 
\hline 
\end{tabular}  
\end{table}

\subsection{LSA of raw document-term matrix}\label{SubS:LSA}

LSA is an application of the mathematical tool SVD, and can take many forms, depending on the matrix analyzed. We start our discussion of LSA with the SVD of a raw document-term matrix $\bm{F}$, having size  $m \times n$, with elements $f_{ij}$, $i=1,...,m$ and $j=1,...,n$  \citep{berry1995using, deisenroth2020mathematics}. Without loss of generality we assume that $n \geq m$ and $\bm{F}$ has full rank. 

SVD can be used to decompose $\bm{F}$ into a product of three matrices: $\bm{U}^f$, $\bm{\Sigma}^f$, and $\bm{V}^f$, namely
\begin{equation}
\label{SVD}
\bm{F} = \bm{U}^f\bm{\Sigma}^f(\bm{V}^f)^T
\end{equation}
Here $\bm{U}^f$ is a $m \times m$ matrix with orthonormal columns called left singular vectors so that $(\bm{U}^f)^T\bm{U}^f=\bm{I}$, $~\bm{V}^f$ is a $n \times m$ matrix with orthonormal columns called right singular vectors so that  $(\bm{V}^f)^T\bm{V}^f = \bm{I}$, and $\bm{\Sigma}^f$ is a $m \times m$ diagonal matrix with singular values on the diagonal in descending order.

We denote the first $k$ columns of $\bm{U}^f$ as the $m\times k$ matrix $\bm{U}^f_k$, the first $k$ columns of $\bm{V}^f$ as the $n\times k$ matrix $\bm{V}^f_k$, and the $k$ largest singular values on the diagonal of $\bm{\Sigma}^f$ as the $k \times k$ matrix $\bm{\Sigma}^f_k$ ($k\leq m$). Then $\bm{U}^f_k\bm{\Sigma}^f_k(\bm{V}^f_k)^T$ provides the optimal rank-$k$ approximation of $\bm{F}$ in a least-squares sense. That is, $\bm{X} = \bm{U}^f_k\bm{\Sigma}^f_k(\bm{V}^f_k)^T$ minimizes  Equation (\ref{miniSVD}) amongst all matrices $\bm{X}$ of rank $k$:
\begin{equation}
\label{miniSVD}
	||\bm{F}-\bm{X}||^2_F
	=\sum_i\sum_j(f_{ij}-x_{ij})^2
\end{equation}
 The idea is that the matrix $\bm{U}^f_k\bm{\Sigma}^f_k(\bm{V}^f_k)^T$ captures the major associational structure in the matrix and throws out noise  \citep{dumais1988using, dumais1991improving}. The total sum of squared singular values is equal to $\textup{tr}((\bm{\Sigma}^f)^{2})$, where $\textup{tr}$ is the sum of elements on the main diagonal of a square matrix. The proportion of the total sum of squared singular values explained by the rank $k$ approximation is $\textup{tr}((\bm{\Sigma}^f_k)^2)/\textup{tr}((\bm{\Sigma}^f)^2)$.

SVD can also be interpreted geometrically.  As $\bm{F}$ is of size $m \times n$, each row of $\bm{F}$ can be represented as a point in an $n$-dimensional space with the row elements as coordinates, and each column can be represented as a point in an $m$-dimensional space with the column elements as coordinates. In a rank-$k$ approximation, where $k<(m,n)$,  each of the original $m$ documents and $n$ terms are approximated by only $k$ coordinates. Thus SVD projects the sum of squared Euclidean distances from these row (column) points to the origin in the $n$ ($m$)-dimensional space as much as possible to a lower, a $k$-dimensional space.
The Euclidean distances between the rows of $\bm{F}$ are approximated by the Euclidean distances between the rows of $\bm{U}^f_k\bm{\Sigma}^f_k$ from below, and the Euclidean distances between the rows of $\bm{F}^T$ are approximated by the Euclidean distances between the rows of $\bm{V}^f_k\bm{\Sigma}^f_k$ from below.

The choice of $k$ is crucial in many applications \citep{albright2004taming}. A lower rank approximation cannot always express prominent relationships in text, whereas the higher rank approximation may add useless noise. How to choose $k$ is an open issue \citep{deerwester1990indexing}. In practice, the value of $k$ is selected such that a certain criterion is satisfied, for example, the proportion of explained total sum of squared singular values is at least a pre-specified proportion. Also, the use of a scree plot, showing the decline in subsequent squared singular values, can be considered.

As $\bm{F}$ is a non-negative matrix, the first column vectors in $\bm{U}$ and $\bm{V}$ have the special property that the elements of the vectors depart in the same direction from the origin \citep{1907Peron, frobenius1912matrizen, hu2003lsa}. We give an intuitive geometric explanation for the $m$ rows of $\bm{F}$. Each row is a vector in the non-negative $n$-dimensional subspace of $R^n$. As a result, the first singular vector, being in the middle of the $m$ vectors, is also in the non-negative $n$-dimensional subspace of $R^n$. As each vector is the non-negative subspace, the angle between each vector with the first singular vector is between 0 and 90 degrees, and therefore the projection of each of the $m$ vectors on the first singular vector, corresponding to the elements of $\bm{U}_1$, is non-negative (or each is non-positive, as we will discuss now). The same holds for the columns of $\bm{F}$ and the first singular vector $\bm{V}_1$. The reason that the elements of $\bm{U}_1$ and $\bm{V}_1$ are all either non-negative or non-positive is that $\bm{U}^f_1\bm{\Sigma}^f_1(\bm{V}^f_1)^T = -\bm{U}^f_1\bm{\Sigma}^f_1(-\bm{V}^f_1)^T$, as the singular values are defined to be non-negative. As the lengths of the row vectors in $n$-dimensional space to the origin are influenced by the sizes of the documents (i.e. the marginal frequencies), larger documents have larger projections on the first singular vector, and the first dimension mainly displays differences in the sizes of the margins.

As it turns out, the raw document-term matrix $\bm{F}$ in Table~\ref{T6*6dt} does not have full rank; its rank is 5. The SVD of $\bm{F}$ in Table~\ref{T6*6dt} is  
\begin{equation}\footnotesize
\label{Eq:exampleXSVD}
\begin{aligned}
\bm{F}
&= \bm{U}^f\bm{\Sigma}^f(\bm{V}^f)^T 
\\ &=\left[
 \begin{array}{rrrrr}
    -0.411&  0.175&  0.825&  0.252& -0.239\\
  -0.646 & 0.314& -0.562&  0.301& -0.279\\
   -0.232&  0.127&  0.034& -0.099&  0.503\\
   -0.562& -0.203&  0.044& -0.603&  0.333 \\
   -0.099& -0.456& -0.024& -0.404& -0.672\\
  -0.186& -0.778& -0.034&  0.556&  0.223
  \end{array}
  \right]\left[
 \begin{matrix}
 8.425  & 0 &0 &0 &0 \\
 0 & 3.261 &0 &0 &0 \\
 0  & 0 &0.988 &0 &0 \\
 0  & 0 &0 &0.574 &0 \\
 0  & 0 &0 &0 &0.272 \\
  \end{matrix}
  \right] \\&
 ~~~~~\left[
 \begin{array}{rrrrrr}
    -0.412&  0.214&  0.655& -0.344&  0.486\\
   -0.488&  0.311&  0.087&  0.180& -0.540 \\
   -0.440&  0.257& -0.748& -0.259&  0.339\\
   -0.611& -0.369&  0.039&  0.366& -0.148\\
   -0.101& -0.441& -0.014& -0.783& -0.426\\
   -0.123& -0.679& -0.048&  0.186&  0.392  \end{array}
  \right]^T
\end{aligned}
\end{equation}

For the raw matrix,  LSA-RAW in Table~\ref{T6*6dtlsasingular} shows the singular values, the squares of the singular values, and the proportions of explained total sum of squared singular values (denoted as PSSSV). Together, the first two dimensions account for 0.855 + 0.128 = 0.983 of the total sum of squared singular values. Therefore, the documents and the terms can be approximated adequately in a two dimensional representation using $\bm{U}^f_2\bm{\Sigma}^f_2$ and $\bm{V}^f_2\bm{\Sigma}^f_2$ as coordinates. As the Euclidean distances between the documents and between the terms in the two-dimensional representation, i.e., between the rows of $\bm{U}^f_2\bm{\Sigma}^f_2$ and the rows of $\bm{V}^f_2\bm{\Sigma}^f_2$, approximate the Euclidean distances between rows and between columns of the original matrix $\bm{F}$, such a two dimensional representation simplifies the interpretation of the matrix considerably.

\begin{table}[t]
\centering  
\caption{The singular values, the squares of singular values, and the proportion of explained total sum of squared singular values (PSSSV) for each dimension of LSA of $\bm{F}$, of $\bm{F}^{L1}$, of $\bm{F}^{L2}$, and of $\bm{F}^{\text{TF-IDF}}$.} 
\label{T6*6dtlsasingular}
\begin{tabular}{llrrrrr}
\hline
methods&items& dim1 & dim2 & dim3 & dim4 & dim5
\\
\hline
\multirow{3}*{LSA-RAW}&
singular value & 8.425& 3.261& 0.988& 0.574 &0.272  \\
&square of singular value & 70.985&  10.635 &  0.976&   0.330&   0.074  \\
&PSSSV & 0.855& 0.128& 0.012& 0.004& 0.001  \\
\hdashline
\multirow{3}*{LSA-NROWL1}&
singular value &1.070& 0.692& 0.123& 0.114& 0.046 \\
&square of singular value &1.146& 0.479& 0.015& 0.013& 0.002 \\
&PSSSV & 0.692& 0.289& 0.009& 0.008& 0.001\\
\hdashline
\multirow{3}*{LSA-NROWL2}&
singular value &2.095& 1.228& 0.239& 0.198 &0.092 \\
&square of singular value &4.388 &1.507& 0.057& 0.039& 0.009 \\
&PSSSV & 0.731 &0.251&  0.009 & 0.007&  0.001\\
\hdashline
\multirow{3}*{LSA-TFIDF}&
singular value &11.878 & 5.898 & 1.565&  1.017&  0.449\\
&square of singular value & 141.088 & 34.782  & 2.451  & 1.034 &  0.202 \\
&PSSSV & 0.786& 0.194 &0.014& 0.006 &0.001\\
\hline
\end{tabular}
\end{table}

\begin{figure}[htbp] 
    \centering
    \subfigure[]{
      \label{F66biplotnewnod}
        \begin{minipage}[t]{0.41\linewidth}
        \centering
        \includegraphics[width=1\textwidth]{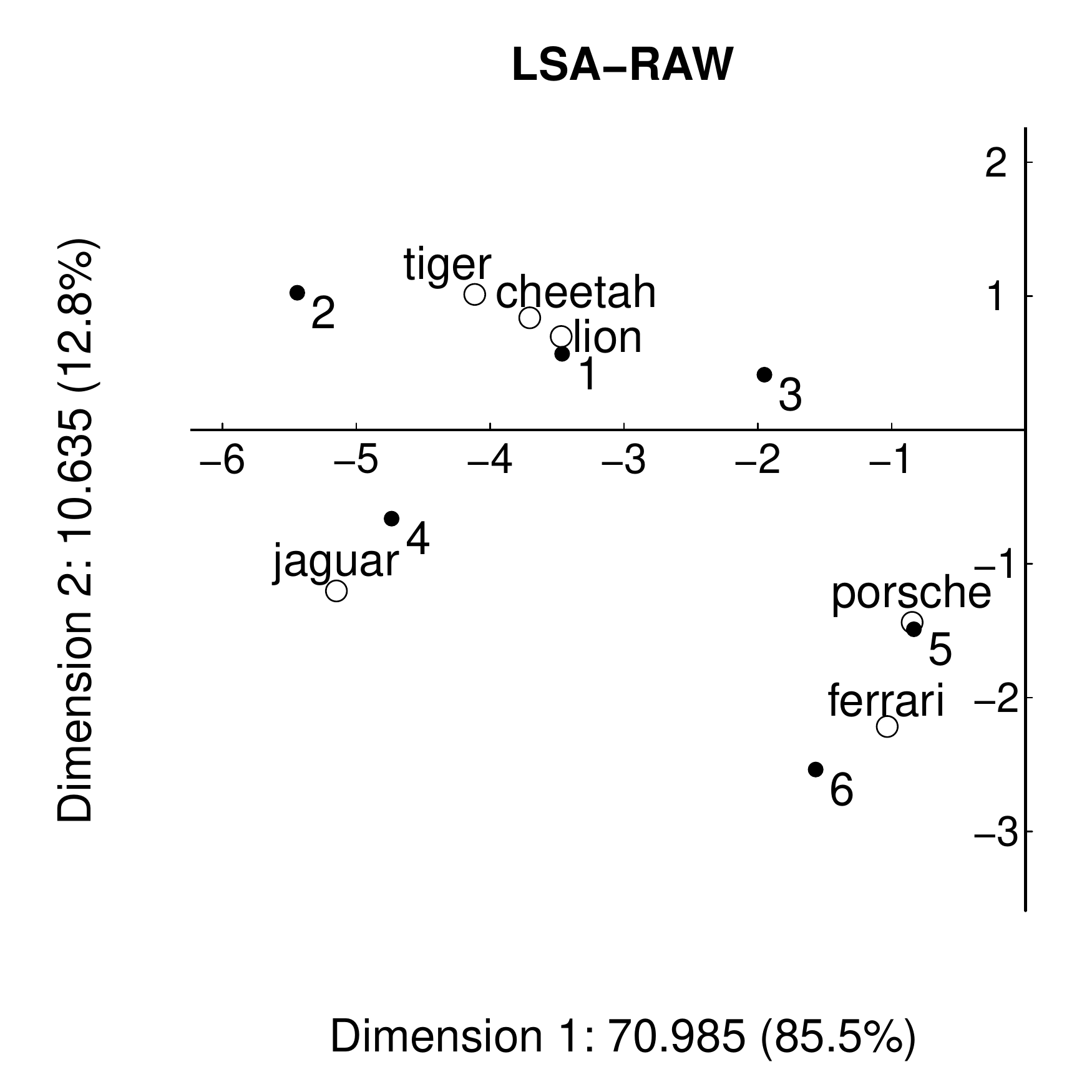}
        \end{minipage}
        }
    \subfigure[]{
       \label{F66biplotFNnewnodL1}
         \begin{minipage}[t]{0.41\linewidth}
         \centering
         \includegraphics[width=1\textwidth]{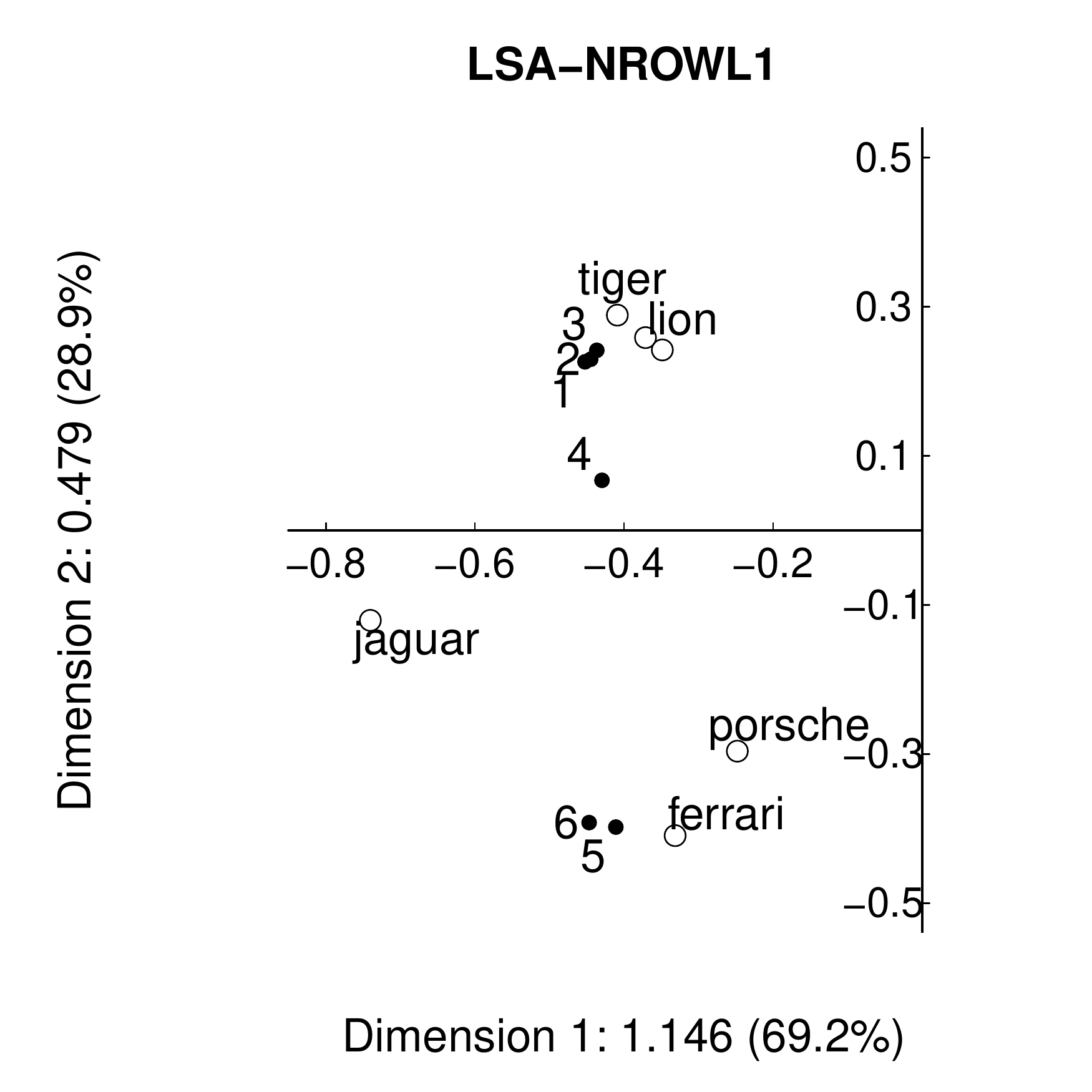}
         \end{minipage}
         }
       \subfigure[]{
       \label{F66biplotFNnewnod}
         \begin{minipage}[t]{0.41\linewidth}
         \centering
         \includegraphics[width=1\textwidth]{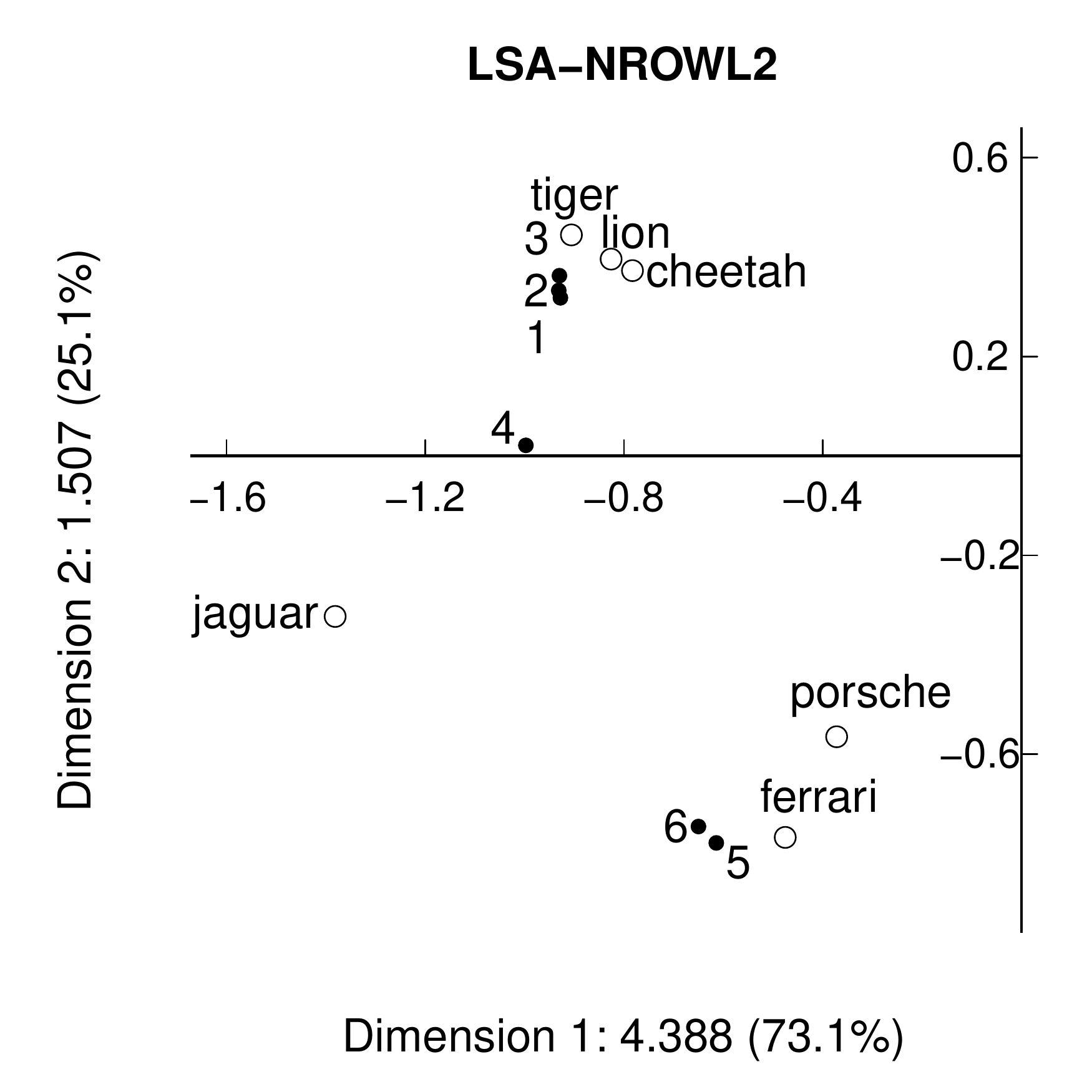}
         \end{minipage}
         }
      \subfigure[]{
       \label{f: TFIDF2base2}
         \begin{minipage}[t]{0.41\linewidth}
         \centering
         \includegraphics[width=1\textwidth]{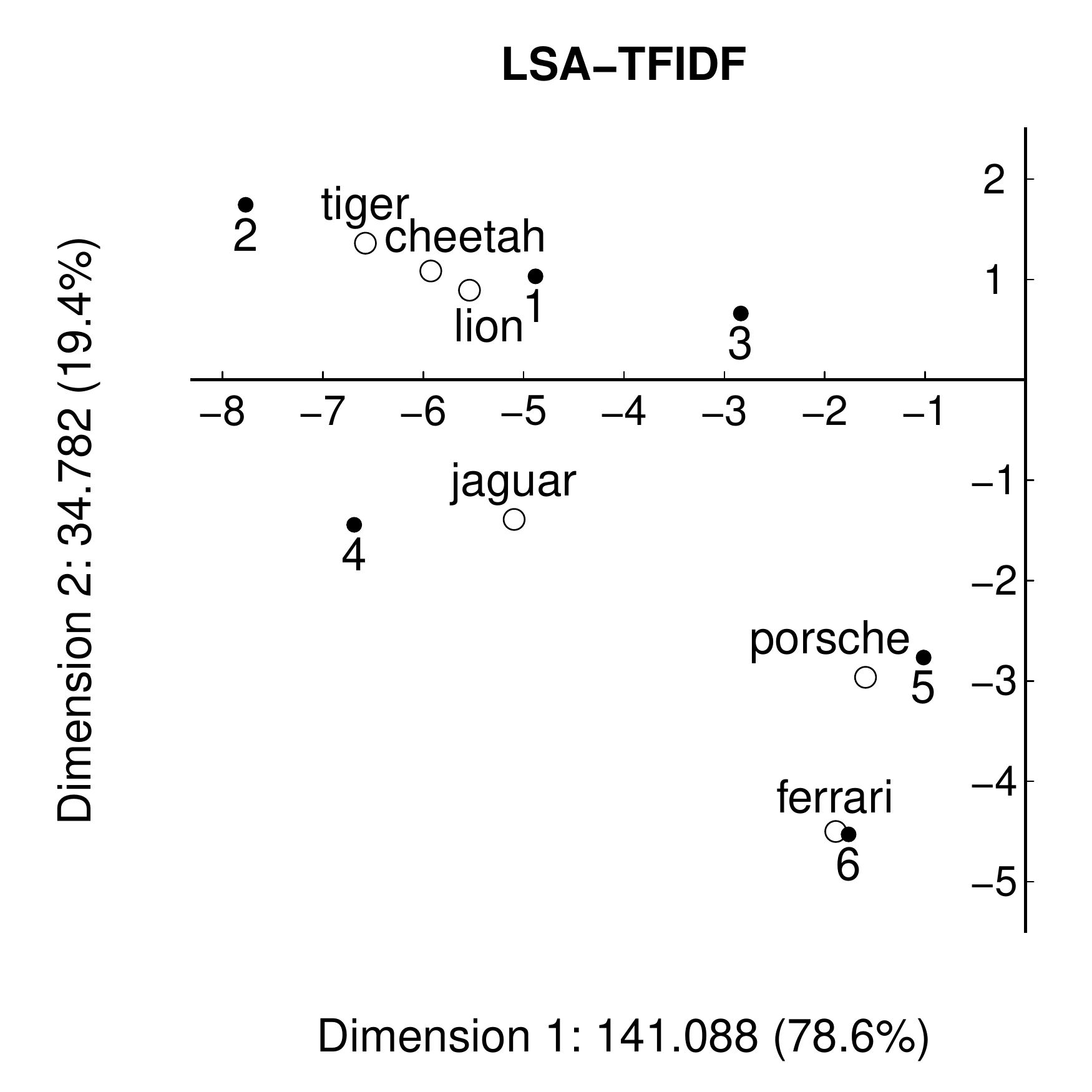}
         \end{minipage}
         }
         \vspace{8mm}
    \caption{A two-dimensional plot of documents and terms (a) for raw matrix $\bm{F}$; (b) for row-normalized data $\bm{F}^{L1}$; (c) for row-normalized data $\bm{F}^{L2}$; (d) for matrix $\bm{F}^{\text{TF-IDF}}$.}
    \label{Fweighted}
\end{figure}

On the other hand, it is somewhat more difficult to examine the relation between a document and a term. The reason is that, by choosing a Euclidean distance-representation both for the documents and for terms, the singular values are used {\em twice} in the coordinates $\bm{U}^f_2\bm{\Sigma}^f_2$ and $\bm{V}^f_2\bm{\Sigma}^f_2$, and the inner product of coordinates of a document and coordinates of a term does not approximate the corresponding value in $\bm{F}$.
Directions from the origin can be interpreted, though, as the double use of the singular values only leads to relatively reduced coordinates on the second dimension in comparison to the coordinates on the first dimension.

The two-dimensional representation of LSA-RAW is shown in Figure~\ref{F66biplotnewnod}. In Figure~\ref{F66biplotnewnod} Euclidean distances between documents, and between terms, reveal the similarity of documents, and terms, respectively. For example, documents 5 and 6 are close, and similar in the sense that their Euclidean distance is small. For these two documents the Euclidean distance in the matrix $\bm{F}$ is 1.414, and in the first two dimensions it is 1.279, so the first two dimensions provide an adequate representation of their similarity. The value 1.279 is much smaller than the Euclidean distances between Documents 5 and 1 (3.338), 5 and 2 (5.248), 5 and 3 (2.205), 5 and 4 (3.988) as well as the Euclidean distances between Documents 6 and 1 (3.638), 6 and 2 (5.262), 6 and 3 (2.975), 6 and 4 (3.681). On the first dimension all documents and terms have a negative coordinate (see above). There is an order of 5, 6, 3, 1, 4, and 2 on the first dimension. This order is related to the row margins of Table~\ref{T6*6dt}, where 2 and 4 have the highest frequencies and therefore are further away from the origin. Overall, the two-dimensional representation of the documents reveals a mix of the sizes of the documents, the row margins $\Sigma_j f_{ij}$, and the relative use of the terms by the documents, i.e., for row $i$ this is the vector of elements $f_{ij}/\Sigma_j f_{ij}$, also known as the {\it row profile} for row $i$. This mix makes the graphic representation difficult to interpret. Similarly, {\it porsche} and {\it ferrari} are lower left but close to the origin, {\it tiger}, {\it cheetah}, and {\it lion} are upper left and further away from the origin, and {\it jaguar} is far away at the lower left. Also there is a mix of the sizes of the terms, i.e., for column $j$ this is column margin $\Sigma_i f_{ij}$, and the relative use of the documents by the terms, i.e., for column $j$ this is the vector of elements $f_{ij}/\Sigma_i f_{ij}$, also known as the {\it column profile} for column $j$. The terms {\it porsche} and {\it ferrari} are related to documents 5 and 6 as they have the same position w.r.t. the origin, and similarly for {\it tiger}, {\it cheetah}, and {\it lion} to documents 1, 2, and 3, and {\it jaguar} to document 4. 

Although the first dimension accounts for $85.5$ per cent of the total sum of squared singular values, it provides little information about the relations among documents and terms. In particular, from Table~\ref{T6*6dt} we expect that documents 1 to 3 are similar, documents 5 and 6 are similar, and document 4 is in-between; term {\em jaguar} is between cat terms ({\em tiger}, {\em cheetah}, and {\em lion}) and car terms ({\em porsche} and {\em ferrari}), but we cannot see that from the first dimension. This is because the margins of Table~\ref{T6*6dt} play a dominant role in the first dimension.  

\subsection{LSA of weighted document-term matrix}\label{lsawdtm}

Weighting can be used to prevent differential lengths of documents from having differential effects on the representation, or be used to impose certain preconceptions of which terms are more important \citep{deerwester1990indexing}. The frequencies $f_{ij}$ in the raw document-term matrix $\bm{F}$ can be transformed with the aim to provide a better approximation of the interrelations between documents and terms \citep{nakov2001weight}. The weight $w_{ij}$ for term $j$ in document $i$ is normally expressed as a product of three components \citep{salton1988term,kolda1998semidiscrete,ab2008term}
\begin{equation}\label{Eqweighfunction}
w_{ij}
=L(i,j) \times G(j) \times N(i)
\end{equation}
where the local weighting $L(i,j)$ is the weight of term $j$ in document $i$, the global weighting $G(j)$ is the weight of the term $j$ in the entire document set, and $N(i)$ is the normalization component for document $i$.

When $L(i,j) = f(i,j)$, $G(j)=1$, and $N(i)=1$, the weighted $\bm{F}$ is equal to $\bm{F}$. In matrix notation, Equation~(\ref{Eqweighfunction}) can be expressed as $\bm{W}
=\bm{N}\bm{L}\bm{G}$, where $\bm{N}$ is a diagonal matrix with diagonal elements $N(i)$ and $\bm{G}$ is a diagonal matrix with diagonal elements $G(j)$. Notice that pre- or post-multiplying by a diagonal matrix leaves the rank of the matrix $\bm{L}$ intact. 

We examine two common ways to weight $f_{ij}$. One is row normalization \citep{salton1988term, ab2008term} with L1 and L2. The other is TF-IDF \citep{dumais1991improving}.

\subsubsection{SVD of matrix with row-normalized elements with L1}

In row-normalized weighting with L1, we use Equation~(\ref{Eqweighfunction}) with $L(i,j)=f_{ij}$, $G(j)=1$, and $N(i)=1/\sum_{j=1}^n{f_{ij}}$, and apply an SVD to this transformed matrix that we denote as $\bm{F}^{L1}$, which consists of the row profiles of $\bm{F}$. See Table \ref{T6*6dtrp}. The last row, the average row profile, is the row profile of the column margins of Table~\ref{T6*6dt}.

\begin{table}[htbp]
\centering  
\caption{Row profiles of $\bm{F}$} 
\label{T6*6dtrp}
\begin{tabular}{cccccccc}    
\hline
 & lion & tiger & cheetah &jaguar & porsche & ferrari&total\\  
\hline  
   doc1 & 0.286 & 0.286 & 0.143 & 0.286 & 0.000 & 0.000 &1.000\\
   doc2 & 0.182 & 0.273 & 0.273 & 0.273 & 0.000 & 0.000 &1.000\\
   doc3 & 0.250 & 0.250 & 0.250 & 0.250 & 0.000 & 0.000 &1.000\\
   doc4 & 0.182 & 0.182 & 0.182 & 0.273 & 0.091 & 0.091 &1.000\\
   doc5 & 0.000 & 0.000 & 0.000 & 0.333 & 0.333 & 0.333 &1.000\\
   doc6 & 0.000 & 0.000 & 0.000 & 0.400 & 0.200 & 0.400 &1.000\\
   \hline
   average row profile & 0.171 & 0.195 & 0.171 & 0.293 & 0.073 & 0.098&1.000\\
\end{tabular}  
\end{table}

We perform LSA of $\bm{F}^{L1}$ and find Table~\ref{T6*6dtlsasingular}, part LSA-NROWL1. This shows that a rank 2 matrix approximates the data well as 0.692 + 0.289 = 0.981 of the total sum of squared singular values is explained by these two dimensions. The first two columns of LSA of $\bm{F}^{L1}$ can be used to approximate $\bm{F}^{L1}$, see Equation~(\ref{Eq:exampleXFNSVDL1}).
\begin{equation}
\label{Eq:exampleXFNSVDL1}
\begin{aligned}
\bm{F}^{L1}
&\approx \bm{U}^{L1}_2\bm{\Sigma}^{L1}_2(\bm{V}^{L1}_2)^T  \\&= \left[
 \begin{array}{rr}
    -0.423&  0.327 \\
    -0.415&  0.332\\
   -0.408&  0.349\\
   -0.401&  0.097\\
   -0.384& -0.575\\
   -0.417& -0.567
  \end{array}
  \right]\left[
 \begin{array}{cc}
 1.070  & 0\\
 0 & 0.692\\
  \end{array}
  \right] 
 \left[
 \begin{array}{rr}
   -0.347&  0.374\\
   -0.382 & 0.417\\
   -0.326&  0.350 \\
   -0.692 &-0.174\\
   -0.232& -0.428\\
   -0.310& -0.592
  \end{array}
  \right]^T  
\end{aligned}
\end{equation}

Documents and terms can be projected on a two dimensional space using $\bm{U}^{L1}_2\bm{\Sigma}^{L1}_2$ and $\bm{V}^{L1}_2\bm{\Sigma}^{L1}_2$ as coordinates, see Figure~\ref{F66biplotFNnewnodL1}. In this representation documents 1, 2, and 3 are quite close, and so are 5 and 6. Also, the terms {\it ferrari} and {\it porsche} are close and related to 5 and 6, {\it tiger}, {\it lion}, and {\it cheetah} are close and related to 1, 2, and 3. 

Although the first dimension accounts for $69.2$ per cent of the total sum of squared singular values, this dimension does not provide information about different use of terms by the documents as all documents have a similar coordinate. This is caused by the same marginal value  1 for each of the documents in $\bm{F}^{L1}$, which leads to almost the same distance from the origin. Also, we would expect {\em jaguar} to be in between cat terms ({\em tiger}, {\em cheetah}, and {\em lion}) and car terms ({\em porsche} and {\em ferrari}), but on the first dimension it appears as a separate, third group. This is caused by the high values in its column in $\bm{F}^{L1}$, which lead to a larger distance from the origin.

\subsubsection{SVD of matrix with row-normalized elements with L2}

In row-normalized weighting with L2, we use Equation~(\ref{Eqweighfunction}) with $L(i,j)=f_{ij}$, $G(j)=1$, and $N(i)=1/\sqrt{\sum_{j=1}^n{f_{ij}^2}}$. 
The transformed matrix, denoted as $\bm{F}^{L2}$, is shown in Table~\ref{T6*6dtFN}. We then perform LSA
on Table \ref{T6*6dtFN}. Table~\ref{T6*6dtlsasingular}, part LSA-NROWL2, indicates that a rank 2 matrix approximates the data well, as the sum of the PSSSV of the first two dimensions 0.731 + 0.251 = 0.982 contributes to 98.2 per cent of the total sum of squared singular values. The first two columns of LSA of $\bm{F}^{L2}$ can be used to approximate $\bm{F}^{L2}$, see Equation~(\ref{Eq:exampleXFNSVD}).
\begin{equation}
\label{Eq:exampleXFNSVD}
\begin{aligned}
\bm{F}^{L2}
&\approx \bm{U}^{L2}_2\bm{\Sigma}^{L2}_2(\bm{V}^{L2}_2)^T  \\&= \left[
 \begin{array}{rr}
    -0.443&  0.259\\
    -0.445&  0.271\\
   -0.444&  0.295\\
   -0.476&  0.017\\
   -0.293& -0.635\\
   -0.310& -0.608
  \end{array}
  \right]\left[
 \begin{array}{cc}
 2.095  & 0\\
 0 & 1.228\\
  \end{array}
  \right] 
 \left[
 \begin{array}{rr}
   -0.394&  0.323\\
   -0.432&  0.362\\
   -0.374&  0.304 \\
   -0.659& -0.263\\
   -0.178& -0.460\\
   -0.227& -0.625
  \end{array}
  \right]^T  
\end{aligned}
\end{equation}

\begin{table}[htbp]
\centering  
\caption{A row-normalized document-term matrix $\bm{F}^{L2}$} 
\label{T6*6dtFN}
\begin{tabular}{ccccccc}    
\hline
 & lion & tiger & cheetah &jaguar & porsche & ferrari\\  
\hline  
doc1 & 0.555& 0.555& 0.277& 0.555& 0.000& 0.000\\ 
doc2 &  0.359& 0.539& 0.539& 0.539& 0.000& 0.000\\  
doc3 & 0.500& 0.500 &0.500 &0.500& 0.000& 0.000\\ 
doc4 &  0.417& 0.417& 0.417& 0.626& 0.209& 0.209\\ 
doc5 & 0.000& 0.000 &0.000& 0.577& 0.577& 0.577\\ 
doc6 & 0.000& 0.000 &0.000& 0.667& 0.333& 0.667\\ 
\hline 
\end{tabular}  
\end{table}

Documents and terms can be projected on a two dimensional space using $\bm{U}^{L2}_2\bm{\Sigma}^{L2}_2$ and $\bm{V}^{L2}_2\bm{\Sigma}^{L2}_2$ as coordinates, see Figure~\ref{F66biplotFNnewnod}. In this representation documents 1, 2, and 3 are quite close, and so are 5 and 6. Also, the terms {\em ferrari} and {\em porsche} are close and related to 5 and 6, {\em tiger}, {\em lion}, and {\em cheetah} are close and related to 1, 2, and 3. 

Although the first dimension accounts for $73.1$ per cent of the total sum of squared singular values, and so, a major portion of the information in the matrix, we do not find the important aspect in the data that document 4 should be in between documents 1-3 on the one hand and documents 5-6 on the other hand on this dimension. This is caused by the high values in the row for doc4 in Table~\ref{T6*6dtFN}, which lead to a larger distance from the origin than the other documents have. Also, we would expect {\em jaguar} to be in between cat terms ({\em tiger}, {\em cheetah}, and {\em lion}) and car terms ({\em porsche} and {\em ferrari}), but on the first dimension it appears as a separate, third group. This is caused by the high values in its column in Table~\ref{T6*6dtFN}, which lead to a larger distance from the origin.

\subsubsection{SVD of the term frequency-inverse document frequency matrix}\label{S:TFIDF2}

TF-IDF is one commonly used transformation of text data. We use equation~(\ref{Eqweighfunction}) with $L(i,j)=f_{ij}$,
$G(j)=1+\text{log}(\frac{n\text{docs}}{df_j})$, and $N(i)=1$, one form of TF-IDF, where $n$docs is the number of documents in the set and $df_j$ is the number of documents where term $j$ appears, and then apply an SVD to this transformed matrix that we denote as $\bm{F}^{\text{TF-IDF}}$, see Table~\ref{TFIDF2base2}. As is common in the literature, here we choose 2 as the base of the logarithmic function.

\begin{table}[htbp]
\centering  
\caption{A document-term matrix $\bm{F}^{\text{TF-IDF}}$} 
\label{TFIDF2base2}
\begin{tabular}{ccccccc}    
\hline
 & lion & tiger & cheetah &jaguar & porsche & ferrari\\  
\hline  
doc1 & 3.170  &3.170  &1.585   &  2   &  0   &  0\\ 
doc2 & 3.170  &4.755 & 4.755   &  3  &   0  &   0\\  
doc3 & 1.585  &1.585  &1.585  &   1    & 0    & 0\\ 
doc4 & 3.170  &3.170  &3.170   &  3   &  2    & 2\\ 
doc5 & 0.000  &0.000 & 0.000    & 1  &   2    & 2\\ 
doc6 & 0.000  &0.000  &0.000     &2  &   2   &  4\\ 
\hline 
\end{tabular}  
\end{table}

We perform LSA of Table~\ref{TFIDF2base2} and find Table~\ref{T6*6dtlsasingular}, part LSA-TFIDF. This shows that a rank 2 matrix approximates the data well as 0.786 + 0.194 = 0.980 of the total sum of squared singular values is explained by these two dimensions. The matrix $\bm{F}^{\text{TF-IDF}}$ in Table~\ref{TFIDF2base2} is approximated in the first two dimensions as follows:
\begin{equation}
\label{Eq:exampleXTFIDF2SVDbase2}
\begin{aligned}
\bm{F}^{\text{TF-IDF}}
&\approx \bm{U}^{\text{TF-IDF}}_2\bm{\Sigma}^{\text{TF-IDF}}_2(\bm{V}^{\text{TF-IDF}}_2)^T  \\& =\left[
 \begin{array}{rr}
     -0.411 & 0.175\\
    -0.654 & 0.296\\
   -0.239 & 0.112\\
   -0.563& -0.245\\
   -0.086& -0.469\\
   -0.148& -0.768
  \end{array}
  \right]\left[
 \begin{array}{cc}
 11.878  & 0\\
 0 & 5.898 \\
  \end{array}
  \right] 
 \left[
 \begin{array}{rr}
  -0.466&  0.151 \\
  -0.554&  0.231 \\
  -0.499 & 0.184\\
   -0.429& -0.236\\
   -0.134& -0.502\\
    -0.159& -0.763
  \end{array}
  \right]^T  
\end{aligned}
\end{equation}

Figure~\ref{f: TFIDF2base2} is a two-dimensional plot of the documents and terms using $\bm{U}^{\text{TF-IDF}}_2\bm{\Sigma}^{\text{TF-IDF}}_2$ and $\bm{V}^{\text{TF-IDF}}_2\bm{\Sigma}^{\text{TF-IDF}}_2$ as coordinates for the $6 \times 6$ sample document-term matrix $\bm{F}^{\text{TF-IDF}}$. The configuration of documents in Figure~\ref{f: TFIDF2base2} is very similar to that in Figure~\ref{F66biplotnewnod}. The configuration of terms in Figure~\ref{f: TFIDF2base2} is different from that of terms in Figure~\ref{F66biplotnewnod}. In Figure~\ref{f: TFIDF2base2}, there is an order of {\em porsche}, {\em ferrari}, {\em jaguar}, {\em lion}, {\em cheetah}, and {\em tiger} on the first dimension, whereas in Figure~\ref{F66biplotnewnod}, there is an order of {\em porsche}, {\em ferrari}, {\em lion}, {\em cheetah}, {\em tiger}, and {\em jaguar} on the first dimension. Compared with Figure~\ref{F66biplotnewnod}, the first dimension of Figure~\ref{f: TFIDF2base2} shows that {\em jaguar} is in between cat terms ({\em tiger}, {\em cheetah}, and {\em lion}) and car terms ({\em porsche} and {\em ferrari}).

\subsubsection{Out-of-sample documents}\label{Sub: OSD}

Representing out-of-sample documents in the $k$-dimensional subspace of LSA is important for many applications. Suppose a out-of-sample document $\bm{d}$ is a row vector. To represent $\bm{d}$ in lower dimensional space, first the out-of-sample document $\bm{d}$ can be transformed in the same way as the original documents \citep{dumais1991improving}. Transformations for the above four applications of LSA are $\bm{d}_w^f = \bm{d}$,  $\bm{d}_w^{L1} = \bm{d}/\sum_{j=1}^nd_j$,  $\bm{d}_w^{L2} = \bm{d}/\sqrt{\sum_{j=1}^n{d_{j}^2}}$, and $\bm{d}_w^{\text{TF-IDF}} = [d_1G(1), \cdots, d_nG(n)]$. The coordinates of the out-of-sample document $\bm{d}$ in LSA-RAW, LSA-NROWL1, LSA-NROWL2, and LSA-TFIDF are then calculated by $\bm{d}_w^f\bm{V}^f$, $\bm{d}_w^{L1}\bm{V}^{L1}$, $\bm{d}_w^{L2}\bm{V}^{L2}$, and $\bm{d}_w^{\text{TF-IDF}}\bm{V}^{\text{TF-IDF}}$, respectively \citep{aggarwal2018machine}.

\subsection{Conclusions regarding LSA of different matrices}\label{LSADM}

In the raw document-term matrix the relationships among the documents and terms is blurred by differences in margins arising from differing document-lengths and marginal term-frequencies. Thus LSA of the raw matrix leads to a mix of margins, and relationships among documents and terms. In order to provide a better approximation of the interrelations between documents and terms, weighting schemes were used.

Normalizations of the documents have a beneficial effect. Yet, the properties of the frequencies that are evident from Table~\ref{T6*6dt} where we expect, for example, that {\em jaguar} lies in between {\em porsche} and {\em ferrari} on the one hand and {\em tiger}, {\em cheetah}, and {\em lion} on the other hand, are not fully represented on the first dimension. This is due to the fact that the column margins of Tables \ref{T6*6dtrp} and \ref{T6*6dtFN} still play a role on the first dimension. The TF-IDF transformation also has a positive effect. Yet LSA is not successful. For example, we expect that documents 1 to 3 are similar, 5 and 6 are similar, and document 4 is in-between, but this order is not found in the first dimension. This is due to the fact that the row margins of Table~\ref{TFIDF2base2} still play a role on the first dimension. 

Generally,  solutions of LSA have the drawback that they include the effect of the margins as well as the dependence. In the first dimension these margins play a dominant role as all points depart in the same direction from the origin. We can try to repair this property of LSA, by applying transformations of the rows and columns of Table~\ref{T6*6dt} simultaneously. However, the transformations appear ad hoc. Instead we present in the next section a different technique, which better fits the properties of the data: CA.

\section{Correspondence analysis}\label{S:CA}

CA provides a low-dimensional representation of the interaction or dependence between the rows and columns of the contingency table \citep{greenacre1987geometric}, which can be used to reveal the structure in the data \citep{hayashi1992quantification}. CA has been proposed multiple times, apparently independently, emphasizing different properties of the technique \citep{gifi1990nonlinear}. Some important contributions are provided in the Japanese literature, by \citet{hayashi1956theory, hayashi1992quantification}, who emphasizes the property of CA that it maximizes the correlation coefficient between the row and column variable by assigning numerical scores to these variables; in the French literature, by \citet{benzecri1973analyse}, who emphasizes a distance interpretation, where \citet{greenacre1984theory} expressed Benz{\'e}cri's work in a more conveniential mathematical notation; and in the Dutch literature, by \citet{gifi1990nonlinear} and \citet{michailidis1998gifi}, who emphasize optimal scaling properties. We present CA here mainly from the French perspective.

The aim of CA as developed by Benz{\'e}cri is to find a representation of the rows (columns) of frequency matrix $\bm{F}$ in such a way that Euclidean distances between the rows (columns) in the representation correspond to so-called $\chi^2$-distances between rows (columns) of $\bm{F}$ \citep{gifi1990nonlinear}. We work with $\bm{P}$ with elements $p_{ij} = f_{ij}/f_{++}$, where $f_{++}$ is the sum of all elements of $\bm{F}$. In the $\chi^2$-distance profiles play an important role. The squared $\chi^2$-distance between the $k$th row profile with elements $p_{kj}/r_k$ and the $l$th row profile with elements $p_{lj}/r_l$ is
\begin{equation}
\label{chidistace}
\delta_{kl}^2 = \sum_j{\frac{\left(p_{kj}/r_k - p_{lj}/r_l\right)^2}{c_j}}
\end{equation}
where $r_{i}$ (also called the average column profile) and $c_{j}$ (the average row profile) are the row and column sums of $\bm{P}$ respectively. Thus the difference between the $j$th elements of the two profiles is weighted by column margin (i.e. the last row of Table~\ref{T6*6dtrp}), $c_j$, so that this difference plays a relatively more important role in the $\chi^2$-distance if it stems from a column having a small value $c_j$.

A representation where Euclidean distances between the rows of the matrix are equal to $\chi^2$-distances is found as follows. In matrix notation, the matrix whose Euclidean distances between the rows are equal to $\chi^2$-distances between rows of $\bm{F}$ is equal to $\bm{D}_r^{-1}\bm{P}\bm{D}_{c}^{-\frac{1}{2}}$, where $\bm{D}_r$ is a diagonal matrix with $r_i$ as diagonal elements and $\bm{D}_c$ is a diagonal matrix with $c_j$ as diagonal elements. Suppose we take the SVD of
\begin{equation}
\label{SP}
\bm{D}_r^{-\frac{1}{2}}\bm{P}\bm{D}_{c}^{-\frac{1}{2}} = \bm{U}^{sp} \bm{\Sigma}^{sp} (\bm{V}^{sp})^T
\end{equation}
Here $\bm{D}_r^{-\frac{1}{2}}\bm{P}\bm{D}_{c}^{-\frac{1}{2}}$ is a matrix with standardized proportions, hence the superscripts $sp$ on the right hand side of the equation. Then, if we pre-multiply both sides of Equation~(\ref{SP}) with $\bm{D}_r^{-\frac{1}{2}}$, we get
\begin{equation}
\label{SP2}
 \bm{D}_r^{-1}\bm{P}\bm{D}_{c}^{-\frac{1}{2}} = \bm{D}_r^{-\frac{1}{2}}\bm{U}^{sp}\bm{\Sigma}^{sp} (\bm{V}^{sp})^T
\end{equation}
Thus a representation using the rows of $\bm{D}_r^{-\frac{1}{2}}\bm{U}^{sp}\bm{\Sigma}^{sp}$ as row coordinates leads to Euclidean distances between these row points being equal to $\chi^2$-distances between rows of $\bm{F}$. Similar to Equation~(\ref{chidistace}) we can also define $\chi^2$-distances between the columns of $\bm{F}$, and in matrix notation this leads to the matrix $\bm{D}_r^{-\frac{1}{2}}\bm{P}\bm{D}_{c}^{-1}$. Then, in a similar way as for the $\chi^2$-distances for the rows, Equation~(\ref{SP}) can be used as an intermediate step to go to a solution for the columns. Post-multiplying the left  and right hand sides in Equation~(\ref{SP}) by $\bm{D}_c^{-\frac{1}{2}}$ provides us with the coordinates for a  representation where Euclidean distances between the column points (the rows of $\bm{D}_c^{-\frac{1}{2}}\bm{V}^{sp}\bm{\Sigma}^{sp}$ as coordinates for these columns) are equal to $\chi^2$-distances between the columns of $\bm{F}$. Notice that Equation~(\ref{SP}) plays the dual role of an intermediate step in going to a solution both for the rows and the columns. 

The matrices $\bm{D}_r^{-\frac{1}{2}}\bm{U}^{sp}\bm{\Sigma}^{sp}$ and $\bm{D}_c^{-\frac{1}{2}}\bm{V}^{sp}\bm{\Sigma}^{sp}$ have a first column being equal to 1, a so-called artificial dimension. This artificial dimension reflects the fact that the row margins of the matrix $\bm{D}_r^{-1}\bm{P}$ with the row profiles of Table~\ref{T6*6dt} are 1 and the column margins of the matrix $\bm{P}\bm{D}_c^{-1}$ with the column profiles of Table~\ref{T6*6dt} are 1. This artificial dimension is eliminated by not taking the SVD of $\bm{D}_r^{-\frac{1}{2}}\bm{P}\bm{D}_{c}^{-\frac{1}{2}}$ but of $\bm{D}_r^{-\frac{1}{2}}(\bm{P}-\bm{E})\bm{D}_{c}^{-\frac{1}{2}}$, where the elements of $\bm{E}$ are defined as the product of the margins $r_i$ and $c_j$. Due to subtracting $\bm{E}$ from $\bm{P}$, the rank of $\bm{D}_r^{-\frac{1}{2}}(\bm{P}-\bm{E})\bm{D}_{c}^{-\frac{1}{2}}$ is $m-1$, which is 1 less than the rank of $\bm{F}$. Notice that the elements of  $\bm{D}_r^{-\frac{1}{2}}(\bm{P}-\bm{E})\bm{D}_{c}^{-\frac{1}{2}}$ are standardized residuals under the independence model, and the sum of squares of these elements yields the so-called total inertia, which is equal to the Pearson $\chi^2$ statistic divided by sample size $f_{++}$. By taking the SVD of the matrix of standardized residuals, we get
\begin{equation}
\label{SP3}
\bm{D}_r^{-\frac{1}{2}}(\bm{P}-\bm{E})\bm{D}_{c}^{-\frac{1}{2}} =  \bm{U}^{sr} \bm{\Sigma}^{sr} (\bm{V}^{sr})^T
\end{equation}
and 
\begin{equation}
\label{SP4}
 \bm{D}_r^{-1}(\bm{P}-\bm{E})\bm{D}_{c}^{-1} =  \bm{\Phi}^{sr}\bm{\Sigma}^{sr} (\bm{\Gamma}^{sr})^T
\end{equation}
where $\bm{\Phi}^{sr}=\bm{D}_r^{-\frac{1}{2}}\bm{U}^{sr}$ and $\bm{\Gamma}^{sr} = \bm{D}_c^{-\frac{1}{2}}\bm{V}^{sr}$.
We use the abbreviation $sr$ for the matrices on the right hand side of Equation~(\ref{SP3}) to refer to the matrix of standardized residuals on the left hand side of the equation. CA simultaneously provides a geometric representation of row profiles and column profiles of Table~\ref{T6*6dt}, where the effects of row margins and column margins of Table~\ref{T6*6dt} are eliminated. $\bm{\Phi}^{sr}$ and $\bm{\Gamma}^{sr}$ are called standard coordinates of rows and columns respectively. They have the property that their weighted average is 0 and weighted sum of squares is 1:
\begin{equation}
\label{restriction1}
\mathbf{1}^T\bm{D}_r\bm{\Phi}^{sr} = \mathbf{0}^T = \mathbf{1}^T\bm{D}_c\bm{\Gamma}^{sr}
\end{equation}
\noindent
and
\begin{equation}
\label{restriction2} 
(\bm{\Phi}^{sr})^T \bm{D}_r \bm{\Phi}^{sr} = \mathbf{I} = (\bm{\Gamma}^{sr})^T \bm{D}_c \bm{\Gamma}^{sr}
\end{equation}
Equation~(\ref{restriction1}) reflects the fact that the row and column margins of
$\bm{P}-\bm{E}$ vanish \citep{van1989combined}.

We can make graphic displays using $\bm{\Phi}^{sr}_k\bm{\Sigma}^{sr}_k$ and $\bm{\Gamma}^{sr}_k\bm{\Sigma}^{sr}_k$ as coordinates, which has the advantage that Euclidean distances between the points approximate $\chi^2$-distances both for the rows of $\bm{F}$ and for the columns of $\bm{F}$, but it has the drawback that $\bm{\Sigma}^{sr}_k$ is used twice. We can also make graphic displays using $\bm{\Phi}^{sr}_k\bm{\Sigma}^{sr}_k$ and $\bm{\Gamma}^{sr}_k$, or $\bm{\Phi}^{sr}_k$ and $\bm{\Gamma}^{sr}_k\bm{\Sigma}^{sr}_k$. Thus, from Equation~(\ref{SP4}), this has the advantage that the inner product of the coordinates of a document and the coordinates of a term approximates the corresponding value in $\bm{D}_r^{-1}(\bm{P}-\bm{E})\bm{D}_{c}^{-1}$. 

If we choose $\bm{\Phi}^{sr}\bm{\Sigma}^{sr}$ for the row points and $\bm{\Gamma}^{sr}$ for the column points, then CA has the property that the row points are in weighted average of the column points, where the weights are the row profile values. Actually, $\bm{\Gamma}^{sr}$ can be seen as coordinates for the extreme row profiles projected onto the subspace. The extreme row profiles are totally concentrated into one of the terms. For example, $[0,0,1,0,0,0]$ represents the row profile of a document that is totally concentrated into {\em cheetah}. At the same time, if we choose $\bm{\Phi}^{sr}$ for the row points and $\bm{\Gamma}^{sr}\bm{\Sigma}^{sr}$ for the column points, column points are in weighted average of row points, where the weights are the column profile values. In a similar way as for the rows, $\bm{\Phi}^{sr}$ provide coordinates for the extreme column profiles projected onto the subspace. The relationship between these row points and column points can be shown by rewriting Equation~(\ref{SP3}) and using Equation~(\ref{restriction1}) as 
\begin{equation}
\label{transitiondecomposition1}
\bm{D}_r^{-1}\bm{P}\bm{\Gamma}^{sr} = \bm{\Phi}^{sr}\bm{\Sigma}^{sr}
\end{equation}
and
\begin{equation}
\label{transitiondecomposition2}
\bm{D}_c^{-1}\bm{P}^T\bm{\Phi}^{sr} = \bm{\Gamma}^{sr}\bm{\Sigma}^{sr}
\end{equation}
These equations are called the transition formulas. In fact, using transition formulas is one of the ways in which the solution of CA can be obtained: starting from arbitrary values for the columns, one first centers and standardizes the column coordinates so that the weighted sum is 0 and the weighted sums of squares is 1, next places the rows in the weighted average of the columns, then places the columns in the weighted average of the rows, and so on, until convergence. This is known as reciprocal averaging \citep{hill1973reciprocal, hill1974correspondence}. Using the transition formula (\ref{transitiondecomposition1}), the coordinates of the out-of-sample document $\bm{d}$ is $(\textbf{\emph{d}}/\sum_{j=1}^nd_j)\boldsymbol{\Gamma}^{sr}$ \citep{greenacre2017correspondence}.

The origin in the graphic representation for the rows stands for the average row profile, which can be seen as follows. Let $\bm{D}_r^{-1}\bm{P}\bm{D}_{c}^{-\frac{1}{2}}$ be the matrix where Euclidean distances between the rows are $\chi^2$-distances between rows of $\bm{F}$. Assume we plot the rows of this matrix using the $n$ elements of each row as coordinates. Then, eliminating the artificial dimension in $\bm{D}_r^{-1}\bm{P}\bm{D}_{c}^{-\frac{1}{2}}$ leads to the subtraction of the average row profile from each row, as $\bm{D}_r^{-1}\bm{E}$ is a matrix with the average row profile in each row. In other words, the cloud of row points is translated to the origin, with the average row profile being exactly in the origin (compare Equation~(\ref{restriction1}): $\mathbf{0}^T = \mathbf{1}^T\bm{D}_c\bm{\Gamma}^{sr}$). When two row points are departing in the same way from the origin, they depart in the same way from the average profile, and when two row points are on opposite sides of the origin, they depart in opposite ways from the average profile. If the documents and terms are statistically independent, then $p_{ij}/r_i = c_j$, and all document profiles would lie in the origin. Thus comparing row profiles with the origin is a way to study the departure from independence and to study the relations between documents and terms. Similarly, the origin in the graphic representation for the columns stands for average column profile.

We now analyze the example discussed in the LSA section. There are three steps to obtain the CA solution. Step 1: make the matrix $\bm{D}_r^{-\frac{1}{2}}(\bm{P}-\bm{E})\bm{D}_{c}^{-\frac{1}{2}}$ of standardized residuals; Step 2: compute the SVD of the matrix; Step 3: derive $\bm{\Phi}^{sr} = \bm{D}_r^{-\frac{1}{2}}\bm{U}^{sr}$ and $\bm{\Gamma}^{sr} = \bm{D}_c^{-\frac{1}{2}}\bm{V}^{sr}$, and post-multiply $\bm{\Phi}^{sr}$ and $\bm{\Gamma}^{sr}$ by $\bm{\Sigma}^{sr}$ to obtain the coordinates. Table~\ref{T6*6dtSR} shows the matrix $\bm{D}_r^{-\frac{1}{2}}(\bm{P}-\bm{E})\bm{D}_{c}^{-\frac{1}{2}}$ of standardized residuals (in lower-case notation, the elements of the matrix are $(p_{ij} - e_{ij})/\sqrt{e_{ij}}$).

\begin{table}[htbp]
\centering  
\caption{The matrix $\bm{D}_r^{-\frac{1}{2}}(\bm{P}-\bm{E})\bm{D}_{c}^{-\frac{1}{2}}$ of standardized residuals} 
\label{T6*6dtSR}
\begin{tabular}{ccccccc}    
\hline
 & lion & tiger & cheetah &jaguar & porsche & ferrari\\  
\hline  
doc1 &  0.115 & 0.085& -0.028 &-0.005& -0.112& -0.129\\ 
doc2 &   0.014&  0.091 & 0.128& -0.019& -0.140& -0.162\\  
doc3 & 0.060 & 0.039&  0.060& -0.025& -0.084& -0.098\\ 
doc4 & 0.014& -0.016&  0.014& -0.019&  0.034& -0.011\\ 
doc5 & -0.112& -0.119& -0.112&  0.020&  0.260&  0.204\\ 
doc6 &  -0.144& -0.154& -0.144&  0.069 & 0.164&  0.338\\ 
\hline 
\end{tabular}  
\end{table}

\begin{table}[htbp]
\centering  
\caption{The singular values,  the inertia, and the proportions of explained total inertia for each dimension of CA.} 
\label{T6*6dtcasingular}
\begin{tabular}{ccccc}    
\hline
 & dim1 & dim2 & dim3 & dim4
 \\
 \hline
singular value &  0.689& 0.131& 0.124& 0.044\\
inertia & 0.475& 0.017& 0.015& 0.002\\
the proportion of inertia & 0.932& 0.034& 0.030& 0.004  \\
\hline 
\end{tabular}  
\end{table}

We perform an SVD of $\bm{D}_r^{-\frac{1}{2}}(\bm{P}-\bm{E})\bm{D}_{c}^{-\frac{1}{2}}$ in Table~\ref{T6*6dtSR} and find Table~\ref{T6*6dtcasingular}. Due to subtracting $\bm{E}$ from $\bm{P}$, the rank of the matrix in Table~\ref{T6*6dtSR} is 4, which is 1 less than that in Table~\ref{T6*6dt}. The proportion of the total inertia explained by only the first dimension accounts for 0.932 of the total inertia. The matrix $\bm{D}_r^{-\frac{1}{2}}(\bm{P}-\bm{E})\bm{D}_{c}^{-\frac{1}{2}}$ in Table~\ref{T6*6dtSR} is approximated in the first two dimensions as follows:
\begin{equation}
\label{Eq:exampleCASVD}
\begin{aligned}
\bm{D}_r^{-\frac{1}{2}}(\bm{P}-\bm{E})\bm{D}_{c}^{-\frac{1}{2}}
&\approx \bm{U}^{sr}_2\bm{\Sigma}^{sr}_2(\bm{V}^{sr}_2)^T  \\&= 
\left[
 \begin{array}{rrrrrr}
    -0.286 & 0.789\\
   -0.368& -0.517 \\
  -0.231& -0.025\\
  0.007& -0.138\\
  0.547& -0.206\\
 0.656&  0.220
  \end{array}
  \right]\left[
 \begin{matrix}
 0.689  & 0\\
 0 & 0.131
  \end{matrix}
  \right] 
 \left[
 \begin{array}{rrrrrr}
    -0.301&  0.544 \\
  -0.338&  0.090\\
  -0.303& -0.761\\
   0.102&  0.152\\
   0.512& -0.275\\
   0.656&  0.136
  \end{array}
  \right]^T  
\end{aligned}
\end{equation}

\begin{figure}[htbp] 
    \centering
    \subfigure[]{
      \label{F66biplotCAnewfs3}
        \begin{minipage}[t]{0.41\linewidth}
        \centering
        \includegraphics[width=1\textwidth]{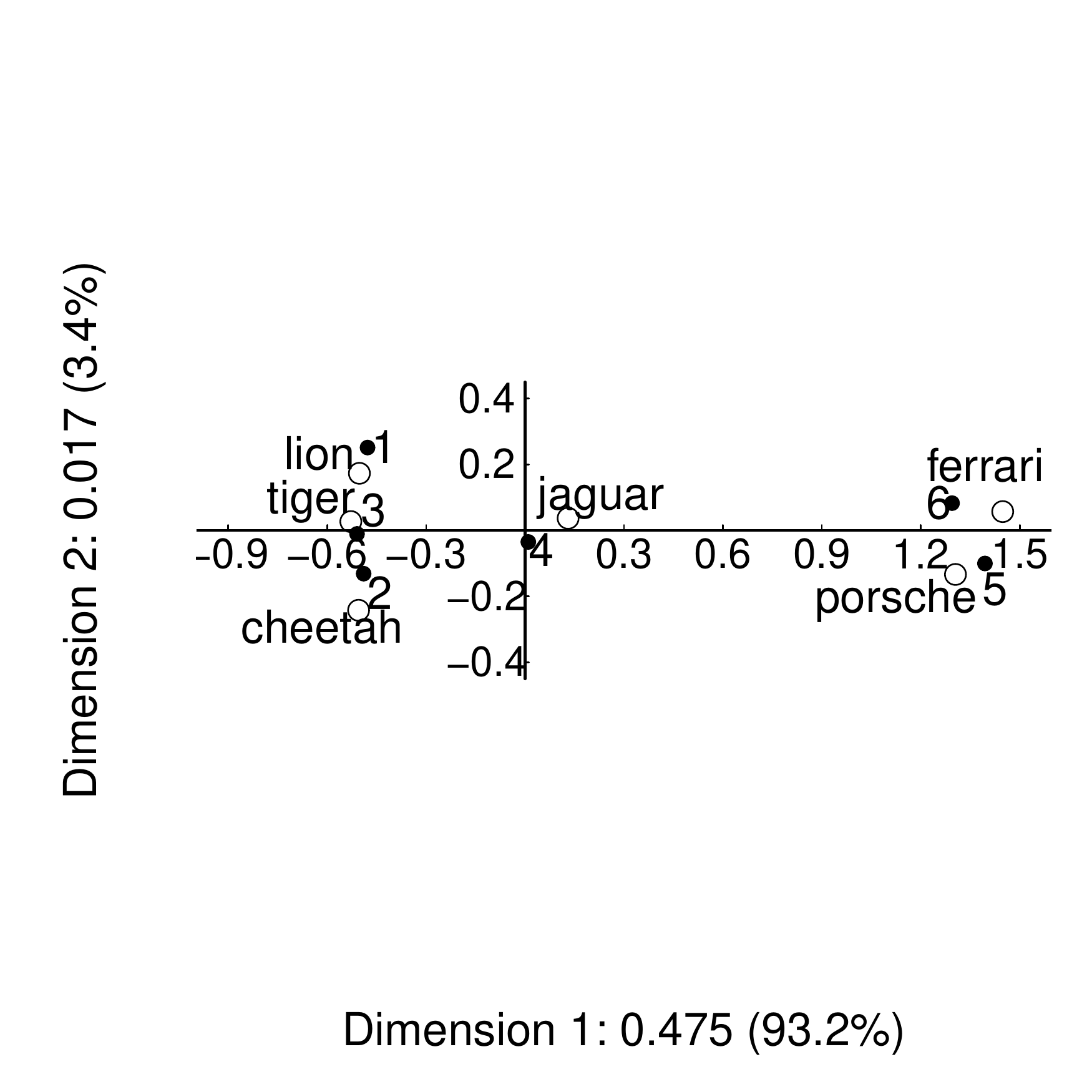}
        \end{minipage}
        }
    \subfigure[]{
       \label{F66biplotCAnewAsymmetric}
         \begin{minipage}[t]{0.41\linewidth}
         \centering
         \includegraphics[width=1\textwidth]{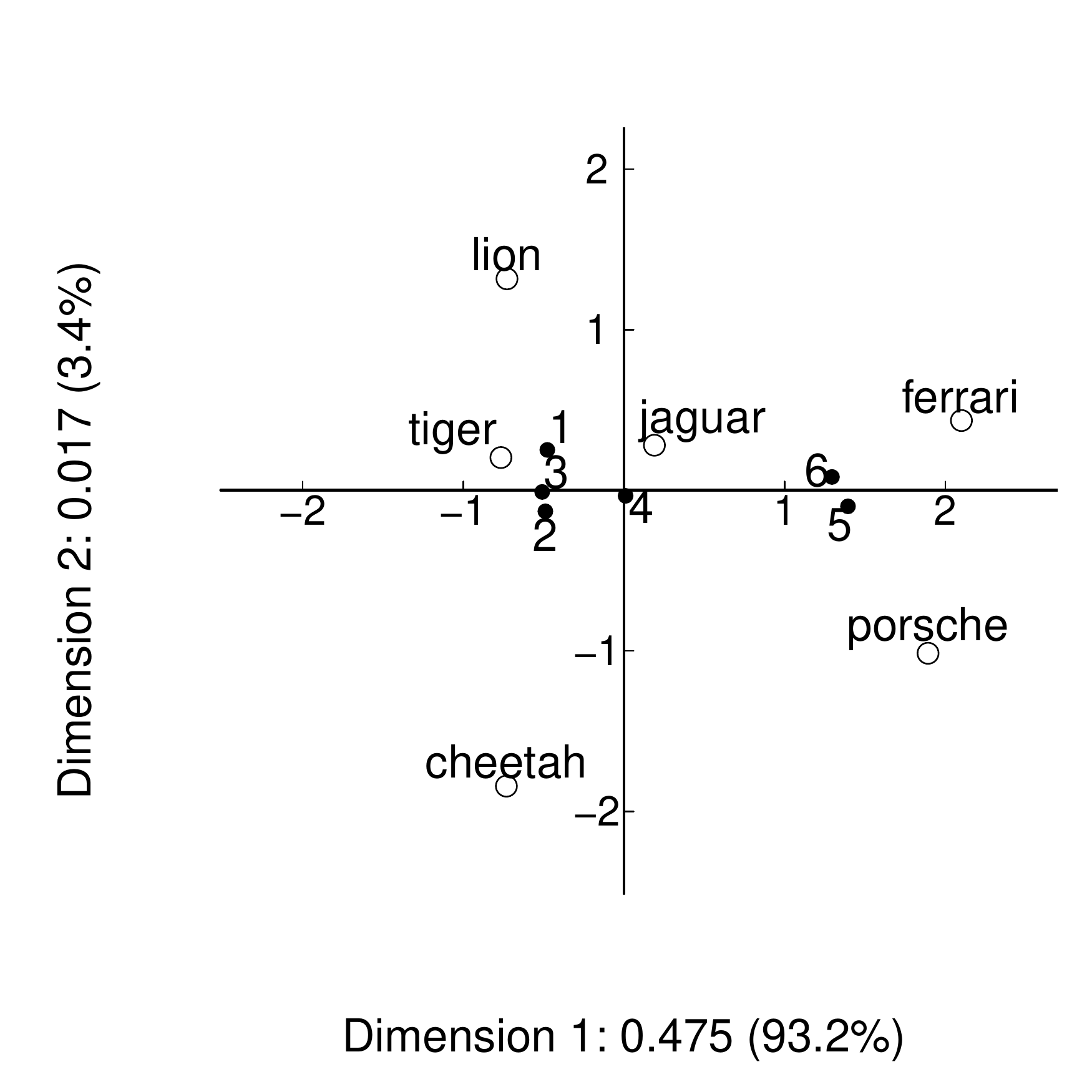}
         \end{minipage}
         }
         \vspace{8mm}
    \caption{The data of Table~\ref{T6*6dt} using CA for (a) symmetric map; (b) asymmetric map. }
    \label{FCAsymeandasyme}
\end{figure}

Figure~\ref{F66biplotCAnewfs3} is the map with a symmetric role for the rows and the columns, having $\bm{\Phi}^{sr}_2\bm{\Sigma}^{sr}_2$ and $\bm{\Gamma}^{sr}_2\bm{\Sigma}^{sr}_2$ as coordinates. The larger the deviations from document (term) points to the origin are, the larger the dependence between documents and terms. Looking only at the first dimension and document profiles’ positions, we can see that the groups furthest apart are documents 1-3 on the left-hand side, opposed to documents 5-6 on the right-hand side. They differ in opposite ways from the average row profile that lies in the origin. For the term points on the first dimension, the cat terms (\textit{tiger}, \textit{cheetah}, and \textit{lion}) lie on the left, and car terms (\textit{porsche} and \textit{ferrari}) on the right. They differ in opposite ways from the average column profile. Importantly, CA clearly displays the properties we see in the the data matrix, as document 4 lies between documents 1-3 and documents 5-6, and the term \textit{jaguar} lies between cat terms and car terms, unlike all four of the LSA based analyses presented in Figure \ref{Fweighted}.

Figure~\ref{F66biplotCAnewAsymmetric}  is the asymmetric map with documents in the weighted average of the terms ($\bm{\Phi}^{sr}_2\bm{\Sigma}^{sr}_2$ and $\bm{\Gamma}^{sr}_2$ as coordinates, notice that the position of the documents is identical as in Figure~\ref{F66biplotCAnewfs3}). From this graphic display we can study the position of the documents as they are in the weighted average of the terms, using the row profile elements as weights. For example, document 1 is closer to \textit{lion} and \textit{tiger} than to \textit{porsche} and \textit{ferrari}, because it has higher profile values than average values on terms \textit{lion} and \textit{tiger} (both 0.286 in comparison with the average profile values 0.171 and 0.195) and lower profile values on the terms \textit{porsche} and \textit{ferrari} (both 0.000 in comparison to 0.073 and 0.098), see Table~\ref{T6*6dtrp}. Thus document 1 is pulled into the direction of \textit{lion} and \textit{tiger}.

\subsection{Conclusions regarding CA}

In CA, an SVD is applied to the matrix $\bm{D}_r^{-\frac{1}{2}}(\bm{P}-\bm{E})\bm{D}_{c}^{-\frac{1}{2}}$ of standardized residuals. Due to $\bm{E}$, in CA the effect of the margins is eliminated—a solution only displays the relationships among documents and terms. In CA all points are scattered around the origin and the origin represents the profile of the row and column margins of $\bm{F}$. 

In comparison, LSA also tries to capture the relationships among documents and terms, which is not easy. The reason is that these relations are blurred by the effect of the margins that are also displayed in the LSA solution. CA does not have this property. Therefore it appears that CA is a better tool for information retrieval, natural language processing, and text mining.

\section{A unifying framework}\label{S:U}

Here we  present a unifying framework that integrates LSA and CA. This section also  serves the purpose of showing their similarities and their differences. 

To first summarize LSA (see section \ref{lsawdtm} for details), a matrix is weighted, and the weighted matrix is decomposed. Assume we start off with the document-term matrix $\bm{F}$, the row weights of $\bm{F}$ are collected in the diagonal matrix $\bm{N}$, the column weights in the diagonal matrix $\bm{G}$, and there may be local weighting of the elements $f_{ij}$ of $\bm{F}$ leading to a locally weighted matrix $\bm{L}$. Thus the weighted matrix $\bm{W}$ can be written as the matrix product 
\begin{equation}
\label{wmatrix}
\bm{W} =\bm{N}\bm{L}\bm{G}
\end{equation}

Subsequently, in LSA the matrix $\bm{W}$ is decomposed using SVD into a product of three matrices: the orthonormal matrix $\bm{U}$, the diagonal matrix $\bm{\Sigma}$ with singular values  in descending order, and the orthonormal matrix $\bm{V}$,  namely
\begin{equation}
\label{SVDint}
\bm{W} = \bm{U}\bm{\Sigma}\bm{V}^T \end{equation}
with
\begin{equation}
\label{ortho}
    \bm{U}^T\bm{U} = \bm{I} = \bm{V}^T\bm{V}
\end{equation}
Graphic representations are usually made using $\bm{U}\bm{\Sigma}$ as coordinates for the rows and $\bm{V}\bm{\Sigma}$ for the columns.

In contrast, in CA we take the SVD of the matrix of standardized residuals. Let $\bm{P}$ be the matrix with proportions $p_{ij} = f_{ij}/f_{++}$, where $f_{++}$ is the sum of all elements of $\bm{F}$; let $\bm{E}$ be the matrix with expected proportions under independence $e_{ij} = r_{i}c_{j}$, where $r_{i}$ and $c_{j}$ are the row and column sums of $\bm{P}$ respectively; let $\bm{D_r}$ and $\bm{D_c}$ be diagonal matrices with row and column sums $r_{i}$ and $c_{j}$ respectively. Thus the matrix of standardized residuals is $\bm{D}_r^{-\frac{1}{2}}(\bm{P}-\bm{E})\bm{D}_{c}^{-\frac{1}{2}}$. If we take the SVD of this matrix we get (\ref{SP3}), 
\begin{equation}
\label{SP3x}
\bm{D}_r^{-\frac{1}{2}}(\bm{P}-\bm{E})\bm{D}_{c}^{-\frac{1}{2}} =  \bm{U} \bm{\Sigma} \bm{V}^T
\end{equation}
In CA the matrices $\bm{U}$ and $\bm{V}$ are further adjusted by
\begin{equation}
\bm{\Phi}=\bm{D}_r^{-\frac{1}{2}}\bm{U}, \\
\bm{\Gamma} = \bm{D}_c^{-\frac{1}{2}}\bm{V}
\end{equation}
so that we can write 
\begin{equation}
\label{SP4a}
 \bm{D}_r^{-1}(\bm{P}-\bm{E})\bm{D}_{c}^{-1} =  \bm{\Phi}\bm{\Sigma} \bm{\Gamma}^T
\end{equation}
with
\begin{equation}
\label{wortho}
    \bm{\Phi}^T \bm{D_r} \bm{\Phi} = \bm{I} = \bm{\Gamma}^T \bm{D_c} \bm{\Gamma}
\end{equation}
Graphic representations are usually made using  $\bm{\Phi} \bm{\Sigma}$ and  $\bm{\Gamma} \bm{\Sigma}$ as coordinates for the rows and columns respectively.

This brings us to the point where we can formulate a unifying framework. We distinguish the matrix to be analyzed and the decomposition of this matrix. For the matrix to be analyzed the weighted matrix defined in (\ref{wmatrix}) can be used by LSA as well as by CA. Equation (\ref{wmatrix}) is sufficiently general for LSA. For CA, using
(\ref{SP3x}), we set $\bm{N} = \bm{D_r}^{-\frac{1}{2}}$, $\bm{L} = \bm{P} - \bm{E}$, and $\bm{G} = \bm{D_c}^{-\frac{1}{2}}$. This shows that the matrix decomposed in CA in (\ref{SP3x}) can be formulated in the LSA framework in (\ref{wmatrix}).

The decomposition used in LSA leads to orthonormal matrices $\bm{U}$ and $\bm{V}$ used for coordinates, see (\ref{ortho}), whereas in CA the decomposition leads to weighted orthonormal matrices $\bm{\Phi}$ and $\bm{\Gamma}$ , see (\ref{wortho}). If we rewrite (\ref{ortho}) as $\bm{U}^T\bm{I}\bm{U} = \bm{I} = \bm{V}^T\bm{I}\bm{V}$, we see this is a difference between using an identity metric $\bm{I}$ and a metric defined by the margins that are collected in $\bm{D_r}$ and in $\bm{D_c}$. The influence of this metric used in CA is most clearly visible in the definition of the chi-squared distances (\ref{chidistace}), that makes that, for example, for row profiles $i$ and $i'$, equally large differences between columns $j$ and $j'$ are weighted by the margins of $j$ and $j'$ in such a way that a column with a smaller margin takes a larger part in the chi-squared distance between $i$ and $i'$.

\section{Text categorization}\label{DCpar}

LSA is widely used in text categorization \citep{zhang2011comparative, elghazel2016ensemble, dzisevivc2019text, phillips2021comparing}. However, to our best knowledge, few papers on text categorization use CA, even though CA is similar to LSA. In this section, we compare the performance of LSA and CA in text categorization of three English datasets: BBCNews, BBCSport, and 20 Newsgroups. These datasets have recently been studied in the evaluation of text categorization, for example \citet{barman2020novel}.

\subsection{Datasets and methods} \label{DCpar-datmet}

The BBCNews dataset \citep{greene2006practical} consists of 2,225 documents that are divided into five categories: “Business” (510 documents), “Entertainment" (386), “Politics" (417), “Sport" (511), and “Technology" (401). The BBCSport dataset \citep{greene2006practical} consists of 737 documents  that are divided into five categories: “athletics" (101), “cricket" (124), “football" (265), “rugby" (147), and “tennis" (100). The 20 Newsgroups dataset, i.e. the 20news-bydata version \citep{newsgroups20}, consists of 18,846 documents that are divided into 20 categories. The dataset is sorted into a training (60 per cent) and a test set (40 per cent). We use a subset of these documents. Specifically, we choose 2,963 documents from three categories: “comp.graphics" (584 documents for training set and 389 documents for test set), “rec.sport.hockey" (600 and 399), and “sci.crypt" (595 and 396). The reason we choose a subset (three categories) of 20 Newsgroups is that we want to explore text categorization for datasets with a different but similar number of categories: six (for \textit{Wilhelmus} dataset in Section~\ref{S:ER}), five (for BBCNews), four (for BBCSport), and three (for a subset of 20 Newsgroups).

To pre-process these datasets we project all characters to lower case, remove punctuation marks, numbers, and stop words, and apply lemmatization. Subsequently, terms with frequencies lower than 10 are ignored. In addition, following \citet{silge2017text}, we remove unwanted parts of the 20 Newsgroups dataset such as headers (including fields like “From:” or “Reply-To:” that describe the message), because these are mostly irrelevant for text categorization.

We use two approaches to compare LSA and CA. One is visualization, where we use LSA and CA to visualize documents by projecting them onto two dimensions. The other is to use distance measures to quantitatively evaluate and compare performance in text categorization. We use four different methods based on Euclidean distance for measuring the distance from a document to a set of documents \citep{guthrie2008unsupervised, koppel2013automatically, KESTEMONT201686}. We choose the Euclidean distance because it plays a central role in the geometric interpretation of LSA and CA (see section \ref{S:LSA} and \ref{S:CA}). 

\begin{description}

\item[Centroid] Euclidean distance between the document and the centroid of the set of documents. The centroid for a set of documents is calculated by averaging the coordinates across all these documents. 

\end{description}
\noindent In the other three methods we first calculate the Euclidean distance between the document and every document of the set of documents. 

\begin{description}

\item[Average] average of these Euclidean distances 
\item[Single] the minimum Euclidean distance among the Euclidean distances

\item[Complete] the maximum Euclidean distance among the Euclidean distances.

\end{description}
These four methods are similar to the procedures of measuring the distance between clusters in hierarchical clustering analysis, using the centroid, average, single, and complete linkage method respectively \citep{jarman2020hierarchical}.

In line with the foregoing sections, we denote the raw document-term matrix by $\bm{F}$. In the case of LSA we examine four versions: LSA of $\bm{F}$ (LSA-RAW), LSA of the row-normalized matrices $\bm{F}^{L1}$ (LSA-NROWL1) and $\bm{F}^{L2}$ (LSA-NROWL2), and LSA of the TF-IDF matrix $\bm{F}^{\text{TF-IDF}}$ (LSA-TFIDF). In addition, we also compare performance with the raw document-term matrix, denoted as RAW, where no dimensionality reduction has taken place.

\subsection{Visualization}

\begin{figure}[htbp]
    \centering
       \subfigure[]{
       \label{FBBCNewsLSARAW}
         \begin{minipage}[t]{0.41\linewidth}
         \centering
         \includegraphics[width=1\textwidth]{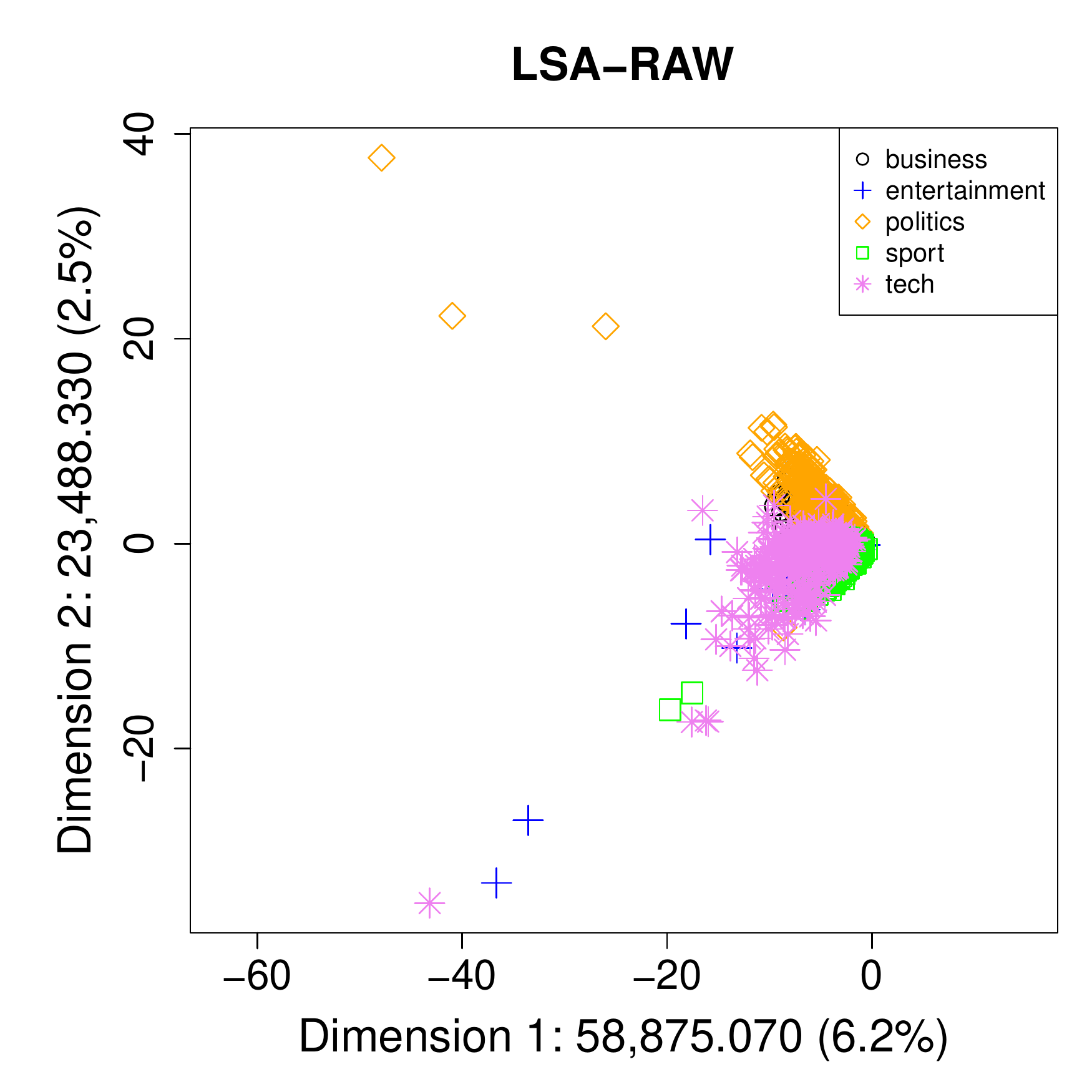}
         \end{minipage}
         }
      \subfigure[]{
       \label{FBBCNewsLSANROWL1}
         \begin{minipage}[t]{0.41\linewidth}
         \centering
         \includegraphics[width=1\textwidth]{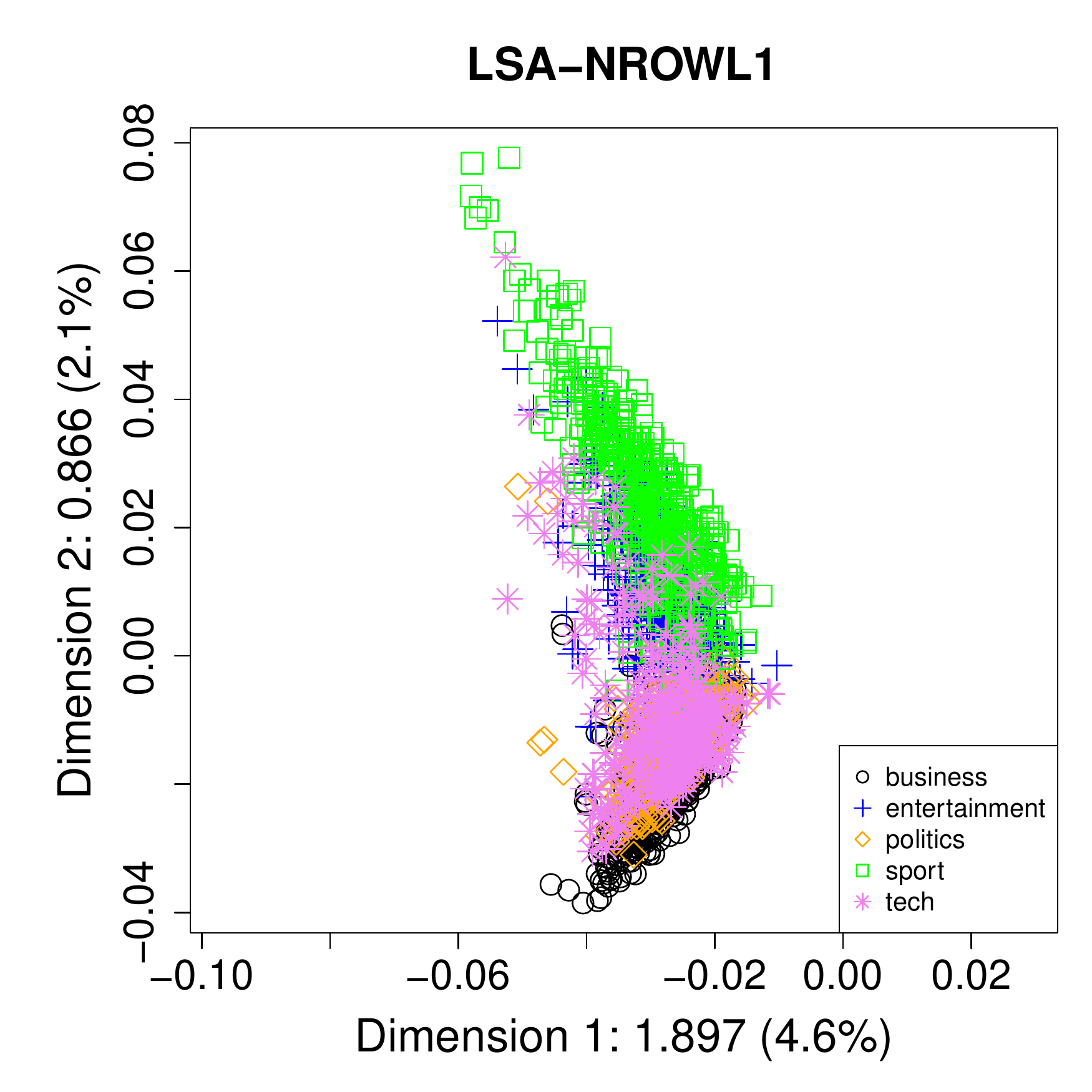}
         \end{minipage}
         }
       \subfigure[]{
       \label{FBBCNewsLSANROWL2}
         \begin{minipage}[t]{0.41\linewidth}
         \centering
         \includegraphics[width=1\textwidth]{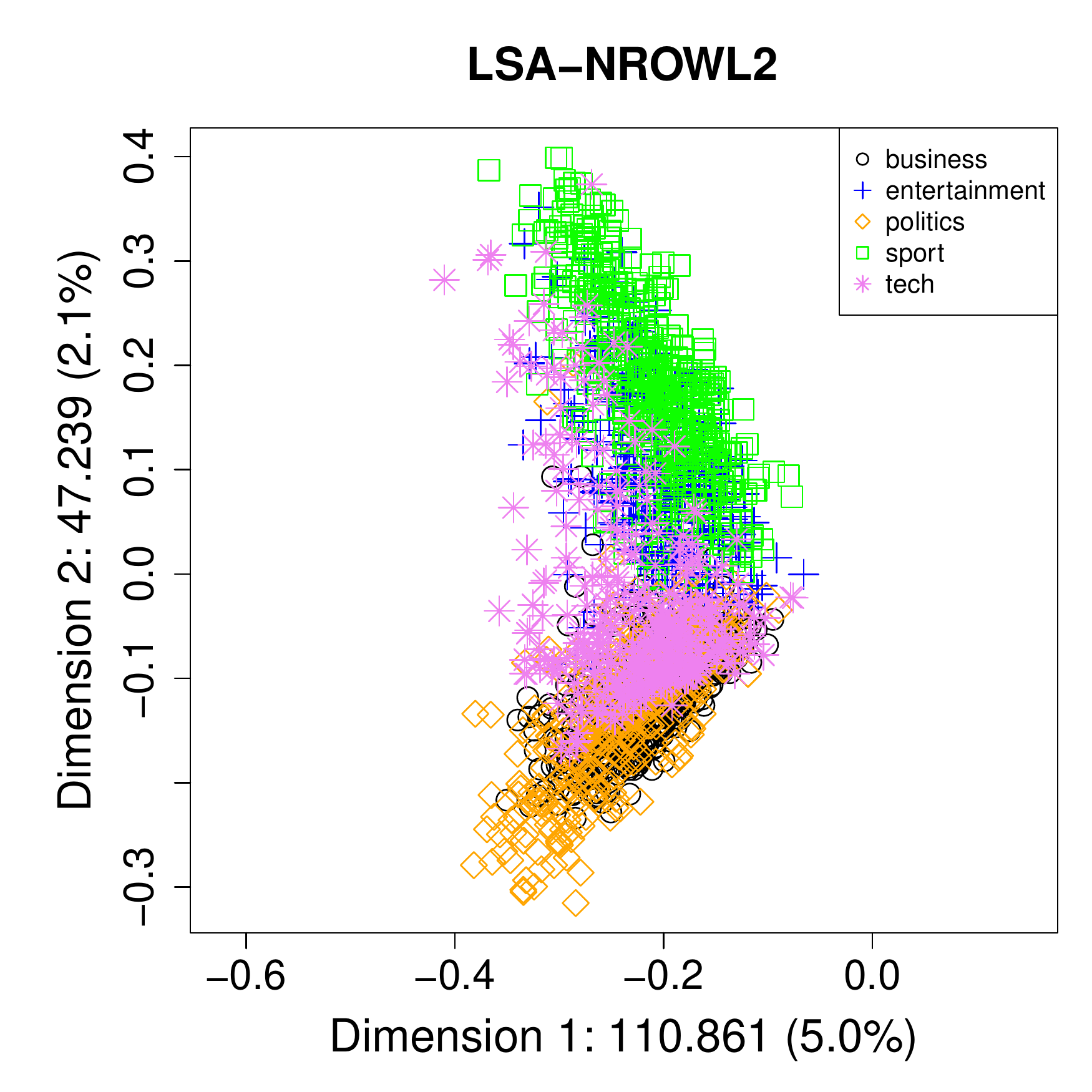}
         \end{minipage}
         }
      \subfigure[]{
       \label{FBBCNewsLSATFIDF}
         \begin{minipage}[t]{0.41\linewidth}
         \centering
         \includegraphics[width=1\textwidth]{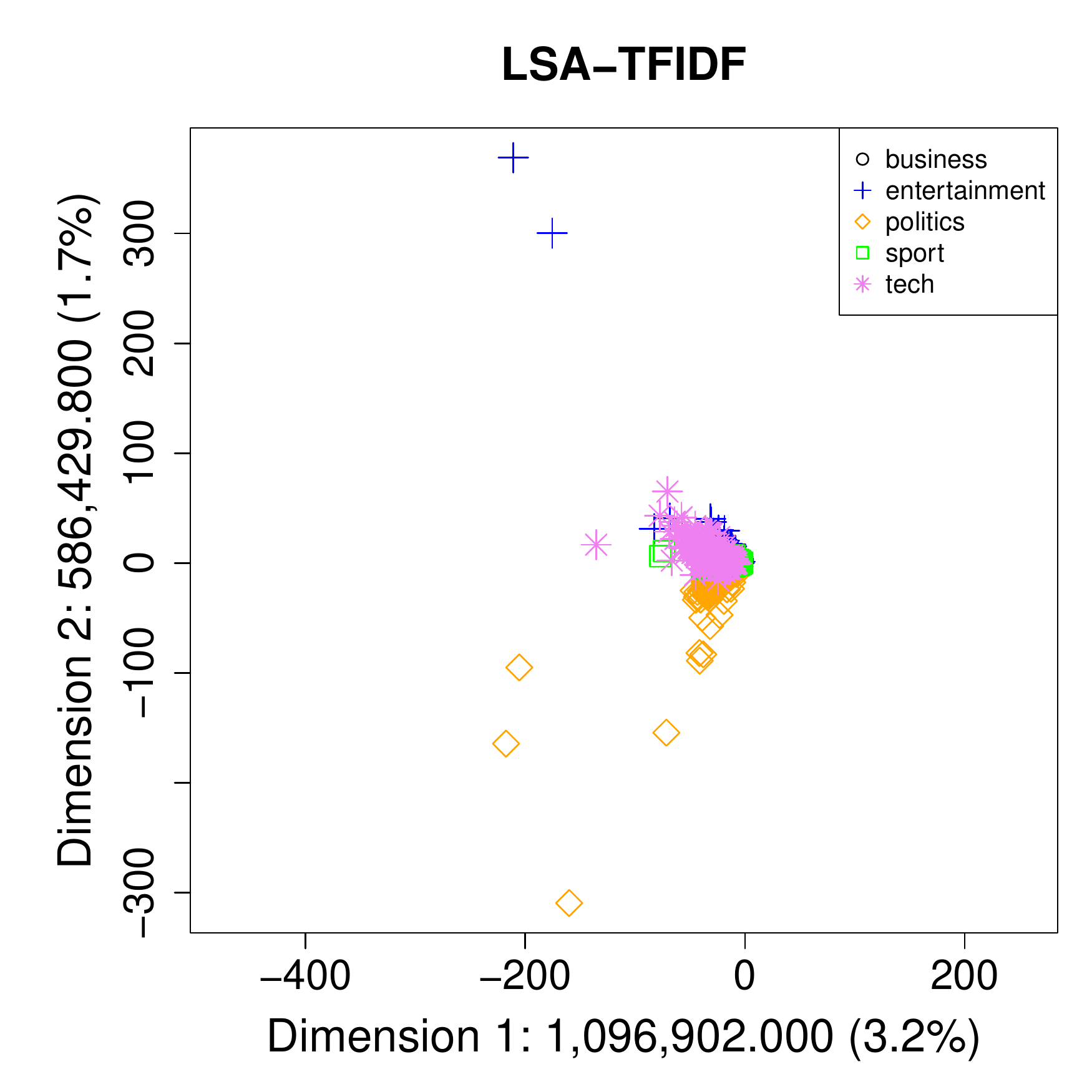}
         \end{minipage}
         }
       \subfigure[]{
       \label{FBBCNewsCA}
         \begin{minipage}[t]{0.41\linewidth}
         \centering
         \includegraphics[width=1\textwidth]{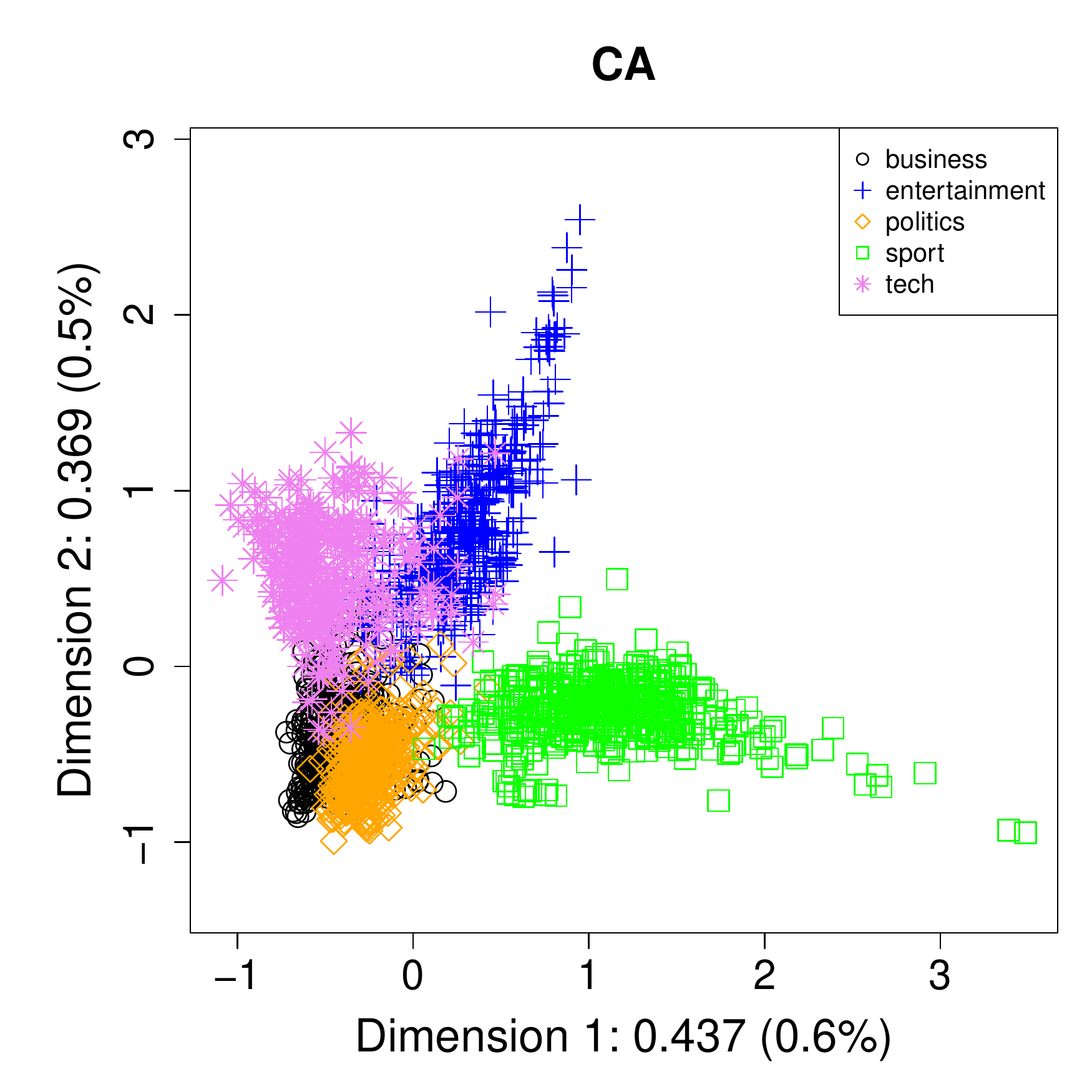}
         \end{minipage}
         }
         \vspace{8mm}
    \caption{The first two dimensions for each document of BBCNews dataset by (a) LSA-RAW; (b) LSA-NROWL1; (c) LSA-NROWL2; (d) LSA-TFIDF; (e) CA.}
    \label{FBBCNews}
\end{figure}

The 2,225 documents of the BBCNews dataset lead to a document-term matrix of size 2,225 $\times$ 5,050. Figure~\ref{FBBCNews} shows the results of an analysis of this document-term matrix by the four LSA methods (LSA-RAW, LSA-NROWL1, LSA-NROWL2, LSA-TFIDF) and CA. On this dataset, we find that, although the percentage of the total sum of squared singular values in the first two dimensions for CA is lower than the four LSA methods, the four LSA methods do not separate the classes well but CA does a reasonably good job. This is because the margins play an important role in the first two dimensions for the four LSA methods and the relations between documents are blurred by these margins.

\begin{figure}[htbp]
    \centering
       \subfigure[]{
       \label{FBBCSportLSARAW}
         \begin{minipage}[t]{0.41\linewidth}
         \centering
         \includegraphics[width=1\textwidth]{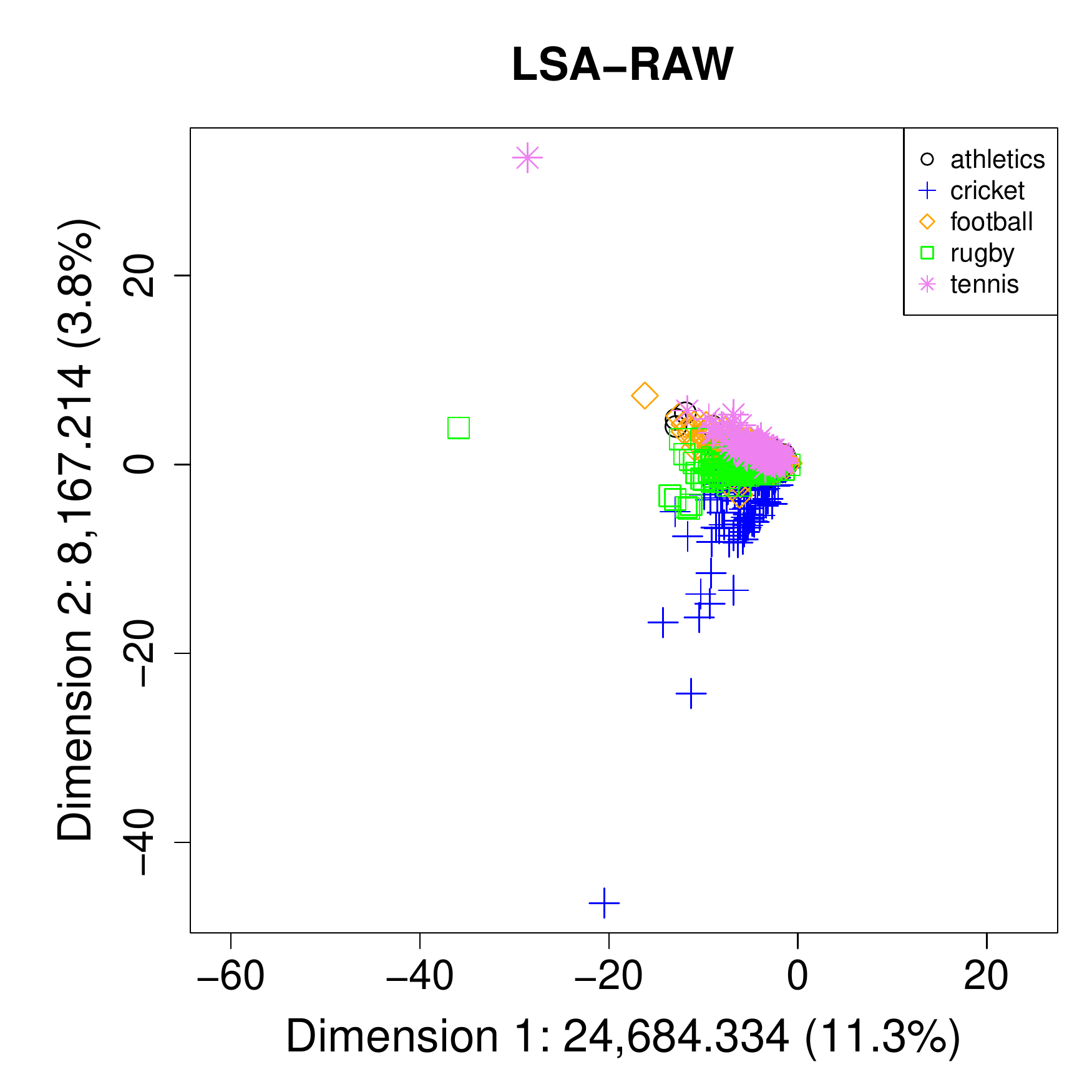}
         \end{minipage}
         }
      \subfigure[]{
       \label{FBBCSportLSANROWL1}
         \begin{minipage}[t]{0.41\linewidth}
         \centering
         \includegraphics[width=1\textwidth]{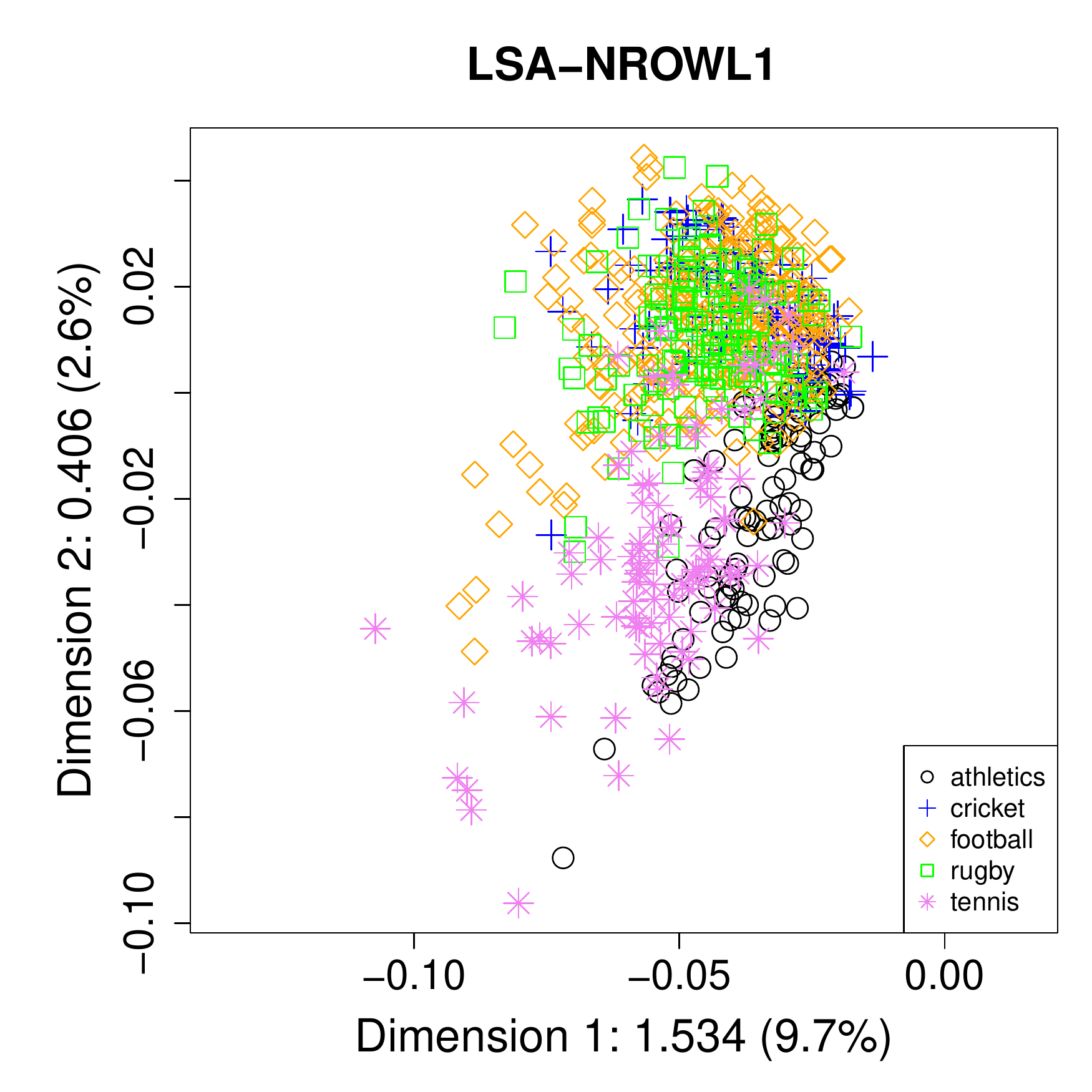}
         \end{minipage}
         }
       \subfigure[]{
       \label{FBBCSportLSANROWL2}
         \begin{minipage}[t]{0.41\linewidth}
         \centering
         \includegraphics[width=1\textwidth]{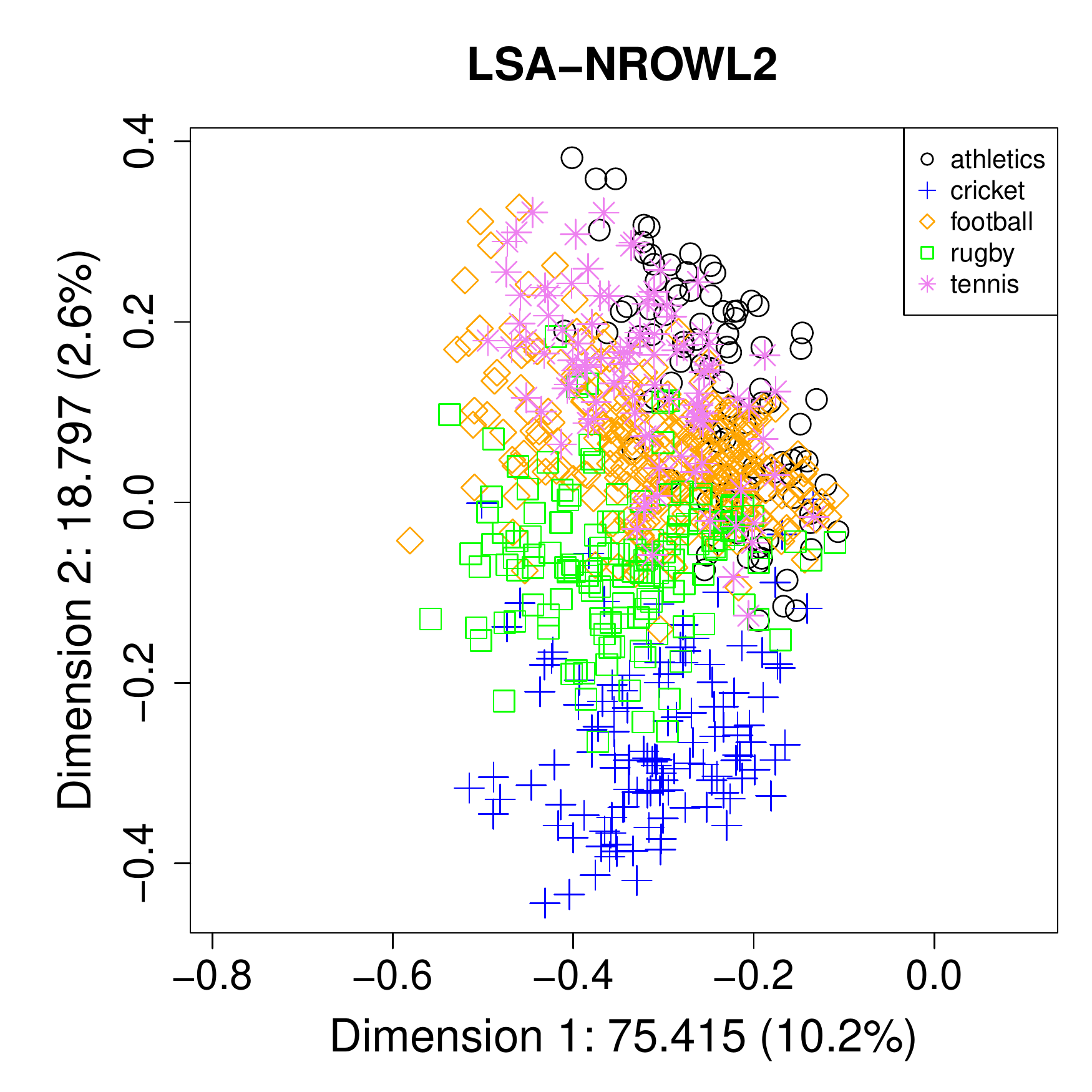}
         \end{minipage}
         }
      \subfigure[]{
       \label{FBBCSportLSATFIDF}
         \begin{minipage}[t]{0.41\linewidth}
         \centering
         \includegraphics[width=1\textwidth]{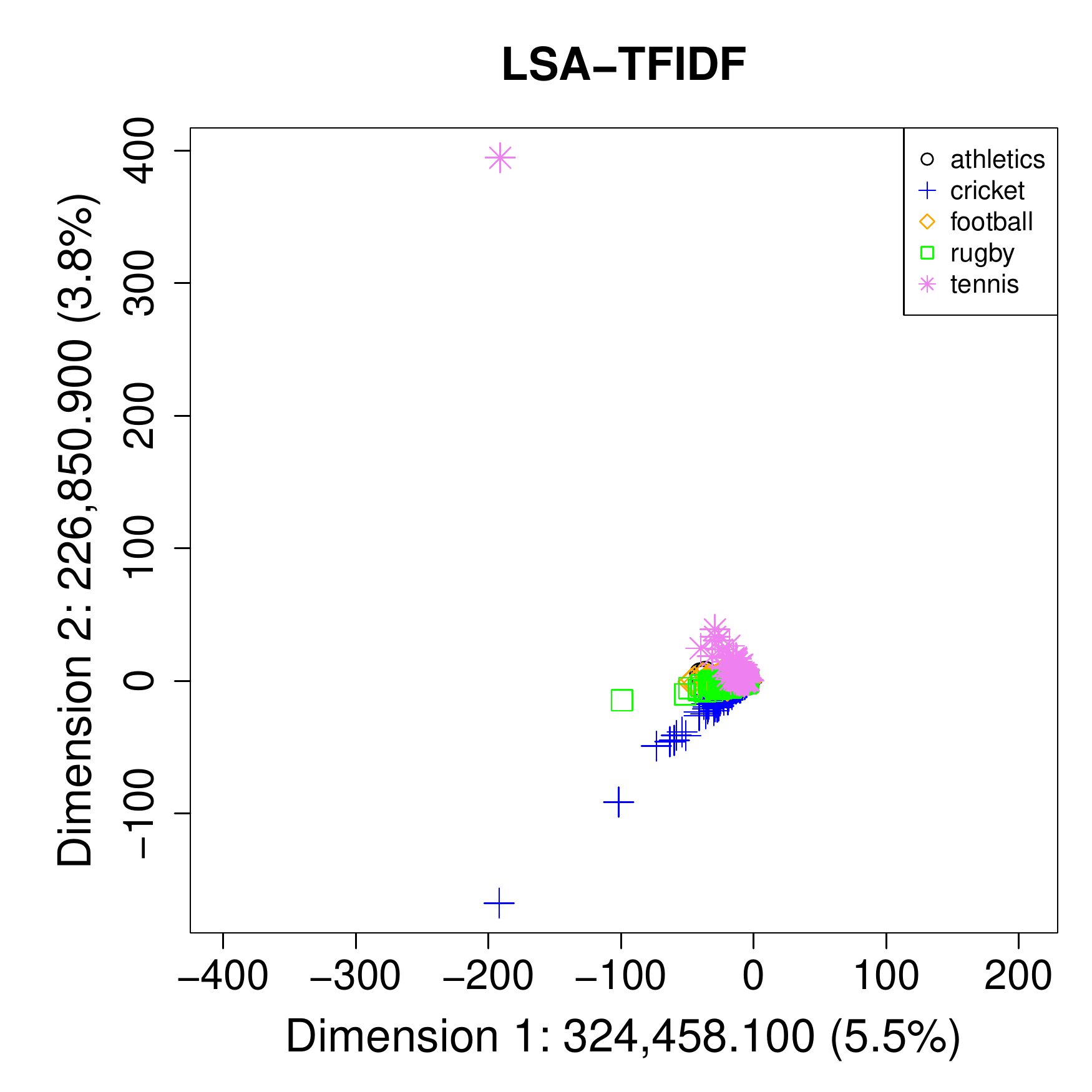}
         \end{minipage}
         }
       \subfigure[]{
       \label{FBBCSportCA}
         \begin{minipage}[t]{0.41\linewidth}
         \centering
         \includegraphics[width=1\textwidth]{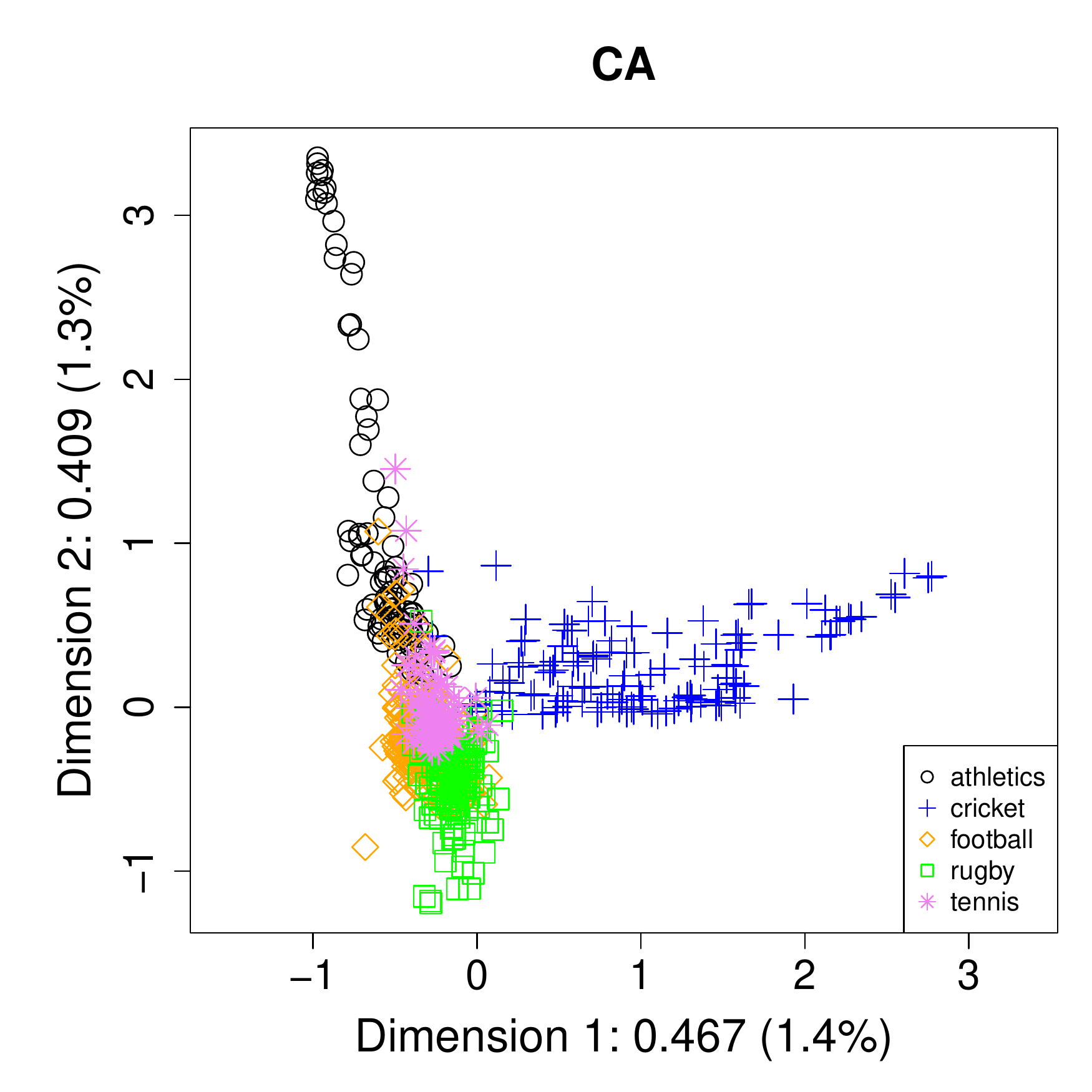}
         \end{minipage}
         }
         \vspace{8mm}
    \caption{The first two dimensions for each document of BBCSport dataset by (a) LSA-RAW; (b) LSA-NROWL1; (c) LSA-NROWL2; (d) LSA-TFIDF; (e) CA.}
    \label{FBBCSport}
\end{figure}

The 737 documents of BBCSport dataset lead to a document-term matrix of size 737 $\times$ 2,071. Figure~\ref{FBBCSport} shows the results of an analysis of this document-term matrix. Again, we find that the LSA methods do not separate the classes well, but CA does a reasonably good job.

\begin{figure}[htbp]
    \centering
       \subfigure[]{
       \label{F20NewsgroupsLSARAW}
         \begin{minipage}[t]{0.41\linewidth}
         \centering
         \includegraphics[width=1\textwidth]{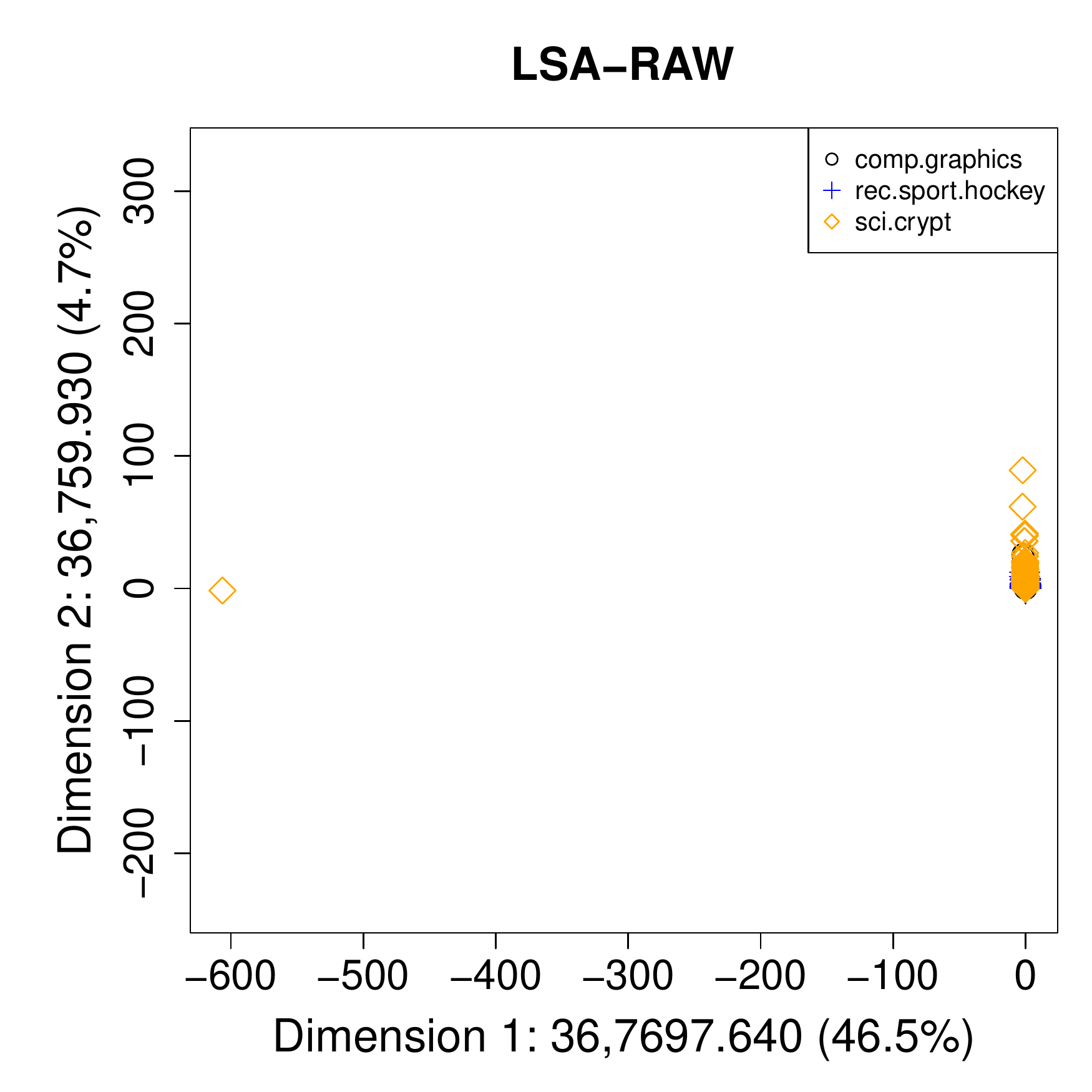}
         \end{minipage}
         }
      \subfigure[]{
       \label{F20NewsgroupsLSANROWL1}
         \begin{minipage}[t]{0.41\linewidth}
         \centering
         \includegraphics[width=1\textwidth]{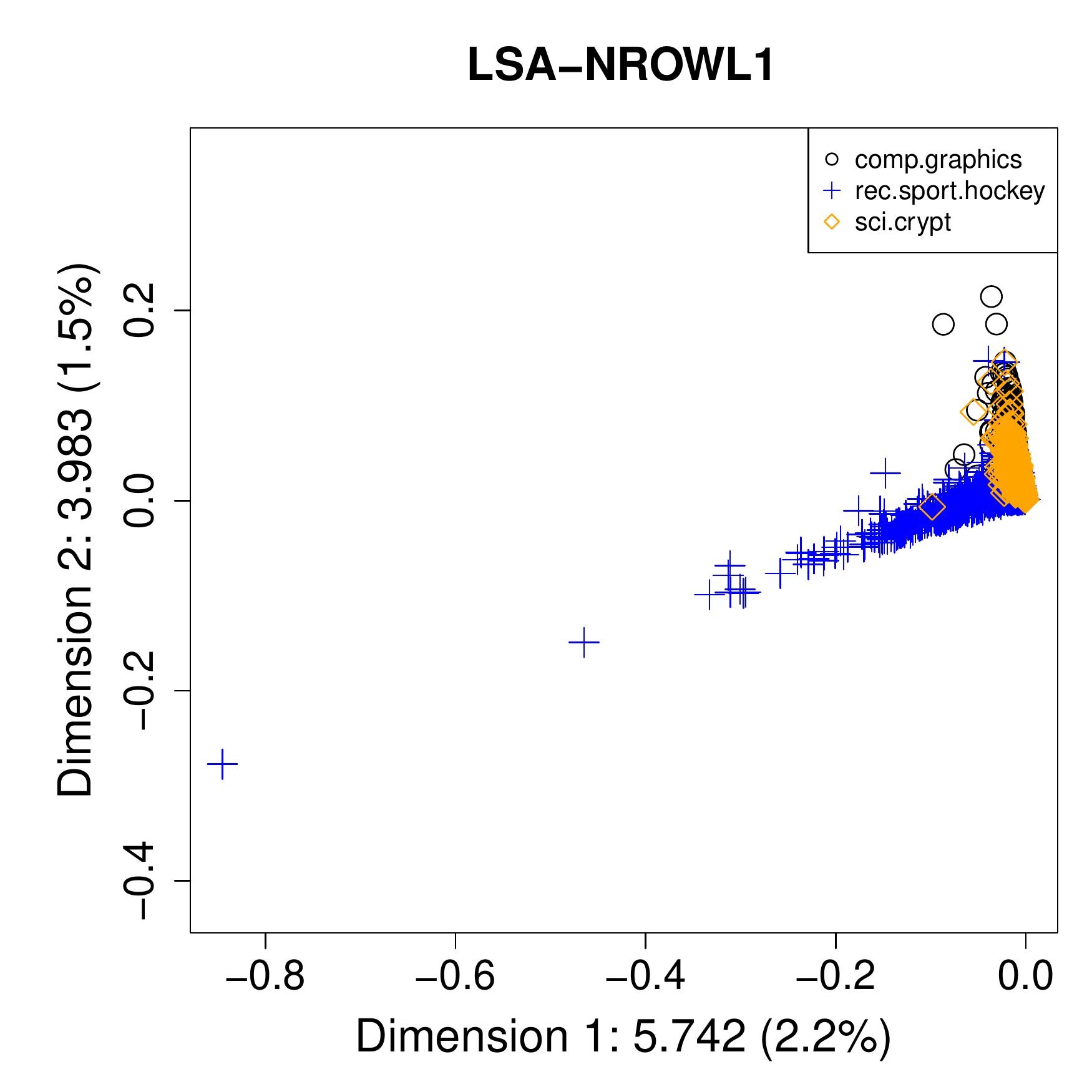}
         \end{minipage}
         }
       \subfigure[]{
       \label{F20NewsgroupsLSANROWL2}
         \begin{minipage}[t]{0.41\linewidth}
         \centering
         \includegraphics[width=1\textwidth]{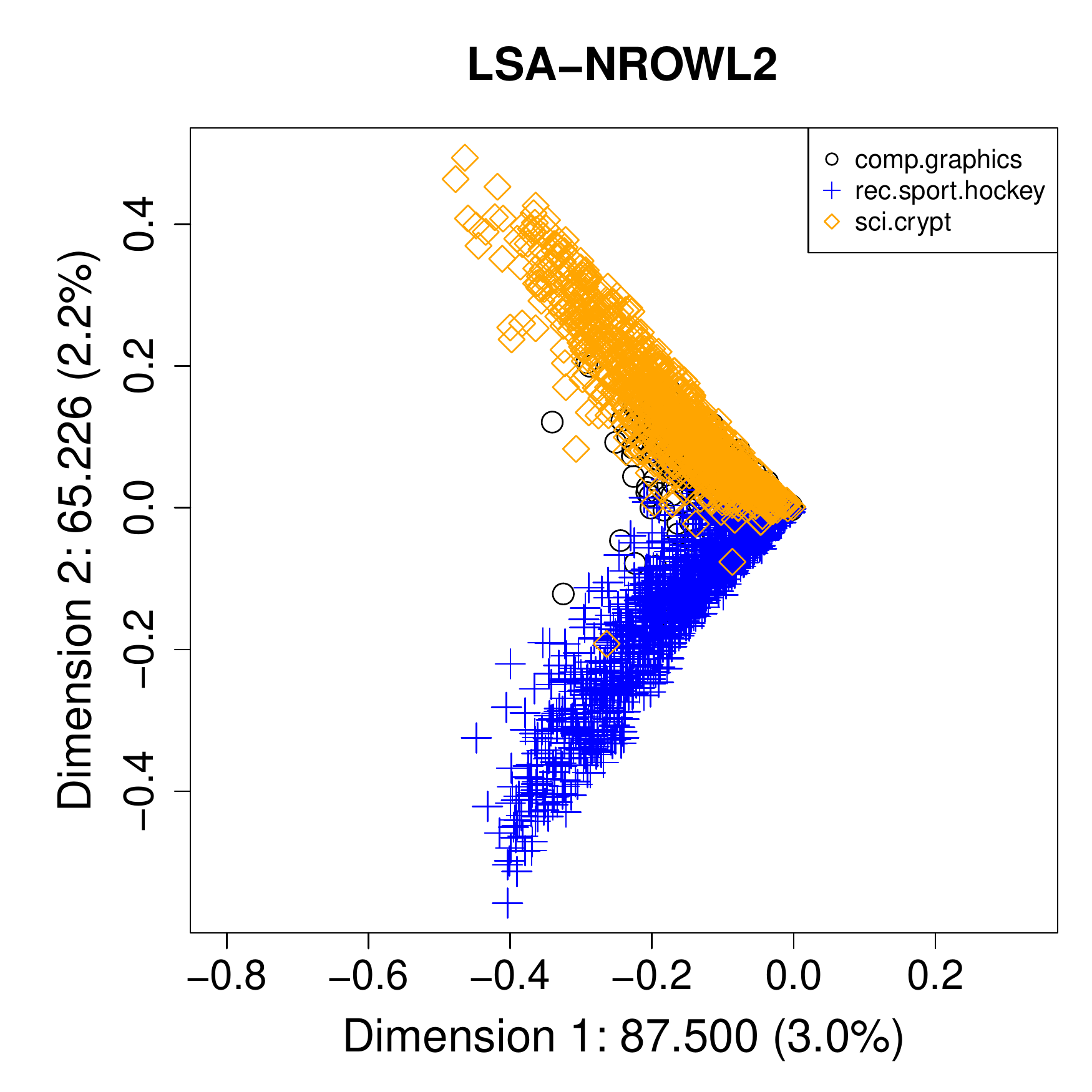}
         \end{minipage}
         }
      \subfigure[]{
       \label{F20NewsgroupsLSATFIDF}
         \begin{minipage}[t]{0.41\linewidth}
         \centering
         \includegraphics[width=1\textwidth]{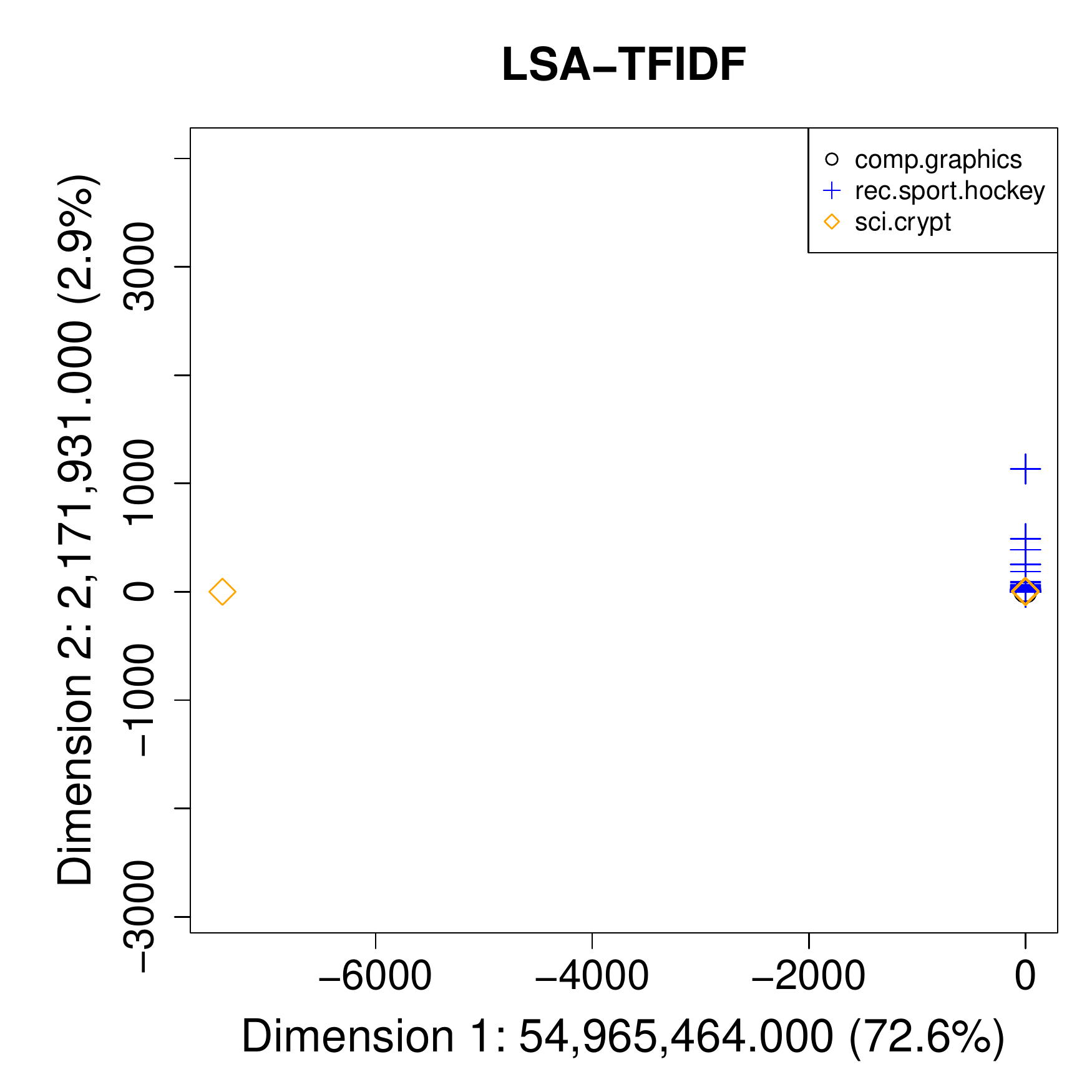}
         \end{minipage}
         }
       \subfigure[]{
       \label{F20NewsgroupsCA}
         \begin{minipage}[t]{0.41\linewidth}
         \centering
         \includegraphics[width=1\textwidth]{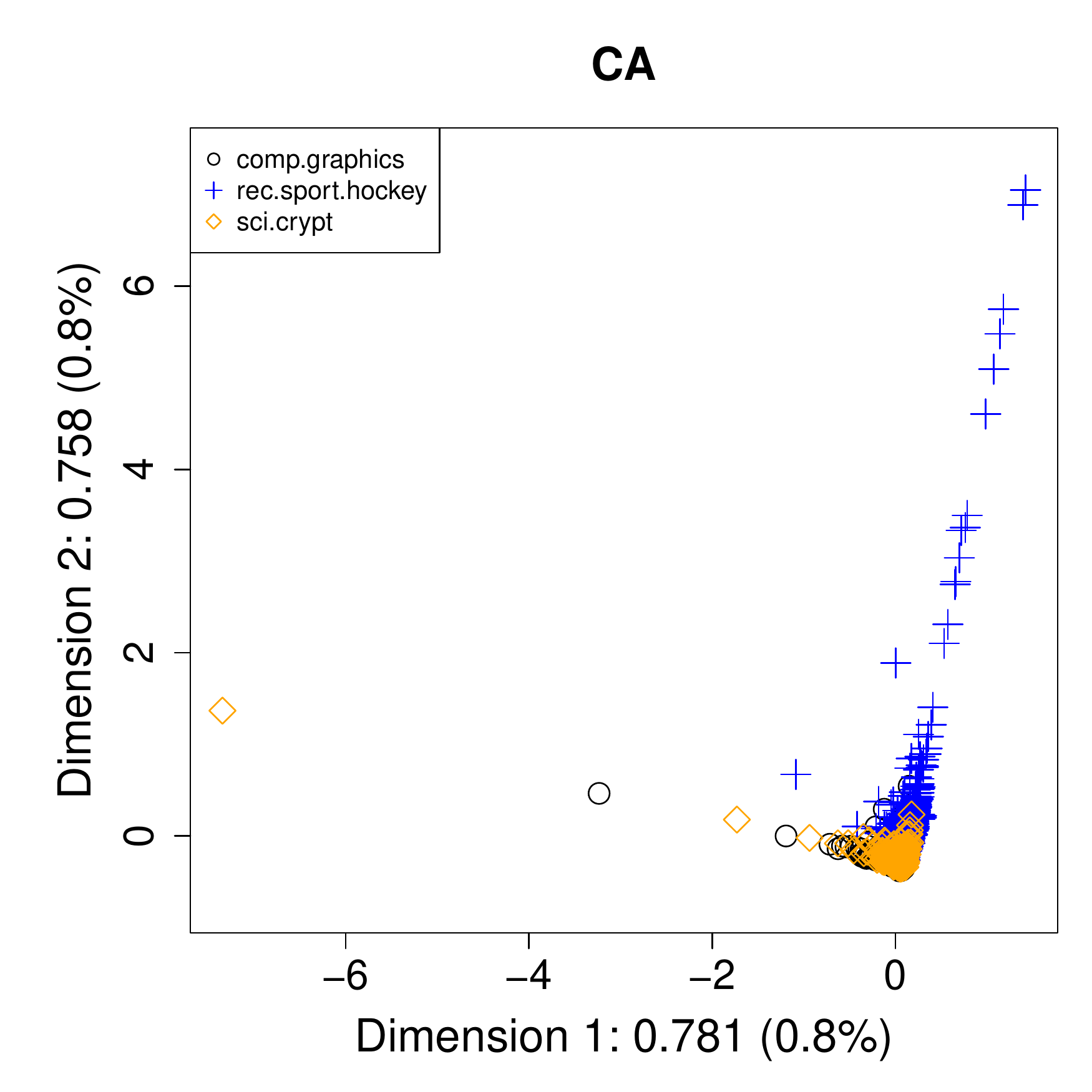}
         \end{minipage}
         }
         \vspace{8mm}
    \caption{The first two dimensions for each document of 20 Newsgroups dataset by (a) LSA-RAW; (b) LSA-NROWL1; (c) LSA-NROWL2; (d) LSA-TFIDF; (e) CA.}
    \label{F20Newsgroups}
\end{figure}

The 2,963 documents of 20 Newsgroups dataset lead to a document-term matrix of size 2,927 $\times$ 2,834.\footnote{After preprocessing, 36 documents out of 2,963 became empty documents and were removed.} Figure~\ref{F20Newsgroups} shows the results of an analysis of this document-term matrix. On this dataset, we find that CA is doing a reasonably good job, and so do LSA-NROWL1 and LSA-NROWL2.

\subsection{Distance measures}

For the 20 Newsgroups dataset, there is a training and a test set, and we assess the accuracy as a measure for the correct classification of the documents of the test set. For the 20 Newsgroups data set there are four steps. First, we apply all four varieties of LSA and CA to all documents of the training set. The documents of the test set are projected into the reduced dimensional space, see Section~\ref{Sub: OSD} and Section~\ref{S:CA}. Second, using the centroid, average, single, and complete method, for each document of the test set, the distance between the document and a set of documents for each of three categories (“comp.graphics”, “rec.sport.hockey”, “sci.crypt”) in the training set is computed. The predicted category for the document is the category with the smallest distance. Third, we compare the predicted category with the true category of the document. Finally, the accuracy is the proportion of correct classifications of all documents of the test set. For BBCNews and BBCSport datasets, in order to evaluate LSA methods and CA, we use five-fold cross validation \citep{gareth2021introduction}. That is, the dataset is randomly divided into five folds. The four folds (80 per cent of the dataset) are used as training set and the remaining one fold  (20 per cent of the dataset) is as validation set. The accuracy of each fold is obtained as in the 20 Newsgroups dataset. Then the accuracy is averaged across five folds.

For each form of LSA and for CA, there is an accuracy for each number of dimensions (for five-fold cross validation, the accuracy is averaged across five folds). The maximum accuracy is the maximum value across these accuracies. Table~\ref{Tcentroidaveragesinglecompletebbcnewssport20nesgroups} shows the maximum accuracy for LSA-RAW, LSA-NROWL1, LSANROWL2,
LSA-TFIDF, and CA for the four distance measures\footnote{For BBCSport dataset, we explore the number of all dimensions of dimensionality reduction methods. For BBCNews and 20 Newsgroups datasets, we vary the number of dimension $k$ from 1 to 450.}, along with the {\it minimum} optimal dimensions $k$ where this maximum accuracy is reached \footnote{There is not one single optimal number of dimensions that provides the maximum accuracy; for reasons of space, we show only the lowest in Tables~\ref{Tcentroidaveragesinglecompletebbcnewssport20nesgroups}, \ref{Tcentroidaveragesinglecomplete}.}. First, if we ignore the complete distance method, considering that it has low accuracy overall, CA yields the maximum accuracy compared to the RAW  method (i.e. without dimensionality reduction) as well as all four LSA methods for each combination of dataset and other distance measurement method, except for the BBCSport dataset with the average method, where CA has the second largest accuracy. Second, for each dataset CA is doing best overall. Specifically, CA with the centroid, the single, and the centroid distance method provides the best accuracy for BBCNews, BBCSport, and 20 Newsgroups datasets, respectively.

\begin{figure}[t] 
    \centering
       \subfigure[]{
       \label{FBBCNewsfoldincentroidfulldimension}
         \begin{minipage}[t]{0.41\linewidth}
         \centering
         \includegraphics[width=1\textwidth]{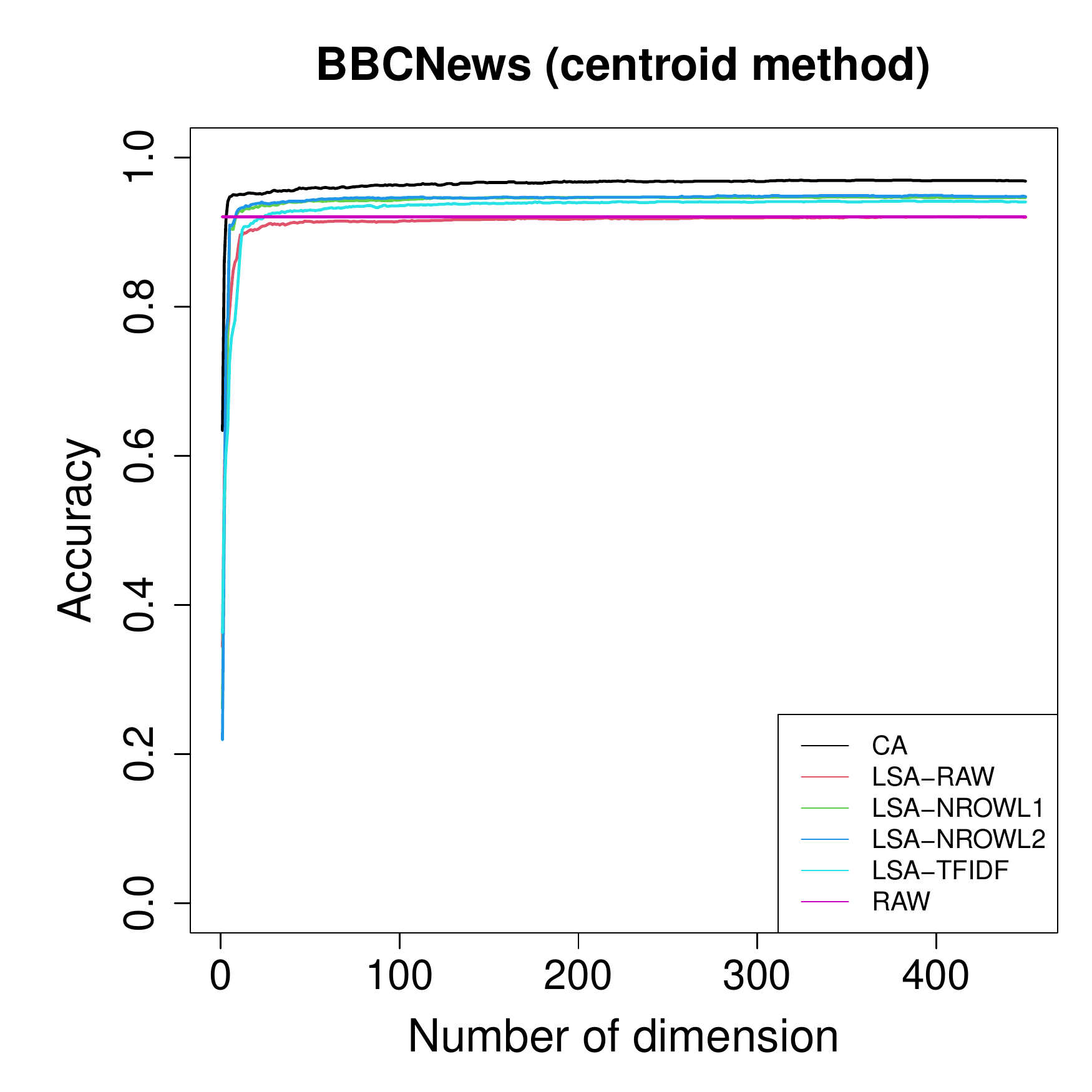}
         \end{minipage}
         }
      \subfigure[]{
       \label{FBBCSportfoldinsinglefulldimension}
         \begin{minipage}[t]{0.41\linewidth}
         \centering
         \includegraphics[width=1\textwidth]{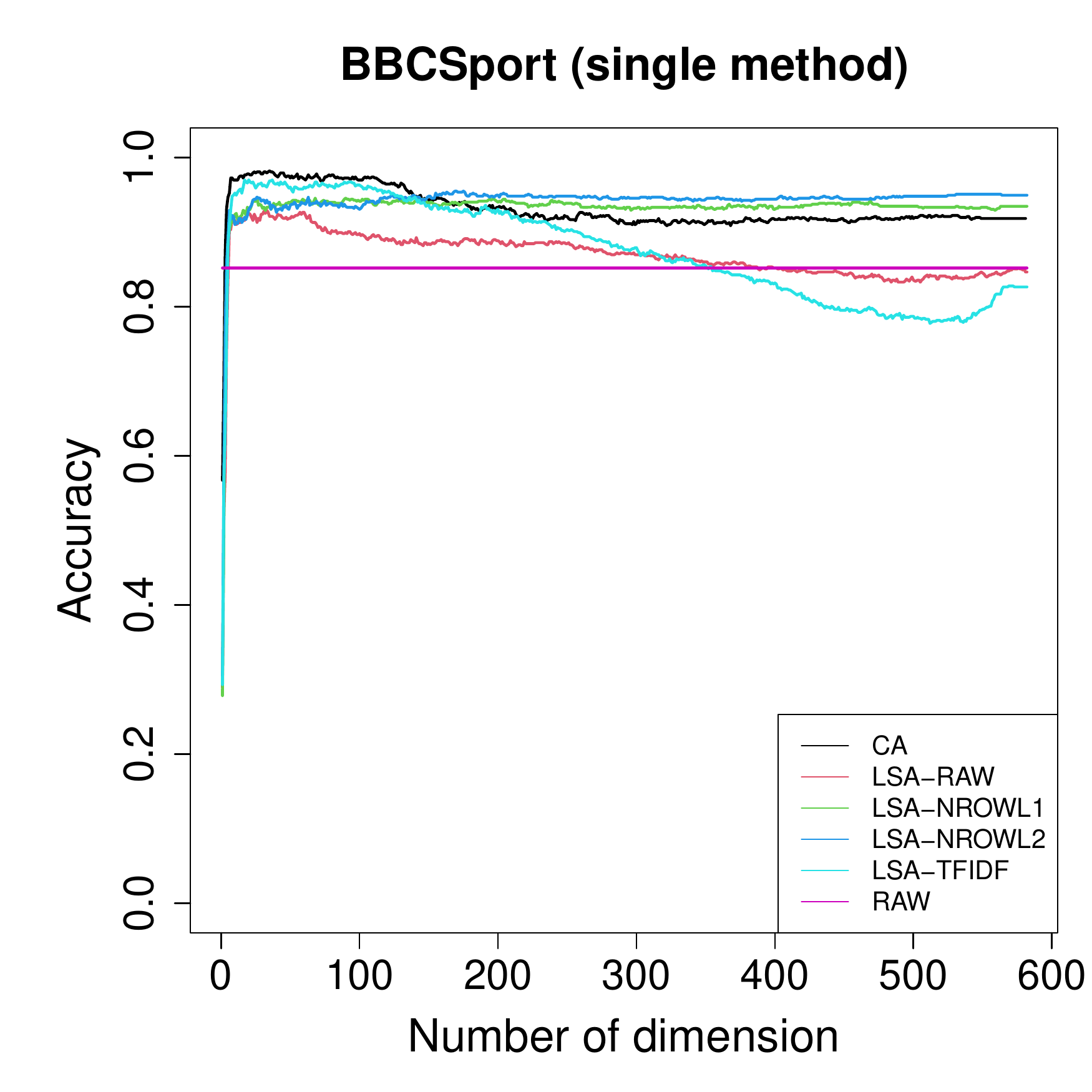}
         \end{minipage}
         }
     \subfigure[]{
       \label{F20Newsgroupsfoldinsinglefulldimension}
         \begin{minipage}[t]{0.41\linewidth}
         \centering
         \includegraphics[width=1\textwidth]{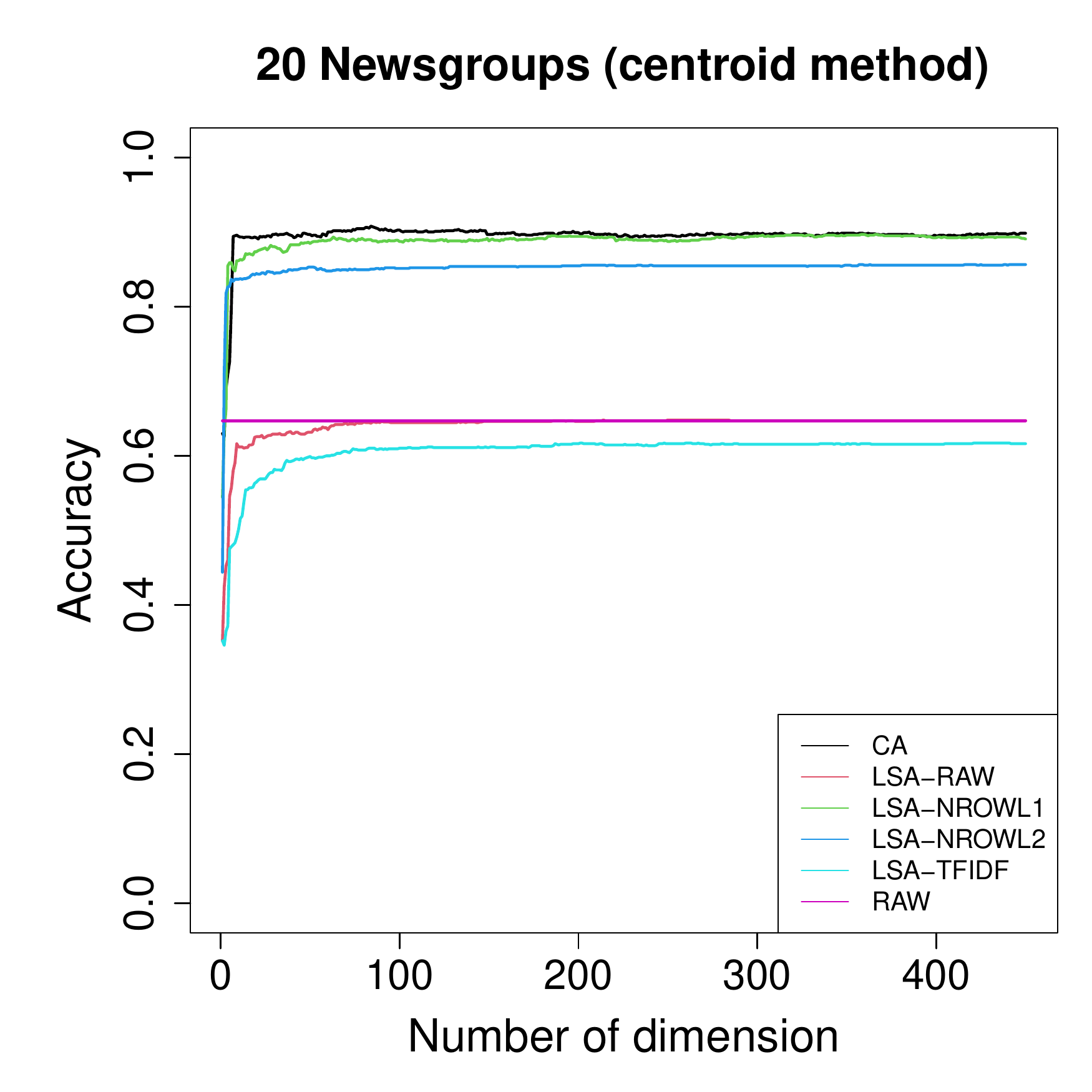}
         \end{minipage}
         }
         \vspace{8mm}
    \caption{Accuracy as a function of dimension for CA, LSA-RAW, LSA-NROWL1, LSA-NROWL2, LSA-TFIDF, and RAW}
    \label{CentroidBBC20NEWS}
\end{figure}

In order to further explore different dimensionality reduction methods under optimal distance measurement method which provides highest accuracy, Figure~\ref{CentroidBBC20NEWS} shows the accuracy as a function of the numbers of dimensions under centroid, single, and centroid methods for BBCNews, BBCSport, and 20 Newsgroups datasets, respectively. CA in combination with the optimal distance measurement method performs better than the other methods over a large range, especially for BBCNews dataset, almost irrespective of dimension.

\begin{table}[t]
\centering  
\caption{The minimum optimal dimensionality $k$ and the accuracy in $k$ for LSA-RAW, LSA-NROWL1, LSA-NROWL2, LSA-TFIDF, and CA, and the accuracy (Acc) for RAW using different distance measurement methods with the BBCNews, BBCSport, and 20 Newsgroups datasets.} 
\label{Tcentroidaveragesinglecompletebbcnewssport20nesgroups}
\begin{tabular}{cccccccccc}    
\hline
\multirow{2}{*}{Datasets}&\multirow{2}{*}{Methods}&\multicolumn{2}{c}{Centroid} &\multicolumn{2}{c}{Average} &\multicolumn{2}{c}{Single}&\multicolumn{2}{c}{Complete}
\\&&$k$&Acc&$k$&Acc&$k$&Acc&$k$&Acc\\\hline 
\multirow{6}{*}{BBCNews}& RAW& 
& 0.921&& 0.339&&0.791&&0.229\\
&LSA-RAW &401 & 0.921& 7&0.714&24&0.942&1&0.237 \\
&LSA-NROWL1&339 & 0.947&5&0.898&30& 0.948&5&0.723\\
&LSA-NROWL2& 385&  0.950&23&0.930&450&0.951&5&\bf{0.829}\\
&LSA-TFIDF&381
& 0.942 & 13&0.725&32&0.953&13&0.253\\
&CA&318&\underline{\bf{0.970}}& 5& \bf{0.943}&22&\bf{0.961}&4&0.647\\
\hdashline
\multirow{6}{*}{BBCSport}&RAW& 
& 0.917&&0.418&&0.852&&0.193\\
&LSA-RAW &72 & 0.919&9&0.843&33&0.930&9&0.332 \\
&LSA-NROWL1&275 & 0.950&10&0.928&129& 0.946&5&0.613\\
&LSA-NROWL2& 96&  0.952&103&\bf{0.950}&175&0.955&5&\bf{0.873}\\
&LSA-TFIDF&486 
& 0.931&9&0.806&20&0.970&7&0.241\\
&CA&565&\bf{0.978}&24&0.936&35& \underline{\bf{0.982}}&4&0.420\\
\hdashline
\multirow{6}{5.2em}{20 Newsgroups}&RAW& 
& 0.647&&0.330&&0.688&&0.328\\
&LSA-RAW &214 & 0.648&9&0.409&26&0.847&2&0.342 \\
&LSA-NROWL1&358 & 0.897&4&0.847&306 & 0.852& 83&0.412\\
&LSA-NROWL2& 357&  0.857& 54&0.885& 6&0.858&3&\bf{0.735}\\
&LSA-TFIDF&201 
& 0.617&1&0.347&70&0.863&1&0.340\\
&CA&84&\underline{\bf{0.908}}&7& \bf{0.888}& 27&\bf{0.902}&11&0.465\\
\hline 
\end{tabular}  
\end{table} 

\section{Authorship attribution}\label{S:ER}

In this section we examine the performance of LSA and CA on a dataset originally set up for authorship attribution. We first use the dataset to see how well LSA and CA are able to assign documents with a known author to the correct author. Second, we assign a document with unknown author to one of the known authors.

Authorship attribution is the process of identifying the authorship of a document; its applications include plagiarism detection and resolving of authorship disputes \citep{bozkurt2007authorship}, and are particularly relevant for historical texts, where other historical records are not sufficient to determine authorship. Both LSA and CA have been used for authorship attribution before. For example, \citet{soboroff1997visualizing} applied LSA with n-grams as terms to visualize authorship among biblical Hebrew texts. \citet{mccarthy2006analyzing} applied LSA to lexical features to automatically detect semantic similarities between words \citep{stamatatos2009survey}. \citet{satyam2014statistical} used LSA on a character n-gram based representation to build a similarity measure between a questioned document and known documents. \citet{mealand1995correspondence} studied the Gospel of Luke using a visualization provided by CA. \citet{mealand1997measuring} also measured genre differences in Mark by CA. \citet{mannion2004sentence} applied CA to study authorship attribution of the case of Oliver Goldsmith by visualization.

The \textit{Wilhelmus} is the national anthem of the Netherlands and its authorship is unknown and much debated. There is a substantive amount of qualitative research attempting to determine the authorship of the \textit{Wilhelmus}, with quantitative or statistical methods being used relatively recently. To the best of our knowledge, the authorship of the \textit{Wilhelmus} was first studied by statistical methods and computational means in \citet{winkel2015deutsches}, whose results on authorship attribution were inconclusive. After that, \citet{kestemont2017did, kestemont2017van}
studied the question using PCA and the General Imposters (GI) method, attributing the \textit{Wilhelmus} to the writer Datheen. \citet{vargas2017information} used the data of \citet{kestemont2017did, kestemont2017van}, and applied the KRIMP compression algorithm \citep{van2006compression} and Kullback-Leibler Divergence --- they tended to agree with \citet{kestemont2017did, kestemont2017van}, even though the KRIMP attributed the \textit{Wilhelmus} to another author when a different feature selection method was used. Thus, the results were inconclusive, with a tendency to prefer Datheen. Our paper  provides further evidence in favour of attributing the authorship to {\it Datheen}.
 
\subsection{Data and methods}\label{ER-data}

We use a total of 186 documents by six writers, consisting of 35 documents written by Datheen, 46 by Marnix, 23  by Heere, 35  by Haecht, 33  by Fruytiers, and 14  by Coornhert. These documents  contain tag-lemma pairs as terms, obtained through part-of-speech tagging and lemmatizing of the texts, and are made publicly available by \citet{KESTEMONT201686, kestemont2017did, kestemont2017van}. The average marginal frequencies range from 406 for documents by Fruytiers to 545 for documents by Haecht. See \citet{mikekestemont2017} for more details regarding the dataset. Similar to Section~\ref{DCpar}, in this section we also use visualization and distance measures to compare LSA and CA.

\subsection{Visualization}\label{Wproject2}

\begin{figure}[htbp]
    \centering
       \subfigure[]{
       \label{FDathMarnpredictwilhelmusLSARAW}
         \begin{minipage}[t]{0.41\linewidth}
         \centering
         \includegraphics[width=1\textwidth]{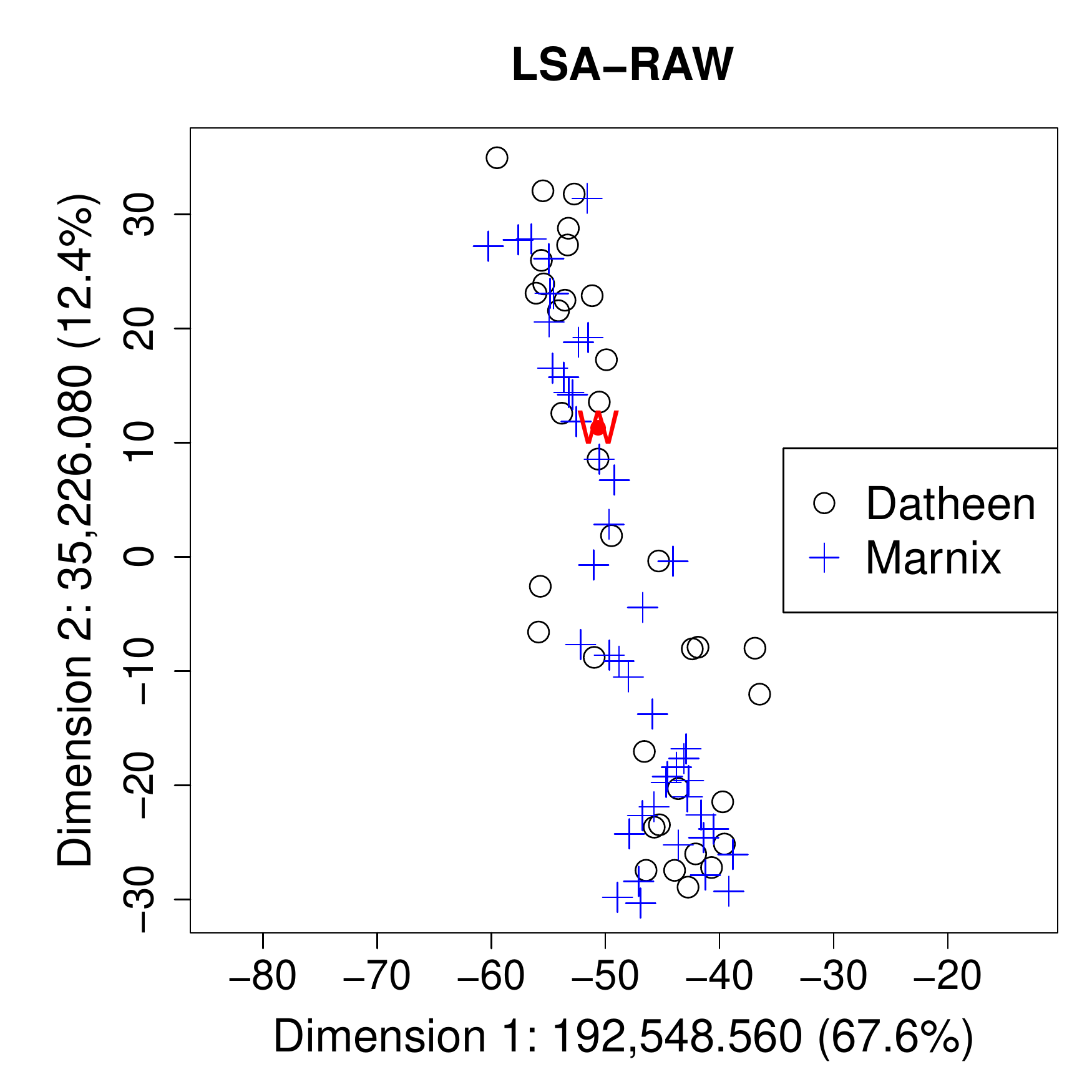}
         \end{minipage}
         }
     \subfigure[]{
       \label{FDathMarnpredictwilhelmusLSANROWL1}
         \begin{minipage}[t]{0.41\linewidth}
         \centering
         \includegraphics[width=1\textwidth]{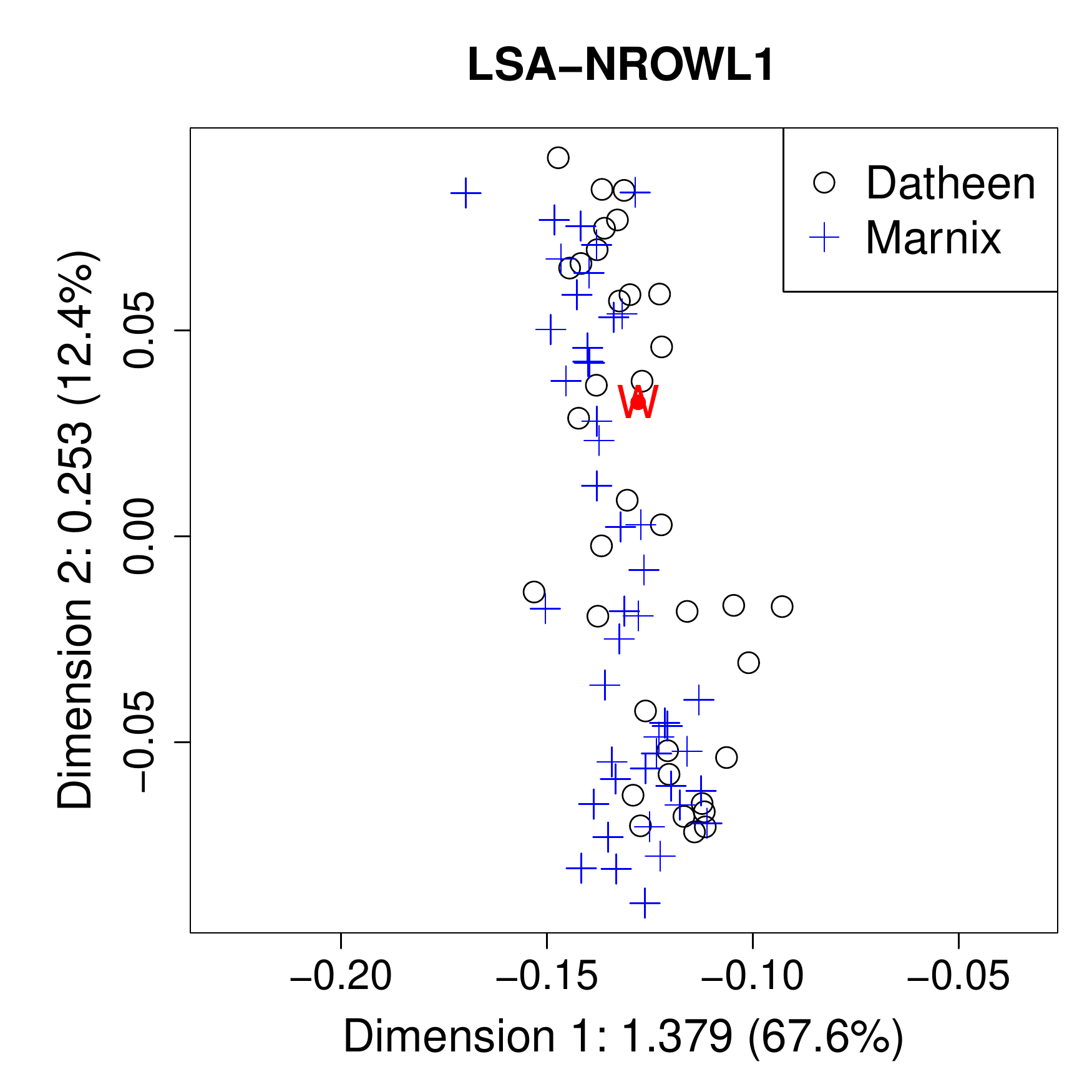}
         \end{minipage}
         }
       \subfigure[]{
       \label{FDathMarnpredictwilhelmusLSANROW}
         \begin{minipage}[t]{0.41\linewidth}
         \centering
         \includegraphics[width=1\textwidth]{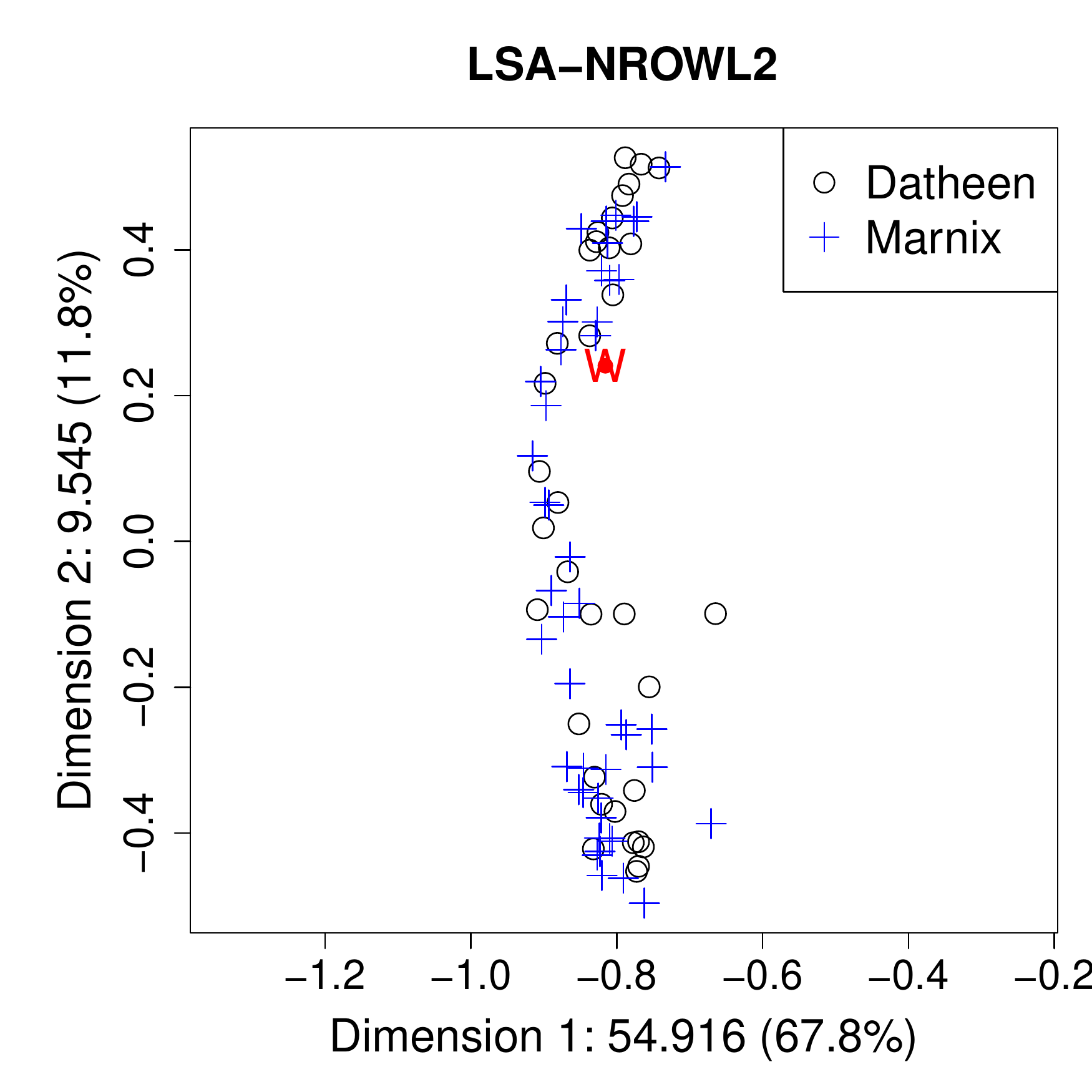}
         \end{minipage}
         }
       \subfigure[]{
       \label{FDathMarnpredictwilhelmusLSATFIDF}
         \begin{minipage}[t]{0.41\linewidth}
         \centering
         \includegraphics[width=1\textwidth]{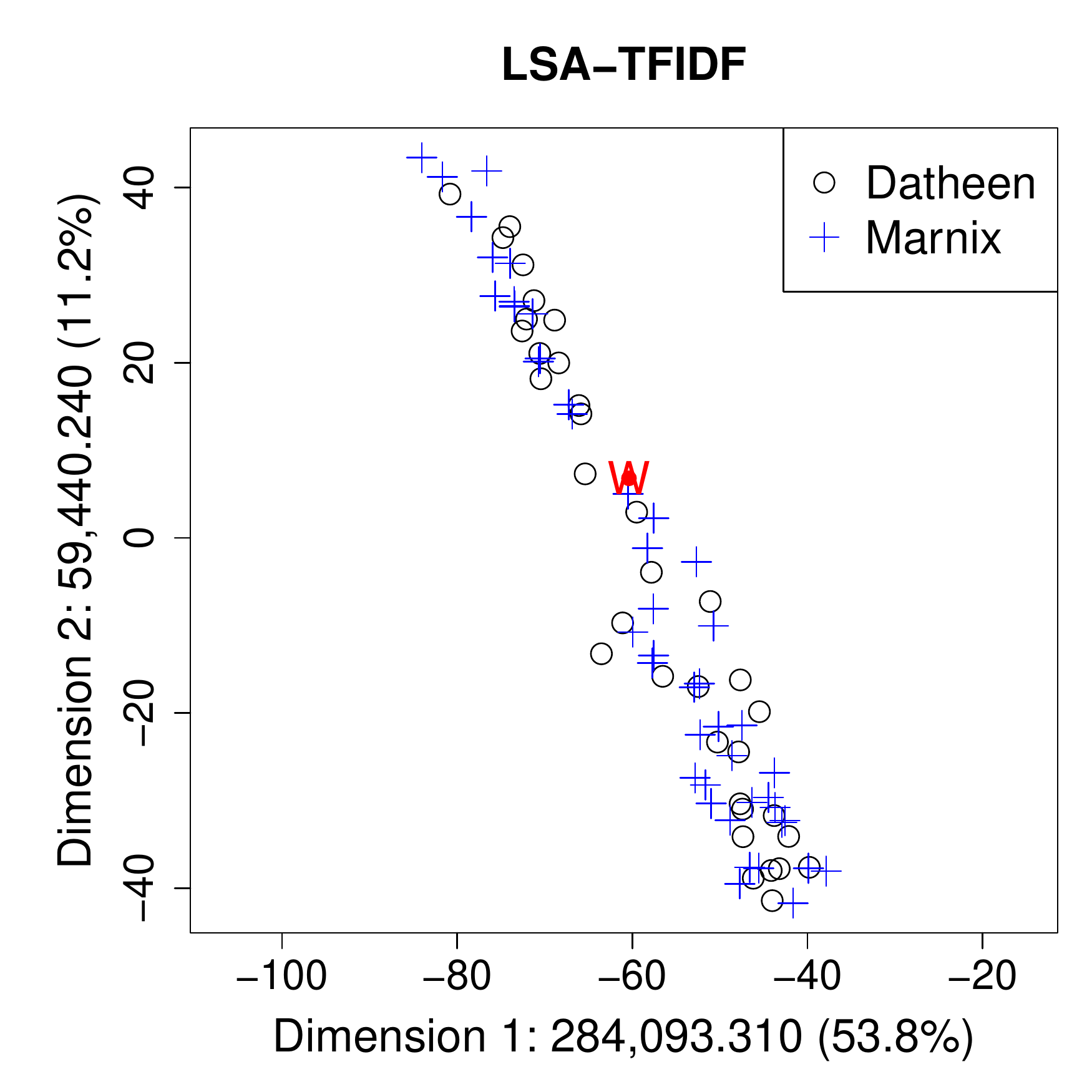}
         \end{minipage}
         }
       \subfigure[]{
       \label{FDathMarnpredictwilhelmusCA}
         \begin{minipage}[t]{0.41\linewidth}
         \centering
         \includegraphics[width=1\textwidth]{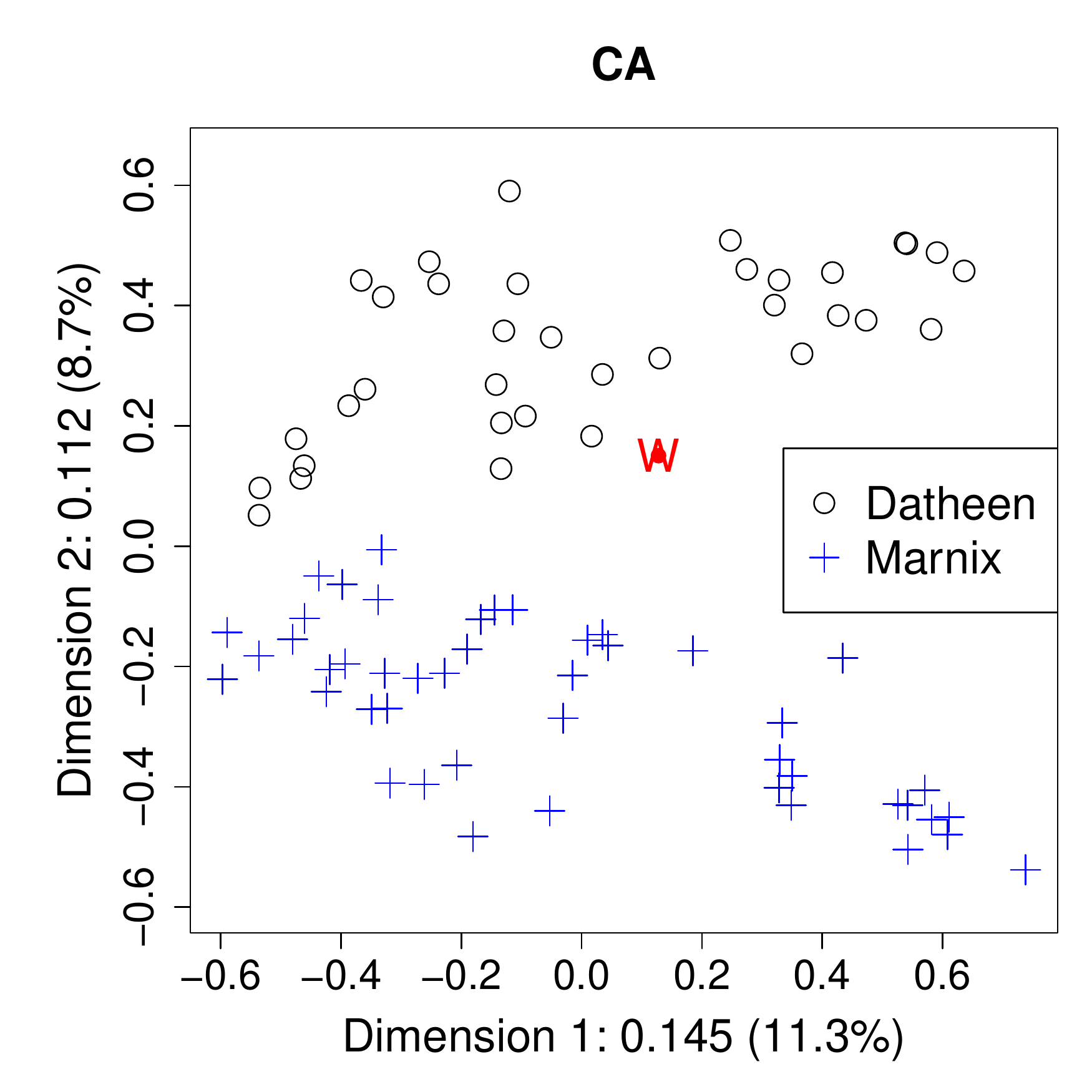}
         \end{minipage}
         }
         \vspace{8mm}
    \caption{The first two dimensions for each document of author Datheen and author Marnix, and the \textit{Wilhelmus} (in red) by (a) LSA-RAW; (b) LSA-NROWL1; (c) LSA-NROWL2; (d) LSA-TFIDF; (e) CA.}
    \label{FWilhelmusbiplotDathMarn}
\end{figure}

We first examine all documents of two authors Marnix and Datheen\footnote{We chose these two authors specifically, out of our dataset, as they are the two main contenders for the authorship of \textit{Wilhelmus} -- Marnix has been the most popular candidate from qualitative analysis, and since the work of \citet{kestemont2017did, kestemont2017van} Datheen is also a serious candidate.}, using the 300 most frequent tag-lemma pairs. These form a document-term matrix of size $81 \times 300$. Figure~\ref{FWilhelmusbiplotDathMarn} shows the results of analyzing this document-term matrix using the four LSA methods (LSA-RAW, LSA-NROWL1, LSA-NROWL2, LSA-TFIDF), and CA. The \textit{Wilhelmus} document is not included in the data matrix but it is projected into the solutions for illustrative purposes by W, in red, see Section~\ref{Sub: OSD} and Section~\ref{S:CA}.
As seen in Figure~\ref{FWilhelmusbiplotDathMarn}, all four varieties of LSA fail to show a clear separation, while CA separates documents by the two authors clearly, even though 
the first 2 dimensions for CA account for a much smaller percentage of the total sum of squared singular values than the first 2 dimensions for the four LSA methods. This is because the margins play an important role in the first two dimensions for the four LSA methods and the relations between documents are blurred by these margins. We also see that in CA the \textit{Wilhelmus} is clearly attributed to Datheen.

\begin{figure}[htbp]
    \centering
       \subfigure[]{
       \label{FDathHeerpredictwilhelmusCA}
         \begin{minipage}[t]{0.41\linewidth}
         \centering
         \includegraphics[width=1\textwidth]{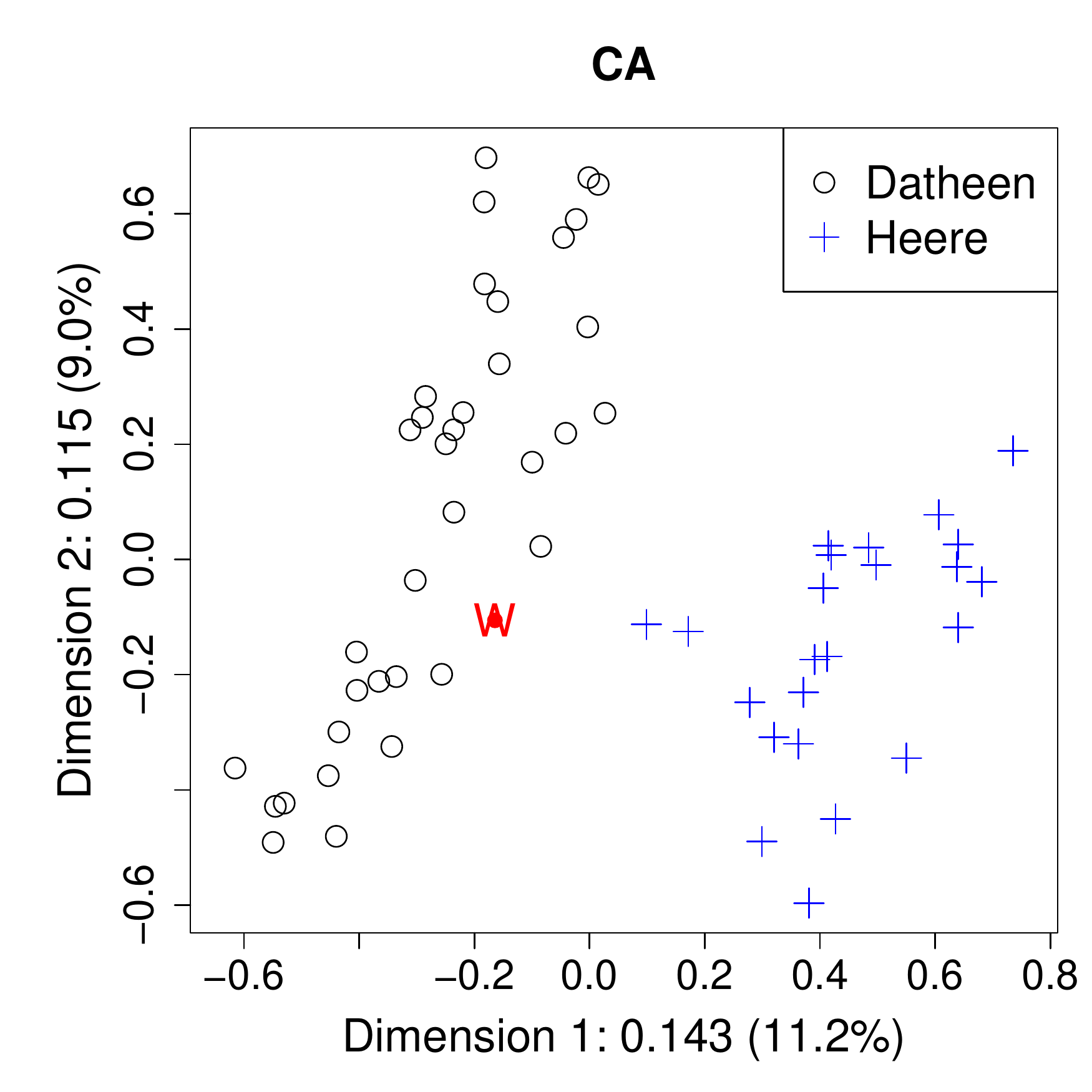}
         \end{minipage}
         }
       \subfigure[]{
       \label{FDathHaecpredictwilhelmusCA}
         \begin{minipage}[t]{0.41\linewidth}
         \centering
         \includegraphics[width=1\textwidth]{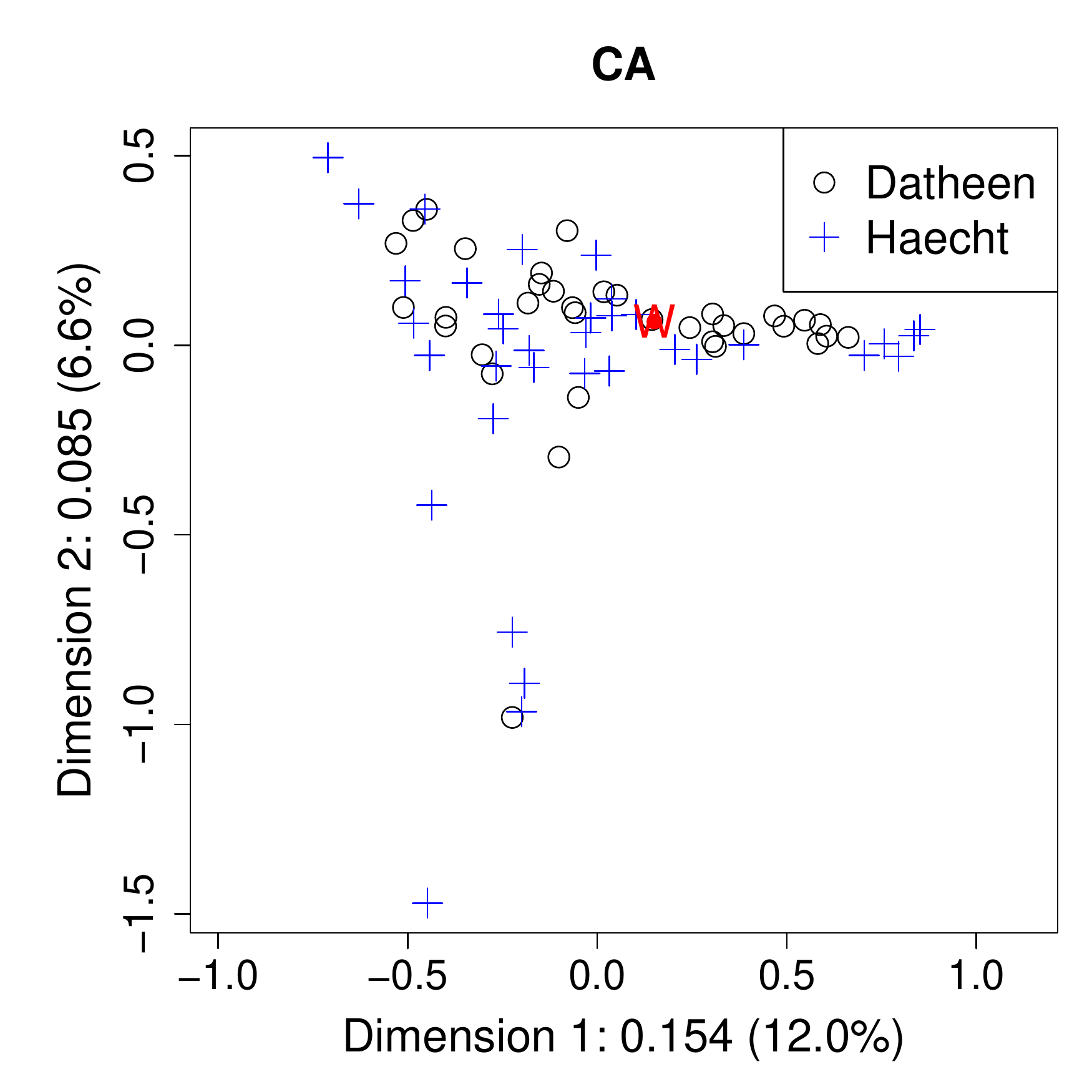}
         \end{minipage}
         }
       \subfigure[]{
       \label{FDathFruypredictwilhelmusCA}
         \begin{minipage}[t]{0.41\linewidth}
         \centering
         \includegraphics[width=1\textwidth]{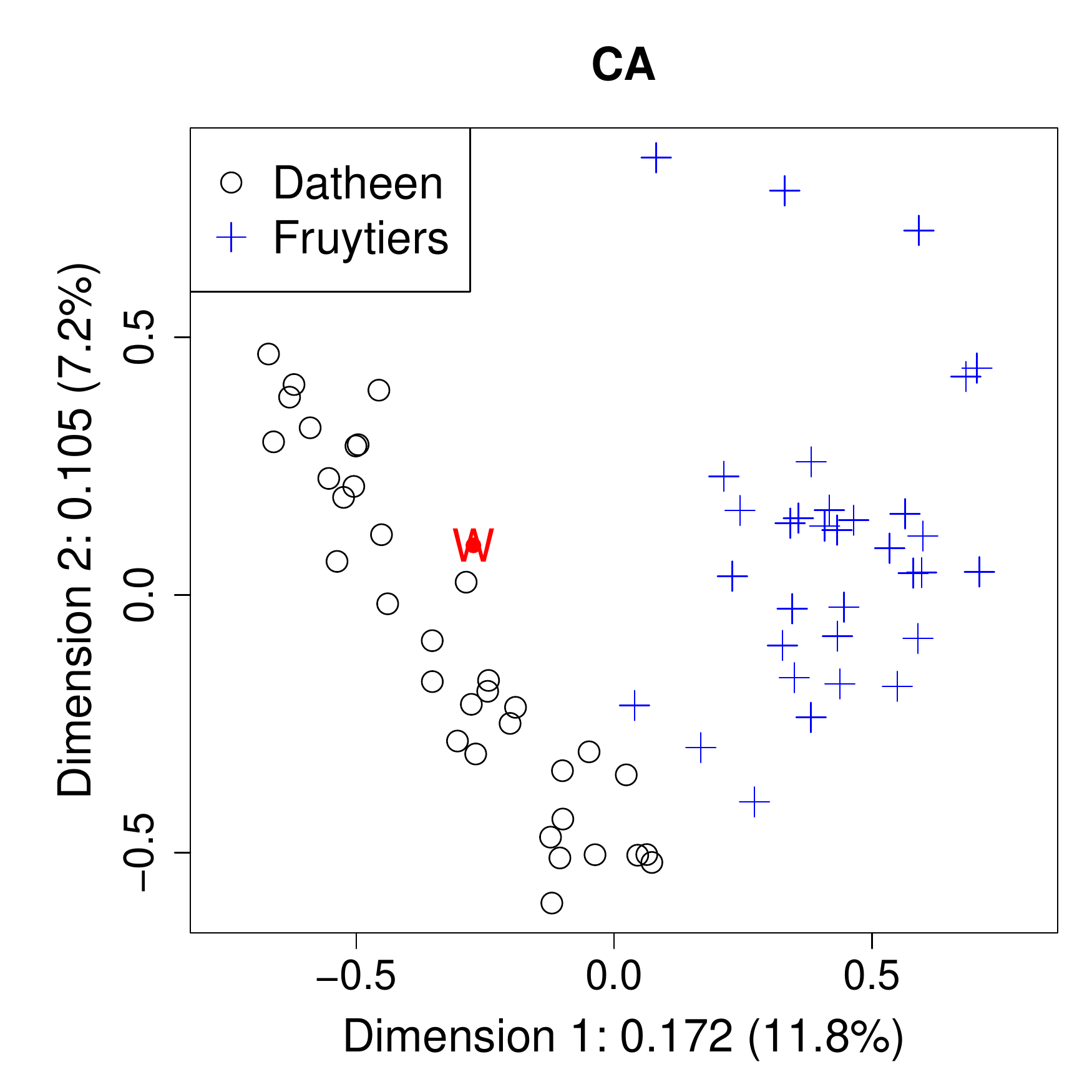}
         \end{minipage}
         }
       \subfigure[]{
       \label{FDathCoorpredictwilhelmusCA}
         \begin{minipage}[t]{0.41\linewidth}
         \centering
         \includegraphics[width=1\textwidth]{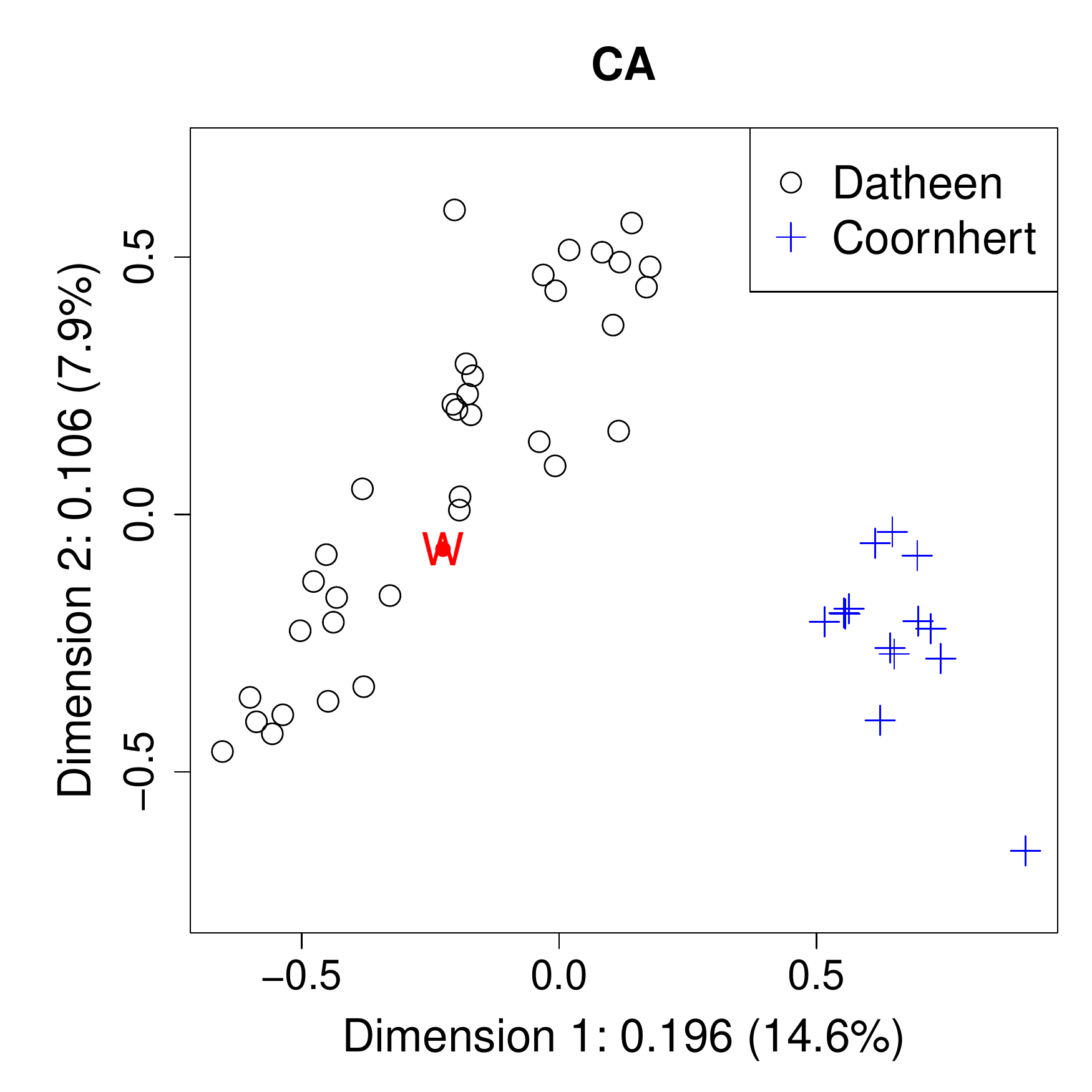}
         \end{minipage}
         }
         \vspace{8mm}
    \caption{The first two dimensions for each document of author Datheen and another author, and the \textit{Wilhelmus} (in red) using CA: (a) Heere; (b) Haecht; (c) Fruytiers; (d) Coornhert.} 
    \label{FWilhelmusbiplotDathCA} 
\end{figure}

Given the effectiveness of CA and the attribution of the \textit{Wilhelmus} to Datheen in the above analysis, we now show visualizations of CA for documents by Datheen and four other authors in turn (Figure~\ref{FWilhelmusbiplotDathCA}).
For three out of four authors, there is a clear separation between that author and Datheen.  In the case Haecht however (sub-figure (b)), there is no clear separation from Datheen. In all three cases where there is a clear separation, \textit{Wilhelmus} is attributed to Datheen, as before.

\begin{figure}[htbp]
    \centering
       \subfigure[]{
       \label{FpredictwilhelmusLSARAW}
         \begin{minipage}[t]{0.41\linewidth}
         \centering
         \includegraphics[width=1\textwidth]{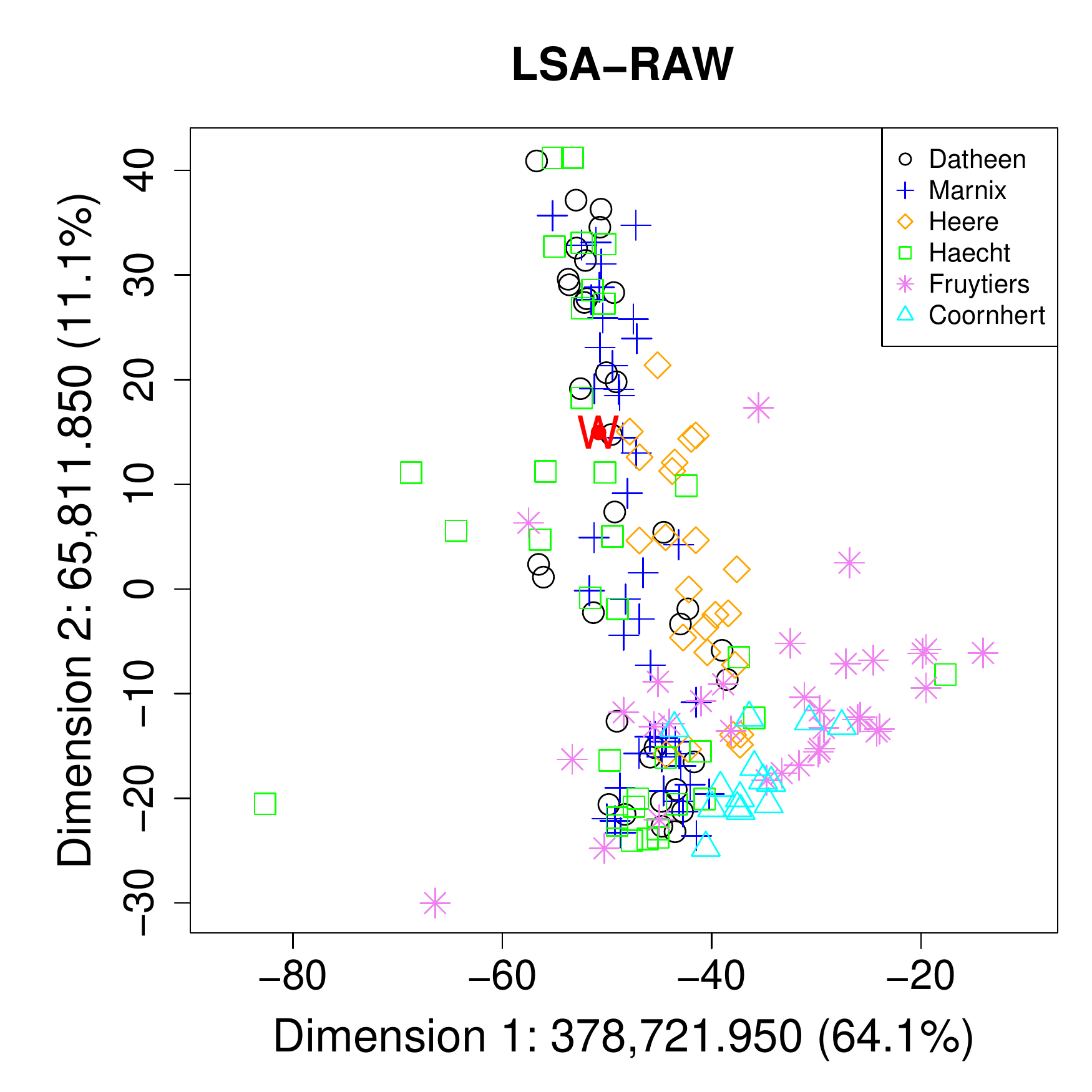}
         \end{minipage}
         }
      \subfigure[]{
      \label{FpredictwilhelmusLSANROWL1}
        \begin{minipage}[t]{0.41\linewidth}
        \centering
        \includegraphics[width=1\textwidth]{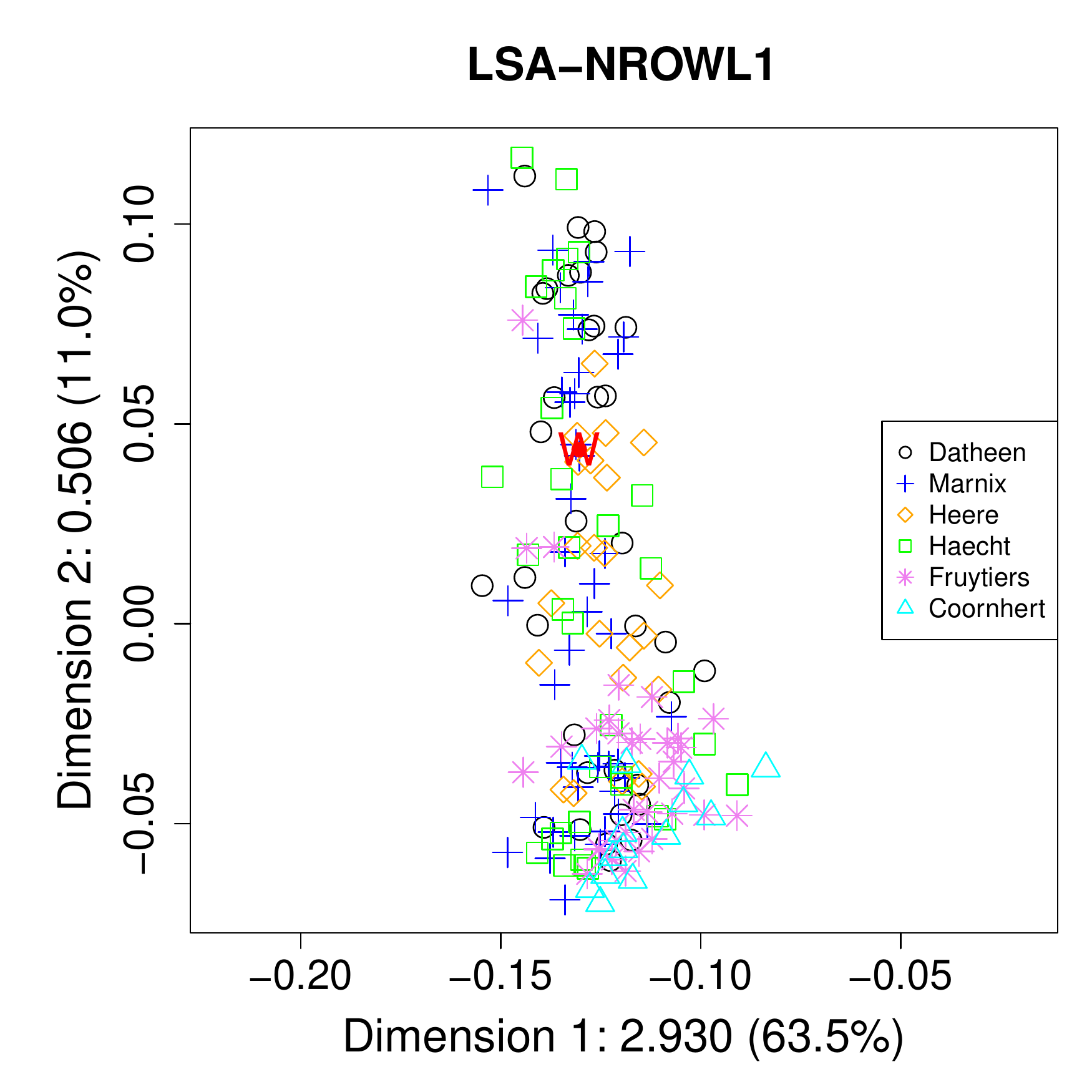}
        \end{minipage}
        }
      \subfigure[]{
       \label{FpredictwilhelmusLSANROW}
         \begin{minipage}[t]{0.41\linewidth}
         \centering
         \includegraphics[width=1\textwidth]{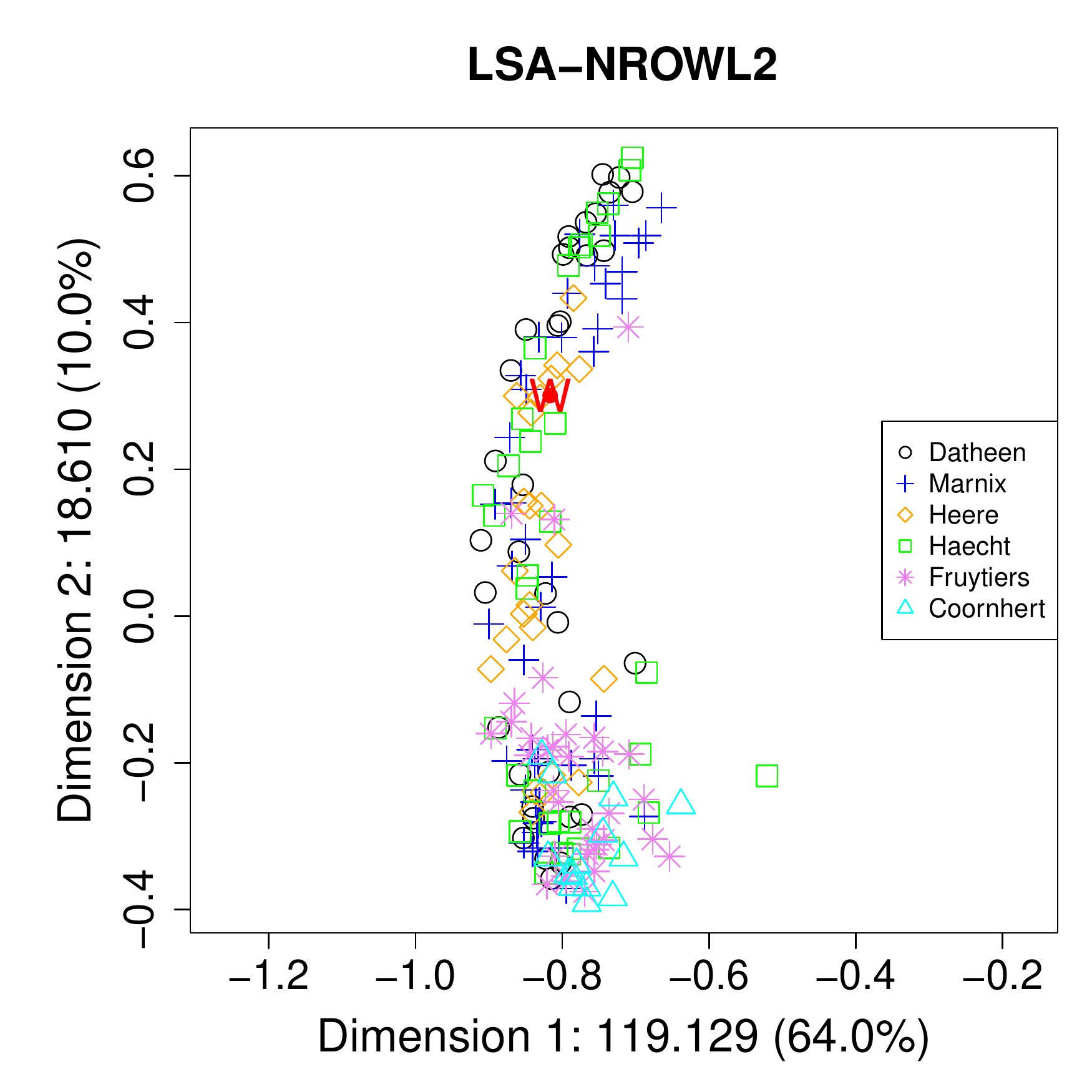}
         \end{minipage}
         }
       \subfigure[]{
       \label{FpredictwilhelmusLSATFIDF}
         \begin{minipage}[t]{0.41\linewidth}
         \centering
         \includegraphics[width=1\textwidth]{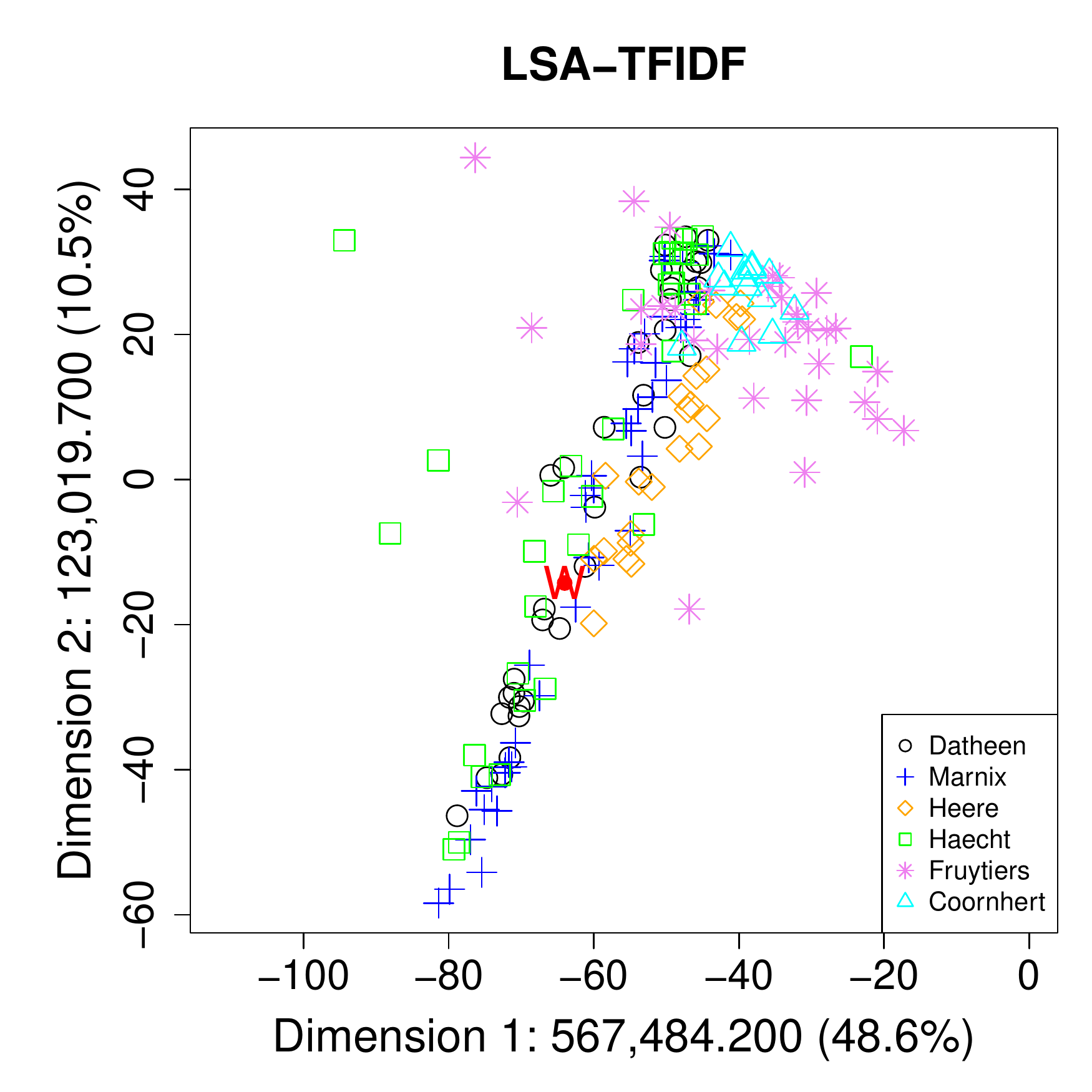}
         \end{minipage}
         }
      \subfigure[]{
       \label{FpredictwilhelmusCA}
         \begin{minipage}[t]{0.41\linewidth}
         \centering
         \includegraphics[width=1\textwidth]{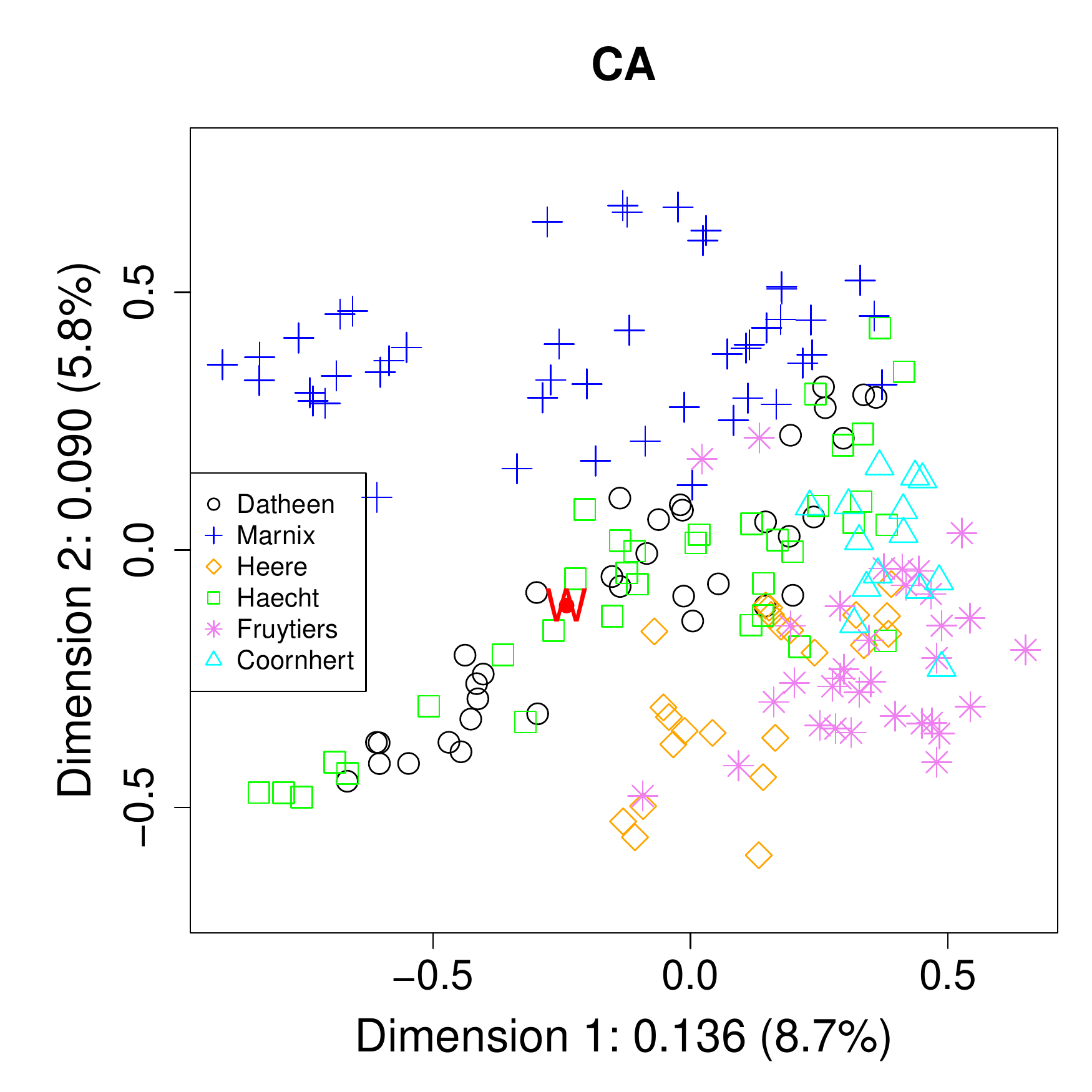}
         \end{minipage}
         }
         \vspace{8mm}
    \caption{The first two dimensions for each document of six authors, and the \textit{Wilhelmus} (in red) by (a) LSA-RAW; (b) LSA-NROWL1; (c) LSA-NROWL2; (d) LSA-TFIDF; (e) CA.}
    \label{FWilhelmusbiplot6authors}
\end{figure}

Finally, we apply all four varieties of LSA and CA to all documents of the six authors, which form a document-term matrix of size $186 \times 300$. Figure~\ref{FWilhelmusbiplot6authors} shows the results of the analysis of this matrix by LSA-RAW, LSA-NROWL1, LSA-NROWL2, LSA-TFIDF, and CA. The \textit{Wilhelmus} is projected into the solutions afterwards. Again we find that, although the percentage of the total sum of squared singular values in the first two dimensions for CA is lower than the four LSA methods, CA separates the documents quite well compared with the four LSA methods. For instance, documents written by Marnix are effectively separated from the documents written by other authors. The documents of the other authors also seem to form much more distinguishable clusters, as compared to LSA, except for Datheen and Haecht.

\subsection{Distance measures}\label{Sub: DM}

To evaluate LSA methods and CA, we use leave-one-out cross-validation (LOOCV) \citep{gareth2021introduction} with the 186 documents of six authors. Using LOOCV, each time we discern the following four steps. At the first step, a single document of the 186 documents is used as the validation set and the remaining 185 documents make up the training set. The 185 documents of training set form a document-term matrix with 185 rows and 300 columns. At step two, we perform LSA-RAW, LSA-NROWL1, LSA-NROWL2, LSA-TFIDF, and CA on this document-term matrix to obtain the coordinates of the 185 documents. The single document of validation set is projected into the solutions, see Section~\ref{Sub: OSD} and Section~\ref{S:CA}. At step three, using the centroid, average, single, and complete method, the distance is computed between the single document and the six author groups of documents. For this single document, the predicted author of the document is the author with the smallest distance. At the final step, we compare the predicted author with the true author of the single document. We repeat this 186 times, once for each single documents. The accuracy is calculated by the ratio: number of times an author is correctly predicted divided by 186.

\begin{table}[t]
\centering  
\caption{The minimum optimal dimensionality $k$ and the accuracy in $k$ for LSA-RAW, LSA-NROWL1, LSA-NROWL2, LSA-TFIDF, and CA, and the accuracy for RAW using different distance measurement methods with {\it Wilhelmus} dataset.} 
\label{Tcentroidaveragesinglecomplete}
\begin{tabular}{ccccccccc}    
\hline
\multirow{2}{*}{Methods}&\multicolumn{2}{c}{Centroid} &\multicolumn{2}{c}{Average} &\multicolumn{2}{c}{Single}&\multicolumn{2}{c}{Complete}
\\
&$k$&Accuracy&$k$&Accuracy&$k$&Accuracy&$k$&Accuracy\\\hline 
RAW& 
 & 0.720&&0.522&&0.672&&0.177\\
LSA-RAW &51& 0.720&70&0.554&14&0.720&1&0.296 \\
LSA-NROWL1&93& 0.731&116&0.645&22&0.710&75&0.226\\
LSA-NROWL2&59 & 0.742&41&0.699&21&0.715&77&0.301\\
LSA-TFIDF&84& 0.720& 90&0.538&23&0.731&1&0.231\\
CA&151&\underline{{\bf 0.930}}&12&{\bf 0.790}&19&{\bf 0.785}&95&{\bf 0.452}\\
\hline 
\end{tabular}  
\end{table} 

Table~\ref{Tcentroidaveragesinglecomplete} shows the maximum accuracy for LSA-RAW, LSA-NROWL1, LSA-NROWL2, LSA-TFIDF, and CA for the four distance measures \footnote{For {\it Wilhelmus} dataset, we explore the number of all dimensions of dimensionality reduction methods}, along with the minimum optimal dimensions $k$. First, CA yields the maximum accuracy for all distance measurement methods as compared to the RAW method as well as all four LSA methods.  Second, CA with the centroid method provides the highest accuracy. 

\begin{figure}[htbp] 
    \centering
       \subfigure[]{
       \label{FCentroidfulldimension}
         \begin{minipage}[t]{0.41\linewidth}
         \centering
         \includegraphics[width=1\textwidth]{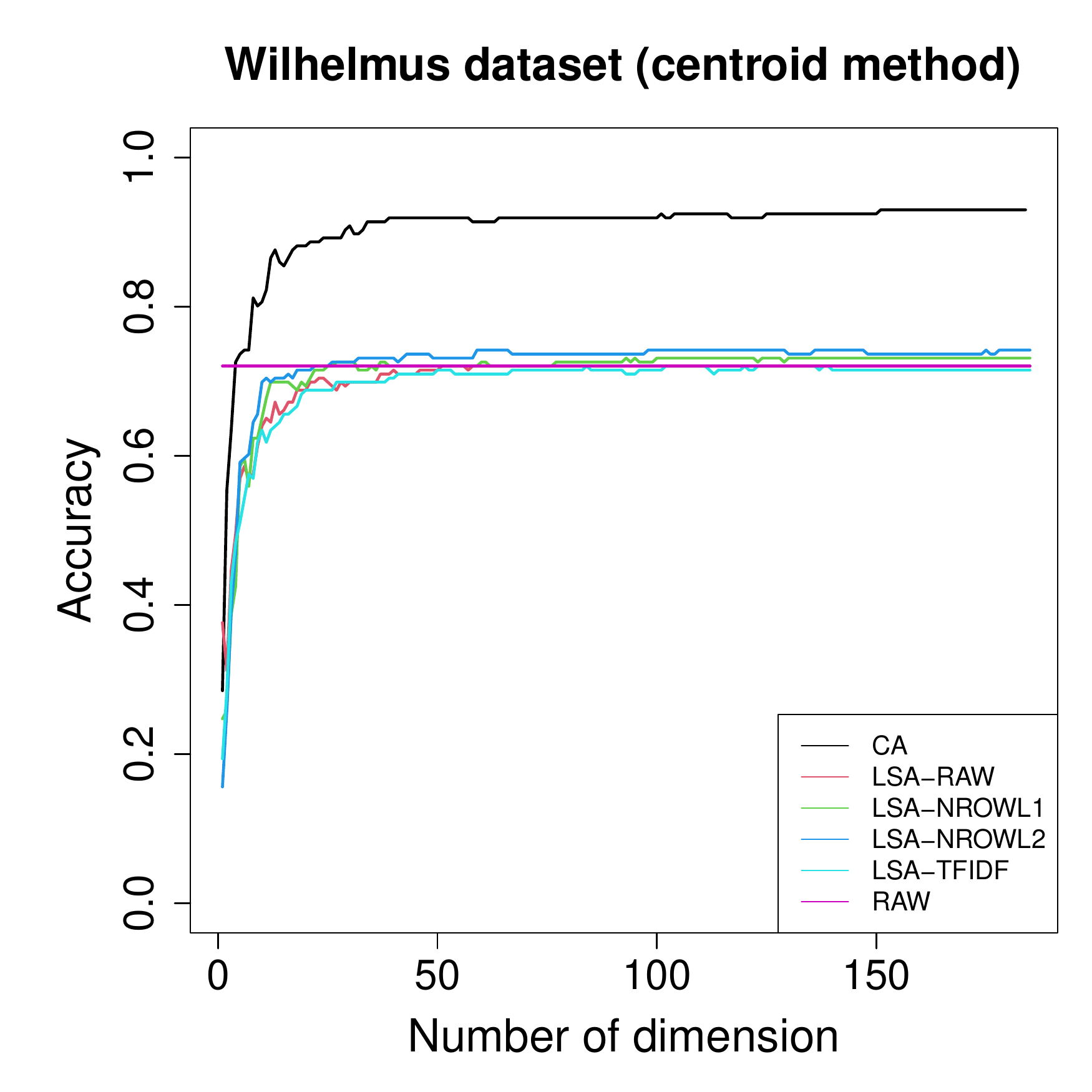}
         \end{minipage}
         }
      \subfigure[]{
       \label{FCentroidpartdimension}
         \begin{minipage}[t]{0.41\linewidth}
         \centering
         \includegraphics[width=1\textwidth]{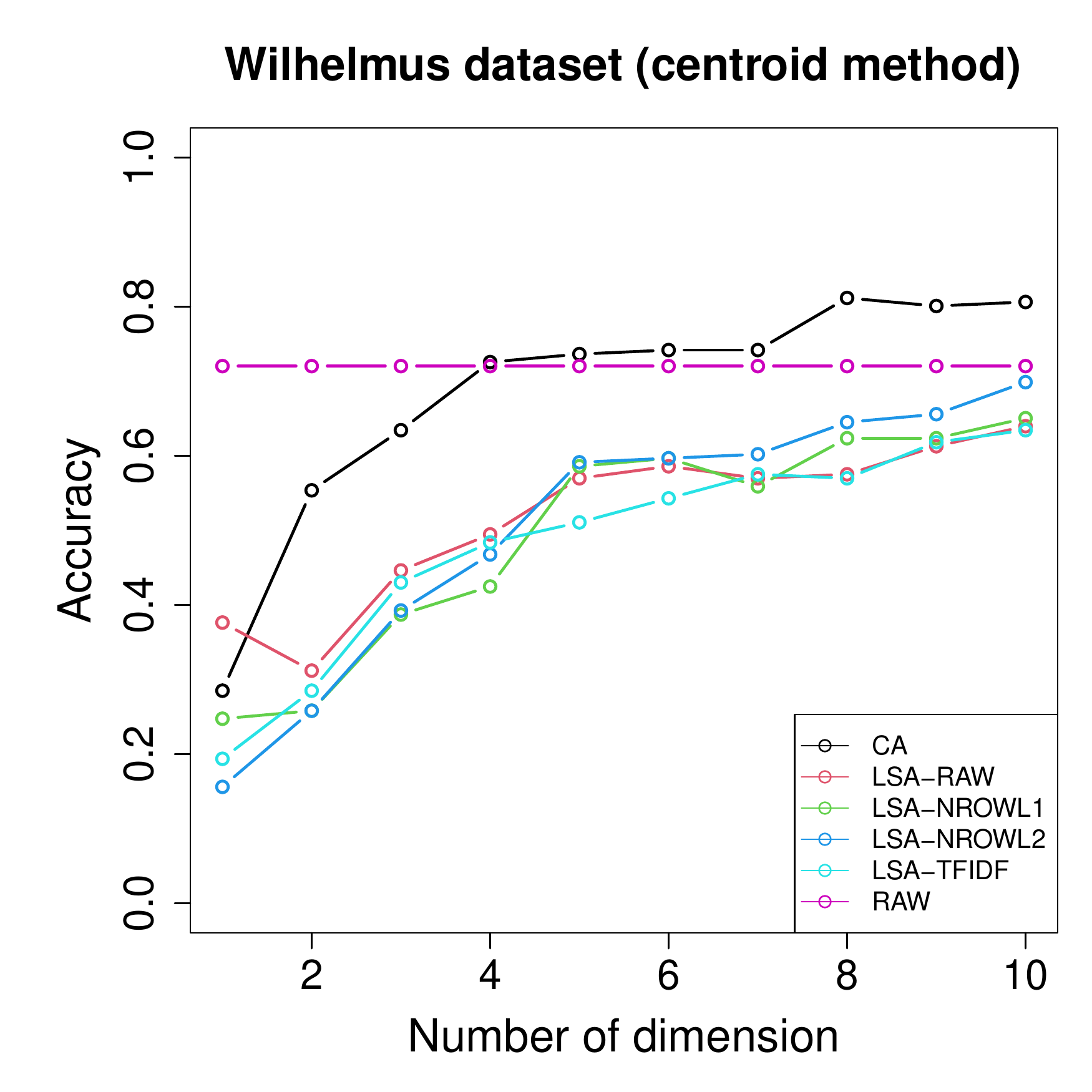}
         \end{minipage}
         }
    \vspace{8mm}
    \caption{Accuracy versus the number of dimensions (centroid method) for CA, RAW, LSA-RAW, LSA-NROWL1, LSA-NROWL2, and LSA-TFIDF with \textit{Wilhelmus} dataset.}
    \label{CentroidWilhelmus}
\end{figure}

In order to further explore the centroid method, Figure~\ref{CentroidWilhelmus} shows the accuracy with different numbers of dimensions for LSA-RAW, LSA-NROWL1, LSA-NROWL2, LSA-TFIDF, and CA. Figure~\ref{FCentroidfulldimension} displays all dimensions on the horizontal axis, and  Figure~\ref{FCentroidpartdimension} focuses on the first 10 dimensions. CA in combination with the centroid method performs better than the other methods almost irrespective of dimension, except for the very first ones. Also, the accuracy of CA in combination with the centroid method is very high over a large range.

\subsection{Authorship attribution of the \textit{Wilhelmus}} 

Since CA in combination with the centroid method appears to be the best overall, we use them to determine the authorship of the \textit{Wilhelmus}. In the 34 optimal dimensions (dimensions 151-184), we find that the \textit{Wilhelmus} is attributed to the author Datheen, while Haecht is the second most likely candidate. The distance of the \textit{Wilhelmus} to the centroid of documents of Datheen averaged across 34 optimal dimensions is 0.825, to Haecht 0.880, to Marnix 0.939, to Heere 1.015, to Fruytiers 1.064, and to Coornhert 1.253. Thus, CA attributes \textit{Wilhelmus} to Datheen, and provides more weight using an independent statistical technique, to prior results by \citet{kestemont2017did, kestemont2017van} in resolving this debate.

\section{Conclusion}\label{S:FR} 

LSA and CA both allow for dimensionality reduction by the SVD of a matrix; however the actual matrix analyzed by LSA and CA is different, and therefore LSA and CA capture different kinds of information. In LSA we apply an SVD to $\bm{F}$, or to a weighted $\bm{F}$. In CA, an SVD is applied to the matrix $\bm{D}_r^{-\frac{1}{2}}(\bm{P}-\bm{E})\bm{D}_{c}^{-\frac{1}{2}}$ of standardized residuals. The elements in $\bm{D}_r^{-\frac{1}{2}}(\bm{P}-\bm{E})\bm{D}_{c}^{-\frac{1}{2}}$ display the departure from the margins, that is, departure from the expected frequencies under independence collected in $\bm{E}$. Due to $\bm{E}$, in CA the effect of the margins is eliminated --- a solution only displays the dependence between documents and terms. Concluding, in LSA, the effect of the margins as well as the dependence is part of the matrix that is analyzed and these margins usually play a dominant role in the first dimension of the LSA solution as usually on the first dimension all points depart in the same direction from the origin. On the other hand, in CA all points are scattered around the origin and the origin represents the profile of the row and column margins of $\bm{F}$. 

In summary, although LSA allows a study of the relations between documents, between terms, and between documents and terms, this study is not easy. The reason is that these relations are blurred by the effect of the margins that are also displayed in the LSA solution. CA does not have this property. Therefore it appears that CA is a better tool for studying the relations between documents, between terms, and between documents and terms. Also, discussed in Section~\ref{S:CA}, CA has many nice properties like providing a geometric display where the Euclidean distances approximate
the $\chi^2$-distances between the rows and between the columns of the matrix, and the relation to the Pearson $\chi^2$ statistic. Overall, from a theoretical point of view it appears that CA has more attractive properties than LSA. Empirically, we evaluated and compared the two methods on text categorization in English and authorship attribution in Dutch, and found that CA can both separate documents better visually, and obtain higher accuracies on text categorization and authorship attribution as compared to LSA techniques. 

A document-term matrix is similar to a word-context matrix, commonly used to represent word meanings, in the sense that it is also a matrix of counts. However, in the context of word-context matrices the ways in which the counts are transformed are usually different from the way they are transformed for document-term matrices, and therefore, due to space limitations, we defer a comparison of CA and LSA of word-context matrices to future work. In the future, it is also interesting to compare word embeddings learned by LSA based methods and CA to more recent static word embedding approaches such as Word2Vec and GloVe, or even contextualized word embeddings modes like BERT. And it is interesting to compare LSA based methods and CA on recent classifiers, such as neural network models.

\section*{Acknowledgments}

Author Qianqian Qi is supported by the China Scholarship Council.

\bibliography{references.bib}

\end{document}